\documentclass[a4paper,fleqn]{cas-sc}

\usepackage[authoryear]{natbib}

\usepackage{multicol}
\usepackage{multirow}
\usepackage{booktabs}
\usepackage{hyperref}
\usepackage{color}
\usepackage{algorithm}
\usepackage{algpseudocode}
\usepackage{listings}
\usepackage{xcolor}
\usepackage{amssymb}
\usepackage{bm}
\usepackage{amsmath}
\usepackage{graphicx}
\usepackage{placeins}
\usepackage{subfig}

\usepackage{pifont}
\usepackage[dvipsnames]{xcolor}
\usepackage{tikz}
\usetikzlibrary{arrows.meta,calc,positioning}

\usepackage{comment}
\usepackage{tcolorbox}
\usepackage{rotating}
\usepackage{textpos}

\definecolor{codegreen}{rgb}{0,0.6,0}
\definecolor{codegray}{rgb}{0.5,0.5,0.5}
\definecolor{codepurple}{rgb}{0.58,0,0.82}
\definecolor{backcolour}{rgb}{0.95,0.95,0.92}
\lstdefinestyle{cppstyle}{
    language=C++,
    backgroundcolor=\color{backcolour},   
    commentstyle=\color{codegreen},
    keywordstyle=\color{magenta},
    numberstyle=\tiny\color{codegray},
    stringstyle=\color{codepurple},
    basicstyle=\ttfamily\scriptsize, 
    numbers=left,
    breakatwhitespace=false,         
    breaklines=true,                 
    captionpos=b,                    
    keepspaces=true,                 
    numbersep=5pt,                  
    showspaces=false,                
    showstringspaces=false,
    showtabs=false,                  
    tabsize=2
}

\newcommand{\optionaltikzinput}[1]{%
  \IfFileExists{#1}{%
    \input{#1}%
  }{%
    \fbox{%
      \begin{minipage}{0.78\linewidth}
        \centering
        \vspace{1.2em}
        Missing TikZ figure source\\[0.4em]
        \texttt{\detokenize{#1}}
        \vspace{1.2em}
      \end{minipage}%
    }%
  }%
}

\begin{document}
\let\WriteBookmarks\relax
\def\floatpagepagefraction{1}
\def\textpagefraction{.001}

\shorttitle{TPIC: Energy-Conserving Space Plasma Simulations on GPU Supercomputers}

\shortauthors{L. Pennati et al.}

\title [mode = title]{TPIC: Energy-Conserving Space Plasma Simulations on GPU Supercomputers}

\author[1]{Luca Pennati}[orcid=0009-0009-4901-1716]
\cormark[1]
\ead{pennati@kth.se}

\author[1]{Ivy Peng}[orcid=0000-0003-4158-3583]
\author[1]{Stefano Markidis}[orcid=0000-0003-0639-0639]

\affiliation[1]{organization={KTH Royal Institute of Technology},
city={Stockholm},
country={Sweden}}

\cortext[1]{Corresponding author.}

\begin{abstract}
Artificial heating and numerical damping affect particle energization in kinetic space plasma simulations, even when the numerical scheme remains stable. Implicit and semi implicit energy conserving Particle-in-Cell (PIC) methods eliminate these errors in the discrete energy balance while relaxing explicit stability constraints. We present TPIC, a massively parallel implementation of the Energy Conserving Semi-Implicit Method (ECSIM) for GPU supercomputers. We extend the exact energy conservation in the ECSIM formulation to closed domains with conducting boundaries by deriving and canceling a spurious term in the discrete Poynting flux theorem. TPIC executes the complete particle and field cycle on NVIDIA and AMD GPUs, combining matrix-engine acceleration with asynchronous communication and computation. We validate TPIC in a variety of electrostatic and electrodynamic test cases relevant for space plasma physics applications, including quasineutral plasma expansion, two-stream instability, Whistler waves, Weibel instability, Whistler turbulence, and three-dimensional magnetic reconnection with conducting walls and 12.3 billion particles. We show that in all scenarios, TPIC agrees with the corresponding analytical theory or reference numerical results, and preserves the total system energy to machine precision. On 4,096 AMD GPUs, TPIC advances 1.9 trillion particles with 96.8\% weak scaling efficiency.

\end{abstract}




\maketitle

\section{Introduction}
\label{sec:introduction}
Kinetic processes in space plasmas span a broad hierarchy of spatial and temporal scales, from electron plasma oscillations and Debye shielding to ion gyromotion, magnetic reconnection, and the global dynamics of the solar wind and Earth's magnetosphere~\citep{lapenta2012pic_spaceWeather, Karimabadi2014shocks_turbulence_reconnection}. The electron Debye length and skin depth can be orders of magnitude smaller than the characteristic dimensions of current sheets and other macroscopic plasma structures. Nevertheless, these scales are coupled through collective electromagnetic fields and particle dynamics~\citep{lapenta2006_celeste3d, kiyani2015micromacro}. For instance, magnetic reconnection evolves on ion and macroscopic timescales, while electron kinetics controls important aspects of energy conversion and particle acceleration~\citep{yamada2010magneticReconnection_review, hesse2020reconnection_spaceSciences, burch2016reconnection_electronScales}. Resolving this hierarchy with conventional explicit Particle-in-Cell (PIC) methods is computationally demanding, since numerical stability and finite grid constraints impose restrictions on the time step and spatial resolution associated with the fastest and smallest plasma scales, even when the phenomena of interest evolve on much larger scales~\citep{HockneyEastwood1988,birdsall1991_pic}.

Implicit PIC methods relax these numerical stability restrictions, allowing the temporal and spatial resolution to be chosen according to the physical processes of interest. However, numerical stability alone is insufficient for reliable kinetic simulations. Conventional explicit PIC schemes can exhibit artificial numerical heating associated with particle and grid coupling and finite grid instabilities~\citep{langdon1970spatialGrid, okuda1972spatialGrid_noise, barnes2021finiteGrid}. On the other hand, many implicit PIC formulations, including moment-implicit~\citep{mason1981_implicitpic, brackbill1982_imm} and direct-implicit~\citep{langdon1983directImplicit} schemes, introduce numerical damping that leads to artificial cooling~\citep{cohen1989performance_direct_implicit}. These errors can accumulate over long simulations, modifying the particle energy distributions, plasma temperature, and energy exchange between particles and electromagnetic fields. This is particularly problematic in collisionless space plasmas, where energy conversion and particle energization are often central quantities of interest. Energy conserving implicit methods address this problem by enforcing the balance between particle kinetic energy and electromagnetic field energy at the discrete level, eliminating spurious energy gain or loss associated with their numerical coupling~\citep{markidis2011_energyconserving, chen2011energyconserving, lapenta2017ecsim}.

The practical application of energy conserving implicit PIC methods to large scale space plasma simulations requires the method to accommodate several important issues. These includes accurate boundary conditions, efficiently handling billions of computational particles, and solving the coupled particle and field equations on massively parallel computing systems. Modern GPU supercomputers offer the computational resources required for these simulations. However, exploiting them efficiently demands algorithms and data structures adapted to accelerator architectures and distributed memory execution. Developing a production quality implementation that combines energy conservation, realistic simulation domains, and scalability across thousands of GPUs is the topic of this paper.

Large-scale kinetic simulations of space and astrophysical plasmas have been enabled by massively parallel explicit PIC codes such as VPIC~\citep{Bowers2008vpic,bird2022vpic}, EMPIC~\citep{wang1995empic,Gary2012whistlerTurbulence}, PSC~\citep{germaschewski2016psc}, OSIRIS~\citep{fonseca2002osiris,lee2025osiris_gpu}, WarpX~\citep{vay2018WarpX,fedeli2022warpx_gordonbell}, and TRISTAN-MP~\citep{buneman1993tristan,Spitkovsky2005collisionlessShocks} with its successor Tristan v2~\citep{tristan_v2}. These codes have been used to study magnetic reconnection, turbulence, collisionless shocks, and kinetic instabilities in the magnetosphere, the solar wind, and high-energy astrophysical environments, with simulations reaching trillions of particles.
However, the numerical constraints of explicit schemes remain expensive when the physical processes of interest evolve on ion and macroscopic scales. Implicit and semi-implicit PIC methods offer an alternative~\citep{peng2015formationMagnetosphere,innocenti2017progress, chen2020mhdepic}, although their implementation on GPU supercomputers is less mature. Examples include iPIC3D~\citep{markidis2010_ipic3d}, recently demonstrated at large scale on GPUs~\citep{markidis2025exascale}, the semi-implicit FLEKS~\citep{chen2023fleks}, and implicit electromagnetic solvers in WarpX based on PICNIC~\citep{angus2023implicit_energy_charge_cons}. Energy conserving formulations such as ECSIM provide an attractive foundation for long time kinetic simulations by preserving the discrete energy balance. Further advances are needed to extend energy conserving implicit PIC algorithms to realistic configurations beyond periodic domains and to execute the complete particle and field computations efficiently on distributed GPU systems.

In this work, we present TilePIC~(TPIC) \footnote{https://github.com/lucak17/TPIC}, an energy conserving semi implicit Particle-in-Cell code designed for large scale kinetic plasma simulations on GPU supercomputers. We extend the ECSIM numerical formulation to conducting boundaries by identifying and eliminating a spurious term in the discrete energy balance. TPIC executes particle and field calculations entirely on GPUs, overlapping communication and computation across distributed NVIDIA and AMD systems. We demonstrate energy conservation to near machine precision in several linear and non-linear test cases, including a three-dimensional magnetic reconnection simulation with conducting walls and 12.3 billion particles. TPIC also achieves 96.8\% weak scaling efficiency on 4,096 GPU compute dies, demonstrating its suitability for production scale plasma simulations.

\section{Energy-conserving semi-implicit algorithm}
\label{sec:ecsim_algorithm}
Implicit PIC algorithms discretize implicitly in time the Maxwell equations, yielding a linear (or more generally nonlinear) formulation where the plasma response to the fields is encoded in a numerical operator. In ECSIM~\citep{lapenta2017ecsim,lapenta2023ecsim_advances}, the linear plasma response is represented by the so called \emph{mass-matrix} operator, and the particle-field coupling is retained through a linear field solve. 

Let $p$ identify a computational particle, $s$ its species, $q_s$ ($m_s$) the species charge (mass), $n$ a time step, $\Delta t$ the time step size, $g$ a mesh node, $V_g$ its nodal control volume, and $\mathbf{E}$ ($\mathbf{B}$) the electric (magnetic) field. The computational particle charge and mass, $q_p$ and $m_p$, include its statistical weight, with $q_p/m_p=q_s/m_s$.
The mass-matrix $\mathbf{M}_{gg'}$ and the current density $\mathbf{J}_g^{n+1/2}$ are defined as 
\begin{equation}
 \begin{aligned}
  M_{gg'}^{ij}
  &=\frac{1}{V_g}\sum_s\beta_s\sum_{p\in s}q_p
    \alpha_p^{ij}W_{pg}W_{pg'},
  \,\,\,\, \beta_s=\frac{q_s\Delta t}{2m_s},\\
  \mathbf{J}_g^{n+1/2}
  &=\widehat{\mathbf{J}}_g+
    \sum_{g'}\mathbf{M}_{gg'}\mathbf{E}_{g'}^{n+1/2}.
 \end{aligned}
 \label{eq_ecsim_mass_matrix}
\end{equation}
Here, $\mathbf{M}_{gg'}$ is a $3\times3$ block coupling the electric field at
$g'$ to the current at $g$. The magnetic rotation tensor $\alpha_p^{ij}$ is
\begin{equation}
  \boldsymbol{\alpha}_p^n=
  \frac{\mathbf{I}-[\mathbf{b}_p]_\times
        +\mathbf{b}_p\mathbf{b}_p^{\mathsf T}}
       {1+|\mathbf{b}_p|^2},
  \qquad \mathbf{b}_p=\frac{\beta_s}{c}\mathbf{B}_p^n,
  \label{eq_ecsim_rotation}
\end{equation}
where $[\mathbf{b}]_\times\mathbf{a}=\mathbf{b}\times\mathbf{a}$ and $c$ is
the speed of light. The hatted current $\widehat{\mathbf{J}}$ is the field independent current contribution and is 
\begin{equation}
  \widehat{\mathbf{J}}_g=
  \frac{1}{V_g}\sum_s\sum_{p\in s}
  q_p\boldsymbol{\alpha}_p^n\mathbf{v}_p^n W_{pg}.
  \label{eq_ecsim_hatted_current}
\end{equation}
Substituting the current response into the Maxwell equations
gives the linear system
\begin{equation}
 \Big[\mathbf{I}
 +\underbrace{(c\theta\Delta t)^2\nabla_h\times\nabla_h\times}_{\text{wave propagation}}
 +\underbrace{4\pi\theta\Delta t\,\mathbf{M}}_{\substack{\text{implicit plasma}\\ \text{response}}}\Big]
 \mathbf{E}^{n+1/2}
 =\underbrace{\mathbf{E}^n+c\theta\Delta t\,\nabla_h\times\mathbf{B}^n
 -4\pi\theta\Delta t\,\widehat{\mathbf{J}}}_{\text{explicit Amp\`ere's law}},
 \label{eq_ecsim_linear_system}
\end{equation}
where, $\mathbf{E}^n$ and $\mathbf{B}^n$ are the known grid fields,
$\mathbf{E}^{n+1/2}$ is the unknown electric field,
$\mathbf{I}$ is the identity operator, $\nabla_h\times$ is the discrete curl,
and $\theta=1/2$ is the time decentering factor. Thus,
$\mathbf{E}^{n+1/2}=(\mathbf{E}^{n+1}+\mathbf{E}^n)/2$. Once it is computed,
the fields are advanced by
\begin{equation}
 \begin{aligned}
  \mathbf{E}^{n+1}&=2\mathbf{E}^{n+1/2}-\mathbf{E}^n,\\
  \mathbf{B}^{n+1}&=\mathbf{B}^n-c\Delta t\,\nabla_h\times\mathbf{E}^{n+1/2},
 \end{aligned}
 \label{eq_ecsim_field_update}
\end{equation}
where the second equation is the discrete Faraday law.

The particle mover is
\begin{equation}
    \begin{aligned}
  \mathbf{x}_p^{n+1/2}
  &=\mathbf{x}_p^{n-1/2}+\Delta t\,\mathbf{v}_p^n,\\
  \overline{\mathbf{v}}_p&= \bm{\alpha}_p^{n}(\mathbf{v}_p^n + \beta_s \mathbf{E}^{n+1/2}_p),\\
  \mathbf{v}_p^{n+1} &= 2\overline{\mathbf{v}}_p - \mathbf{v}_p^n,
 \end{aligned}
 \label{ecsim_mover}
\end{equation}
where $\overline{\mathbf{v}}_p=(\mathbf{v}_p^{n+1}+\mathbf{v}_p^n)/2$ is the
mean velocity defining the deposited current,
$\mathbf{J}_g^{n+1/2}=V_g^{-1}\sum_s\sum_{p\in s}q_p\overline{\mathbf{v}}_pW_{pg}$. The current response in Eq.~\ref{eq_ecsim_mass_matrix} follows exactly from the discrete particle mover, assuming the particle positions and magnetic field are fixed.\\

For $\theta=1/2$ and in the absence of particle injection or loss, energy conservation follows from the discrete Poynting theorem since the particle-field energy exchange is balanced exactly, and the change in total energy is determined only by the Poynting flux through the domain boundary,
\begin{equation}
 (\mathcal{K}+\mathcal{U})^{n+1}-(\mathcal{K}+\mathcal{U})^n
 =-\Delta t\,\Phi_{\partial\Omega}^{n+1/2},
 \label{eq_ecsim_poynting}
\end{equation}
where $\Omega$ is the simulation domain and the total particle kinetic, $\mathcal{K}$, and
electromagnetic field, $\mathcal{U}$, energies are
\begin{equation}
 \begin{aligned}
  \mathcal{K}^n&=\frac{1}{2}\sum_p m_p|\mathbf{v}_p^n|^2,\\
  \mathcal{U}^n&=\frac{1}{8\pi}\left[
    \sum_g V_g|\mathbf{E}_g^n|^2+
    V_c\sum_k|\mathbf{B}_k^n|^2\right],
 \end{aligned}
 \label{eq_ecsim_discrete_energies}
\end{equation}
where, $g$ labels electric-field nodes and $k$ labels magnetic-field cell
centers on a uniform mesh of cell volume $V_c$. The nodal control volume
$V_g$ equals $V_c$ in the interior and $V_c/2$, $V_c/4$, or $V_c/8$ on a physical face, edge, or corner, respectively.
The outward Poynting flux $\Phi_{\partial\Omega}^{n+1/2}$ is evaluated with
the midpoint fields $\mathbf{E}^{n+1/2}$ and
$\mathbf{B}^{n+1/2}$, and it is the boundary remainder of the discrete summation by parts,
\begin{equation}
 \Phi_{\partial\Omega}^{n+1/2}=\frac{c}{4\pi}\Big[
   V_c\sum_k \mathbf{B}_k^{n+1/2}\cdot(\nabla_h\times\mathbf{E}^{n+1/2})_k
   -\sum_g V_g\,\mathbf{E}_g^{n+1/2}\cdot(\nabla_h\times\mathbf{B}^{n+1/2})_g\Big].
 \label{eq_poynting_boundary}
\end{equation}
It depends only on boundary values and is the discrete counterpart of \mbox{$\oint_{\partial\Omega}\frac{c}{4\pi}(\mathbf{E}\times\mathbf{B})\cdot\mathrm{d}\mathbf{A}$}.

In a periodic domain, there are no boundaries, the summation by parts has no remainder, thus $\Phi_{\partial\Omega}^{n+1/2}$ is trivially zero. At the same time, identical weights for field interpolation and current deposition make the particle and field work terms cancel, ensuring exact energy conservation.

\begin{figure}[pos=t!]
  \centering
\begingroup
\definecolor{picInk}{HTML}{000000}
\definecolor{picMuted}{HTML}{333333}
\definecolor{picLine}{HTML}{000000}
\definecolor{picWash}{HTML}{FFFFFF}
\definecolor{picAccent}{HTML}{A8631A}

\begin{tikzpicture}[
  font=\sffamily\fontsize{9}{11}\selectfont,
  panel/.style={
    draw=picLine, fill=white, line width=0.45pt,
    rounded corners=3pt, minimum height=1.45cm,
    text width=3.35cm, align=left, text=picMuted,
    inner sep=6pt, outer sep=0pt
  },
  badge/.style={
    circle, fill=picInk, text=white, minimum size=14pt,
    inner sep=0pt, font=\sffamily\fontsize{8}{9}\selectfont\bfseries
  },
  flow/.style={
    -{Stealth[length=2.2mm,width=1.5mm]},
    draw=picInk!75, line width=0.85pt,
    shorten <=3pt, shorten >=3pt
  }
]
  \node[panel] (mover) at (-3.9,0)
    {{\color{picInk}\bfseries\hspace*{19pt}Particle mover}\\[4pt]
     Update $\mathbf{x}_p$ and $\mathbf{v}_p$};

  \node[panel, text width=3.6cm] (deposit) at (-0.5,2.2)
    {{\color{picInk}\bfseries\hspace*{19pt}Particle $\to$ grid}\\[4pt]
     Form $\boldsymbol{\alpha}_p^n$ from $\mathbf{B}_p^n$\\[2pt]
     Deposit $\widehat{\mathbf{J}}_g$ and $\mathbf{M}_{gg'}$};

  \node[panel] (solve) at (3.9,0)
    {{\color{picInk}\bfseries\hspace*{19pt}Field solve}\\[4pt]
     Solve for $\mathbf{E}^{n+1/2}$\\[2pt]
     Update $\mathbf{E}^{n+1}$, $\mathbf{B}^{n+1}$};

  \node[panel, text width=3.6cm] (gather) at (-0.5,-2.2)
    {{\color{picInk}\bfseries\hspace*{19pt}Grid $\to$ particle}\\[4pt]
     Interpolate $\mathbf{E}$ and $\mathbf{B}$};

  \node[badge] at ([xshift=10pt,yshift=-11pt]mover.north west) {1};
  \node[badge]
    at ([xshift=10pt,yshift=-11pt]deposit.north west) {2};
  \node[badge] at ([xshift=10pt,yshift=-11pt]solve.north west) {3};
  \node[badge] at ([xshift=10pt,yshift=-11pt]gather.north west) {4};

  \draw[flow] (mover.north)
    .. controls +(0,0.6) and +(-0.65,0) .. (deposit.west);
  \draw[flow] (deposit.east)
    .. controls +(0.85,0) and +(0,0.65) .. (solve.north);
  \draw[flow] (solve.south)
    .. controls +(0,-0.65) and +(0.85,0) .. (gather.east);
  \draw[flow] (gather.west)
    .. controls +(-0.65,0) and +(0,-0.6) .. (mover.south);

  \node[circle, draw=picLine, fill=picWash, line width=0.55pt,
    minimum size=1.12cm, inner sep=0pt, text=picInk,
    font=\large] at (-0.5,0.06) {$\Delta t$};
  \node[text=picMuted, font=\scriptsize, inner sep=0pt]
    at (-0.5,-0.76) {$n\;\to\;n+1$};

  \node[panel, text width=5.1cm, draw=picAccent,
    fill=picAccent!10, text=picInk] (mass) at (4.9,2.5)
    {{\color{picAccent}\bfseries Mass matrix assembly}\\[4pt]
     $\displaystyle M_{gg'}^{ij}=\frac{1}{V_g}\sum_s\beta_s
       \sum_{p\in s}q_p\alpha_p^{ij}W_{pg}W_{pg'}$};
  \draw[-{Stealth[length=1.8mm,width=1.2mm]}, draw=picAccent!80,
    line width=0.65pt, dash pattern=on 2pt off 2pt,
    shorten <=4pt, shorten >=4pt]
    (deposit.east) -- (mass.west);
\end{tikzpicture}
\endgroup
  \caption{Diagram of the ECSIM~\citep{lapenta2017ecsim} PIC cycle, with mass matrix deposition highlighted.}
  \label{fig_pic_scheme}
\end{figure}
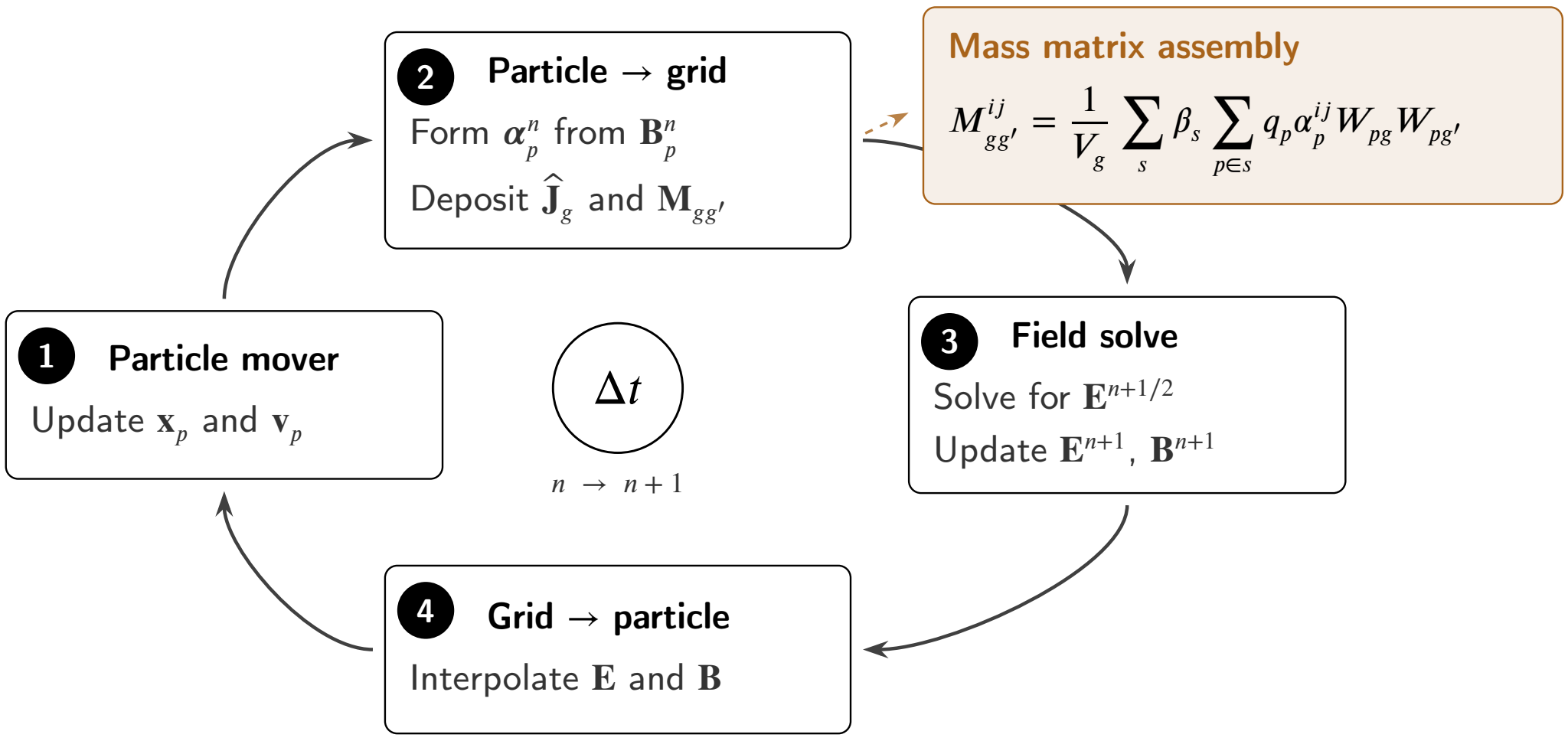

Fig.~\ref{fig_pic_scheme} summarizes the particle and field updates in one
ECSIM step and highlights the mass-matrix assembly used in the field solve.

\section{Exact energy conservation with conductive walls}
\label{sec_wall_energy_conservation}
The original ECSIM~\citep{lapenta2017ecsim} conserves energy exactly only on periodic domains. Here we extend the exact conservation to perfect electric conductor (PEC) and perfect magnetic conductor (PMC) walls, which can be combined with each other and with periodic directions.

TPIC uses a uniform Cartesian mesh with all components of $\mathbf{E}$, $\mathbf{J}$ and $\mathbf{M}$ defined at the nodes $g$, and all components of $\mathbf{B}$ at
the cell centers $k$. The components are not staggered
with respect to one another. Numerical derivatives are implemented as centered finite differences, averaged over the four parallel cell edges, e.g., for the node-to-cell curl $\mathcal{C}$
\begin{equation}
  (\partial^h_x f)_{i+\frac12,j+\frac12,k+\frac12}
  =\frac{1}{4\Delta x}\sum_{a,b\in\{0,1\}}
  \left(f_{i+1,j+a,k+b}-f_{i,j+a,k+b}\right),
  \label{eq_curl_stencil}
\end{equation}
and analogously for the cell-to-node curl, denoted $\mathcal{C}^{\dagger}$
since it is the adjoint of $\mathcal{C}$, (the proof is reported in~\ref{app_discrete_poynting}). 
With the cell volume $V_c=\Delta x\,\Delta y\,\Delta z$ and the nodal control volume $V_g$ defined in Section~\ref{sec:ecsim_algorithm}, nodal and cell inner products are
\begin{equation}
  \langle\mathbf{a},\mathbf{b}\rangle_V=\sum_g V_g\,\mathbf{a}_g\cdot\mathbf{b}_g,
  \qquad
\langle\mathbf{a},\mathbf{b}\rangle_C=V_c\sum_k\mathbf{a}_k\cdot\mathbf{b}_k.
  \label{eq_inner_products}
\end{equation}
This discretization retains mimetic properties of the continuum operators, $\nabla_h\cdot(\nabla_h\times\mathbf{A})=0$ holds exactly, and on periodic meshes $\sum_k V_c\,\mathbf{B}_k\cdot(\mathcal{C}\mathbf{A})_k=\sum_g V_g\,\mathbf{A}_g\cdot(\mathcal{C}^{\dagger}\mathbf{B})_g$, which is the discrete counterpart of $\int_\Omega\nabla\cdot(\mathbf{A}\times\mathbf{B})\,dV=\int_\Omega(\mathbf{B}\cdot\nabla\times\mathbf{A}-\mathbf{A}\cdot\nabla\times\mathbf{B})\,dV=0$, and ensures the Poynting flux $\Phi_{\partial\Omega}^{n+1/2}$ in Eq.~\ref{eq_poynting_boundary} to vanish.
In Eqs.~\ref{eq_ecsim_linear_system} and~\ref{eq_ecsim_field_update}, $\nabla_h\times$ stands for $\mathcal{C}$ acting on $\mathbf{E}$, and $\mathcal{C}^{\dagger}$ acting on $\mathbf{B}$. 

\begin{figure}[pos=t!]
  \centering
\resizebox{0.58\linewidth}{!}{
\begingroup%
\definecolor{wpInk}{HTML}{000000}%
\definecolor{wpMuted}{HTML}{333333}%
\definecolor{wpE}{HTML}{2F6DB5}%
\definecolor{wpB}{HTML}{A8631A}%
\definecolor{wpGhost}{HTML}{EDEDED}%

\def\mjbg{0.3}%
\def\mjbi{0.9}%
\def\mjbn{1.05}%
\pgfmathsetmacro\mjmean{(\mjbg+\mjbi)/2}%

\begin{tikzpicture}[
  font=\sffamily\fontsize{8}{9.5}\selectfont,
  mjaxis/.style={draw=wpInk!70, line width=0.45pt},
  mjface/.style={draw=wpInk!45, line width=0.6pt},
  mjstem/.style={line width=0.8pt},
  mjnode/.style={circle, line width=0.6pt, minimum size=4.6pt, inner sep=0pt},
  mjcell/.style={rectangle, line width=0.6pt, minimum size=4.2pt, inner sep=0pt},
  mjmean/.style={draw=wpInk, fill=white, line width=0.7pt},
  mjguide/.style={draw=wpInk!40, line width=0.45pt, dash pattern=on 1pt off 1.2pt},
  mjjump/.style={{Stealth[length=1.5mm,width=1.1mm]}-{Stealth[length=1.5mm,width=1.1mm]},
    draw=wpInk, line width=0.6pt},
  mjprofile/.style={draw=wpInk!35, line width=0.5pt, dash pattern=on 2pt off 1.5pt}
]
  \fill[wpGhost] (-2.3,-3.0) rectangle (0,1.0);
  \node[text=wpMuted] at (-1.15,0.8) {ghost};
  \node[text=wpMuted] at (1.95,0.8) {domain};

  \draw[mjaxis] (-2.3,0) -- (3.9,0);
  \foreach \xx in {-1.6,0,1.6,3.2}
    \draw[mjface] (\xx,-0.13) -- (\xx,0.13);
  \foreach \xx/\lab/\fl in {-1.6/{-1}/white,0/0/wpE,1.6/1/wpE,3.2/2/wpE} {
    \node[mjnode, draw=wpE, fill=\fl] at (\xx,0) {};
    \node[text=wpE, anchor=south west, inner sep=2pt]
      at (\xx+0.03,0.1) {$\mathbf{E}_{\lab}$};
  }
  \foreach \xx/\lab/\fl in {-0.8/{-1/2}/white,0.8/{1/2}/wpB,2.4/{3/2}/wpB} {
    \node[mjcell, draw=wpB, fill=\fl] at (\xx,0) {};
    \node[text=wpB, anchor=north, inner sep=2pt]
      at (\xx,-0.12) {$\mathbf{B}_{\lab}$};
  }

  \begin{scope}[yshift=-2.6cm]
    \draw[mjaxis] (-2.3,0) -- (3.9,0);
    \foreach \xx in {-1.6,0,1.6,3.2}
      \draw[mjface] (\xx,-0.13) -- (\xx,0.13);
    \foreach \xx/\yy/\lab in {0.8/\mjbi/{1/2},2.4/\mjbn/{3/2}} {
      \draw[mjstem, wpB] (\xx,0) -- (\xx,\yy);
      \node[mjcell, draw=wpB, fill=wpB] at (\xx,\yy) {};
      \node[text=wpB, anchor=south] at (\xx,\yy+0.06) {$\mathbf{B}_{\lab}$};
    }
    \draw[mjstem, wpB, dash pattern=on 1.5pt off 1pt] (-0.8,0) -- (-0.8,\mjbg);
    \node[mjcell, draw=wpB, fill=white] at (-0.8,\mjbg) {};
    \node[text=wpB, anchor=east] at (-0.9,\mjbg) {$\mathbf{B}_{-1/2}$};
    \draw[mjprofile] (-0.8,\mjbg) -- (0.8,\mjbi);
    \draw[mjguide] (-0.8,\mjbg) -- (1.3,\mjbg);
    \draw[mjguide] (0.8,\mjbi) -- (1.3,\mjbi);
    \draw[mjjump] (1.3,\mjbg) -- (1.3,\mjbi);
    \node[text=wpInk, anchor=west] at (1.34,\mjmean) {$[\![\mathbf{B}]\!]$};
  \end{scope}

  \foreach \yy in {0.9,0.7,...,-2.95}
    \draw[wpInk!55, line width=0.35pt] (0,\yy) -- (-0.12,\yy-0.12);
  \draw[wpInk, line width=1.1pt] (0,1.0) -- (0,-3.0);
  \draw[mjmean, shift={(0,-2.6+\mjmean)}]
    (0,0.085) -- (0.085,0) -- (0,-0.085) -- (-0.085,0) -- cycle;
  \node[text=wpInk, anchor=south east, inner sep=2pt]
    at (-0.14,-2.56+\mjmean) {$\overline{\mathbf{B}}$};

\end{tikzpicture}%
\endgroup}
  \caption{Cell and node field values on a 1D discretized domain, including ghost points. The electric field $\mathbf{E}$ lives on nodes, while the magnetic field $B$ is cell-centered. $\overline{\mathbf{B}}$ and $[\![\mathbf{B}]\!]$ are the mean and the jump of $\mathbf{B}$ between the first interior cell and the ghost cell.}
  \label{fig_ghost_bcs}
\end{figure}
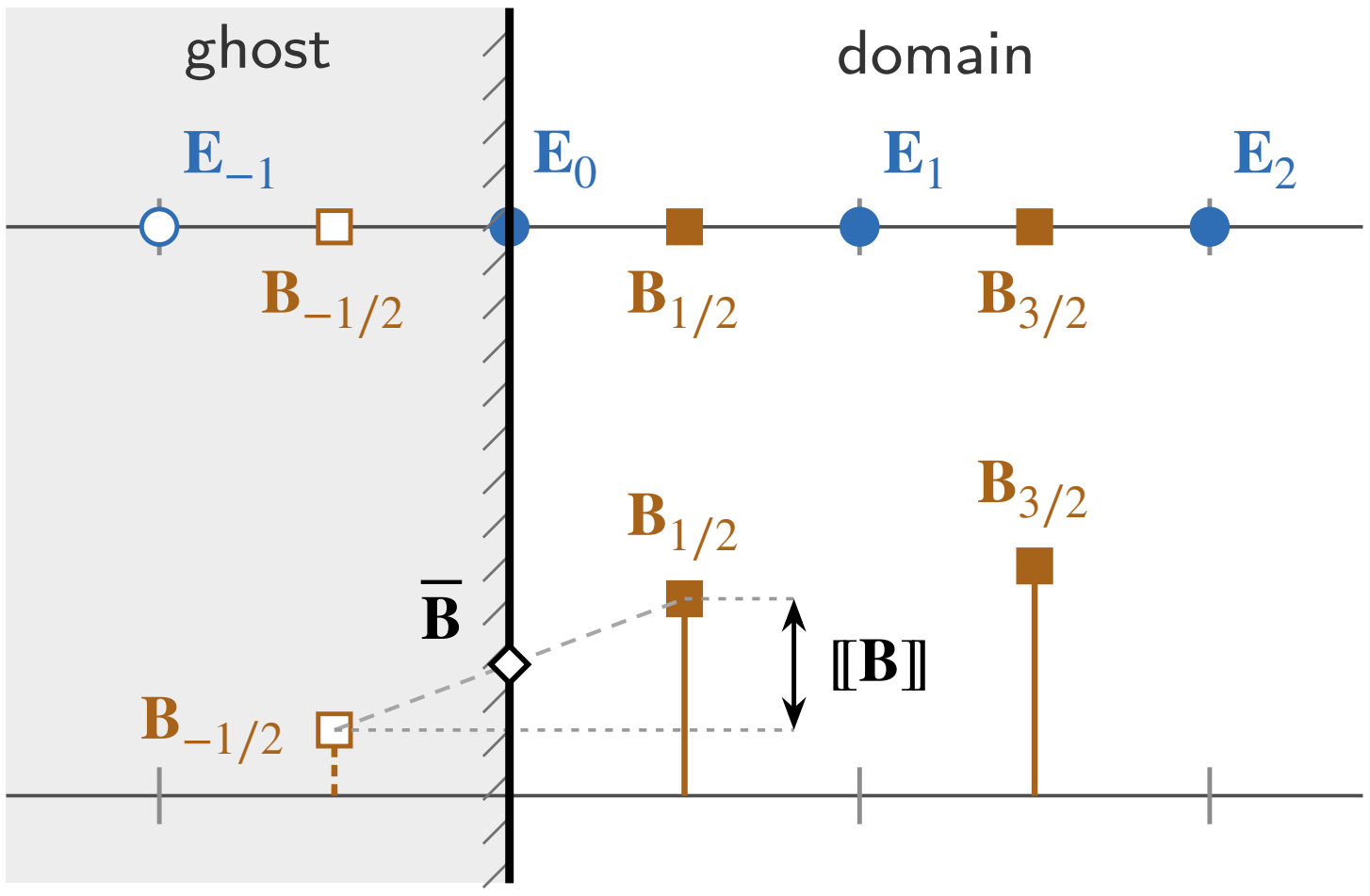

In the discrete scheme, the Poynting flux $\Phi_{\partial\Omega}$ is the summation-by-parts remainder of Eq.~\ref{eq_poynting_boundary}. In a closed domain, at a wall node, $\mathcal{C}^{\dagger}$ uses a ghost cell outside the domain, and, using the numerical differential operator defined in Eq.~\ref{eq_curl_stencil}, we can express Eq.~\ref{eq_poynting_boundary} as 
\begin{equation}
  \Phi_{\partial\Omega} = \frac{c}{4\pi} \left[\langle\mathcal{C}\mathbf{E},\mathbf{B}\rangle_C
  -\langle\mathbf{E},\mathcal{C}^{\dagger}\mathbf{B}\rangle_V \right]
  =\frac{c}{4\pi}\sum_{g\in\partial\Omega}A_g\Big[
   (\mathbf{E}_g\times\overline{\mathbf{B}}_g)\cdot\hat{\mathbf{n}}
   +\frac{\Delta_n}{4}\,\mathbf{E}_g\cdot
   \big(\nabla^h_t\times[\![\mathbf{B}]\!]\big)_g\Big].
  \label{eq_discrete_poynting}
\end{equation}
Here, $\hat{\mathbf{n}}$ is the outward normal, $\Delta_n$ the mesh spacing along it, $A_g=2V_g/\Delta_n$ the wall area of the control volume of node $g$, and $\nabla^h_t\times$ the cell-to-node curl restricted to wall-tangential derivatives. $\overline{\mathbf{B}}$ and $[\![\mathbf{B}]\!]$ are the mean and the jump of $\mathbf{B}$ between the first interior cell and the ghost cell, interpolated to the wall nodes, as shown in Fig.~\ref{fig_ghost_bcs}. We report the full derivation of Eq.~\ref{eq_discrete_poynting} in~\ref{app_discrete_poynting}.

The first term in  Eq.~\ref{eq_discrete_poynting} is the discrete Poynting flux.
The second is a discretization artifact without a continuous counterpart, coupling
$E_n$ to $[\![\mathbf{B}_t]\!]$ and $\mathbf{E}_t$ to $[\![B_n]\!]$. In a closed domain, exact energy conservation requires both terms to vanish.

TPIC cancels them with two discrete constraints. First, we set to zero at the wall nodes the electric components that vanish on the wall, $\mathbf{E}_t$ for PEC and $E_n$ for PMC. Second, following the image field theory~\citep{Taflove2005comp_electrodynamics_fdtd,ricci2002implicit_solver}, each ghost cell holds the reflection of the adjacent interior cell,
$\mathbf{B}_{-1/2}=\sigma\mathbf{B}_{1/2}$ componentwise, with parity
$\sigma=\pm1$, so that
\begin{equation}
  \overline{\mathbf{B}}=\frac{1+\sigma}{2}\,\mathbf{B}_{1/2},
  \qquad
  [\![\mathbf{B}]\!]=(1-\sigma)\,\mathbf{B}_{1/2}.
  \label{eq_wall_mean_jump}
\end{equation}
An odd parity cancels the mean, and an even parity cancels the jump. The
parities in Table~\ref{tab_wall_parities} cancel the terms that the electric constraint leaves. On a PEC wall, $\mathbf{E}_t=0$ removes the flux and the $[\![B_n]\!]$ term, and an even $\mathbf{B}_t$ removes the $[\![\mathbf{B}_t]\!]$
term. On a PMC wall, $E_n=0$ removes the $[\![\mathbf{B}_t]\!]$ term, and since
$\mathbf{E}_t\neq0$, an odd $\mathbf{B}_t$ removes the flux and an even $B_n$
the $[\![B_n]\!]$ term. The odd $B_n$ of PEC is not required by
Eq.~\ref{eq_discrete_poynting}, but it makes $\overline{B}_n=0$, so that both
walls satisfy the discrete counterparts of $\hat{\mathbf{n}}\times\mathbf{E}=
\hat{\mathbf{n}}\cdot\mathbf{B}=0$ (PEC) and $\hat{\mathbf{n}}\times\mathbf{B}=
\hat{\mathbf{n}}\cdot\mathbf{E}=0$ (PMC). The same closure applies to
$\mathcal{C}\mathbf{v}$. Nodal ghosts use the parities of $\mathbf{E}$, and edge and corner ghosts take the product of the parities of the adjacent walls.

\begin{table}[pos=t!]
  \centering
  \caption{Ghost parities $\sigma$ of the wall-tangential ($t$) and normal
  ($n$) field components. Electric components marked $0$ are set to zero at
  the wall nodes.}
  \label{tab_wall_parities}
  \small
  \begin{tabular}{@{}lcccc@{}}
    \hline
    Wall & $\mathbf{E}_t$ & $E_n$ & $\mathbf{B}_t$ & $B_n$ \\
    \hline
    PEC & $-1$, $0$ & $+1$ & $+1$ & $-1$ \\
    PMC & $+1$ & $-1$, $0$ & $-1$ & $+1$ \\
    \hline
  \end{tabular}
\end{table}

With the residual of Eq.~\ref{eq_discrete_poynting} cancelled,
$\langle\mathbf{E},\mathcal{C}^{\dagger}\mathbf{B}\rangle_V
=\langle\mathcal{C}\mathbf{E},\mathbf{B}\rangle_C$, and the discrete Poynting flux through the domain is exactly zero, thus the system energy is conserved to the linear solver tolerance and round-off.

With non periodic boundary conditions, particles are specularly reflected at the walls, which leaves $|\mathbf{v}_p|$ unchanged and thus preserves the energy.

\section{Numerical methods in TPIC}
\label{sec_numerical_methods}
In this section we detail the numerical methods implemented in TPIC, including the accuracy of the discretization, the energy-conserving smoothing, and the matrix-free field solver.

\paragraph{Discretization and accuracy}
The centered differences of Eq.~\ref{eq_curl_stencil} are second-order accurate. Particles use trilinear cloud-in-cell (CIC) shape functions $W_{pg}$ for interpolation operations, and $\mathbf{B}$ is averaged from the eight neighboring cells to the nodes before the gather. 
TPIC spatial discretization is therefore second order accurate.

For $\theta=1/2$, the field update is the implicit-midpoint (Crank-Nicolson) rule, which is second-order accurate, and the particle push of Eq.~\ref{ecsim_mover} is time-centred in $\mathbf{E}$ and $\mathbf{x}$. 
The rotation $\boldsymbol{\alpha}_p$ uses $\mathbf{B}^n$ instead of $\mathbf{B}^{n+1/2}$. This half-step lag keeps Eq.~\ref{eq_ecsim_linear_system} linear and does not affect energy conservation since the magnetic force does no work, but it introduces an error proportional to $\Delta t\,\partial_t\mathbf{B}$. Thus, the coupled scheme is second-order time accurate for static magnetic fields and first-order otherwise.
The implicit field solve removes the CFL and $\omega_{pe}\Delta t$ stability limits, and
energy conservation suppresses the finite-grid instability, so $\Delta x$ does not need to resolve the Debye length.

\paragraph{Energy-conserving smoothing}
Following~\citet{lapenta2023ecsim_advances}, we implement smoothing in TPIC to filter particle and grid noise. Critically, to retain exact energy conservation, smoothing must be applied both to particle moments and fields. Denoting the smoothing operator by $\mathcal{S}$, the smoothed $\mathbf{E}$ field is $\mathcal{S}\mathbf{E}^{n+1/2}$, used in the particles push. By Eq.~\ref{eq_ecsim_mass_matrix}, this yields the current \mbox{$\mathbf{J}^{n+1/2}=\widehat{\mathbf{J}}+\mathbf{M}\mathcal{S}\mathbf{E}^{n+1/2}$}, and Amp\`ere's law then uses $\mathcal{S}\mathbf{J}^{n+1/2}$.
The work done on the particles, $\Delta t\langle\mathcal{S}\mathbf{E}^{n+1/2},\mathbf{J}^{n+1/2}\rangle_V$, equals the work done by the field,
$\Delta t\langle\mathbf{E}^{n+1/2},\mathcal{S}\mathbf{J}^{n+1/2}\rangle_V$, provided that $\mathcal{S}$ is self-adjoint in the nodal inner product of Eq.~\ref{eq_inner_products}.

We implement two smoothing operators in TPIC. The
seven-point operator
\begin{equation}
  (\mathcal{S}_7f)_g=\alpha f_g+\frac{1-\alpha}{6}\sum_{g'\in\mathcal{F}_g}f_{g'},
  \label{eq_smoothing_s6}
\end{equation}
where $\mathcal{F}_g$ contains the six face neighbors of node $g$, has a variable center weight $0\le\alpha<1$. The 27-point operator is the tensor
product of one-dimensional binomial filters,
\begin{equation}
  \mathcal{S}_{27}=\mathcal{S}^{x}_{121}\mathcal{S}^{y}_{121}\mathcal{S}^{z}_{121},
  \qquad
  (\mathcal{S}^{x}_{121}f)_{i,j,k}=\tfrac14\left(f_{i-1,j,k}+2f_{i,j,k}+f_{i+1,j,k}\right),
  \label{eq_smoothing_s27}
\end{equation}
with $\mathcal{S}^{y}_{121}$ and $\mathcal{S}^{z}_{121}$ defined in similar way.
Expanding the product, $\mathcal{S}_{27}$ assigns the fixed weights $8/64$ to node $g$ and $4/64$, $2/64$, and $1/64$ to each of its 6 face, 12 edge, and 8
corner neighbors, respectively. 
Both operators have symmetric stencils, are self-adjoint, and are second-order accurate.
Both operators are applied $N_s$ times.

\paragraph{Matrix-free GMRES}
With the smoothing operator $\mathcal{S}$ previously defined ($\mathcal{S}=\mathcal{I}$
without smoothing), the Maxwell system in Eq.~\ref{eq_ecsim_linear_system} takes the form of $\mathcal{A}\mathbf{E}^{n+1/2}=\mathbf{b}$, with
\begin{equation}
 \begin{aligned}
  \mathcal{A}&=\mathcal{I}+(c\theta\Delta t)^2\,\mathcal{C}^{\dagger}\mathcal{C}
   +4\pi\theta\Delta t\,\mathcal{S}\mathbf{M}\mathcal{S},\\
  \mathbf{b}&=\mathbf{E}^n+c\theta\Delta t\,\mathcal{C}^{\dagger}\mathbf{B}^n
   -4\pi\theta\Delta t\,\mathcal{S}\widehat{\mathbf{J}}.
 \end{aligned}
 \label{eq_maxwell_operator}
\end{equation}
The antisymmetric part of $\boldsymbol{\alpha}_p$ makes $\mathbf{M}$, and hence
$\mathcal{A}$, non-symmetric, so TPIC solves it with a matrix-free restarted GMRES, where the operator $\mathcal{A}$ is applied without assembling the corresponding matrix.
The Krylov basis, the Hessenberg matrix, the Givens rotations, and the least-squares solve reside on the device. A solution is accepted only if the true residual satisfies $\|\mathbf{b}-\mathcal{A}\mathbf{E}^{n+1/2}\|_2\le\varepsilon\|\mathbf{b}\|_2$, with $\epsilon$ the required tolerance.
Boundary conditions are enforced by solving the linear system in the constrained subspace, where the right-hand side and the initial guess are projected, and the constrained rows of the operator are replaced by the
identity.
The mass matrix $\mathbf{M}$ is diagonally dominant, making the operator $\mathcal{A}$ well conditioned, thus the solver generally converges in $10-30$ iterations and no preconditioner is needed.

\section{TPIC architecture}
\label{sec:implementation}
TPIC is a modern C++20 code designed for heterogeneous architectures and large-scale systems. It is GPU-native, with the entire core algorithm executed on the device. This design keeps data fully device-resident throughout the simulation, minimizing expensive host-device data transfers. The host has primarily an orchestration role and is responsible for I/O operations.
Parallelism is exploited at both the data and task levels. Data-level parallelism is achieved through on-node GPU execution, while task-level parallelism is implemented using C++ threads to coordinate multiple tasks concurrently across multiple device streams. For distributed-memory simulations, the computational domain is partitioned into subdomains, each assigned to a distinct process, and TPIC relies on communication libraries such as MPI, NCCL, and RCCL. 

TPIC builds on our custom Light Particle Field Accelerator (LPFx) library, which provides performance portability across NVIDIA and AMD GPU backends and multiple communication stacks, by implementing the distributed-memory data structures and execution framework TPIC uses.

In this section we describe TPIC's design, and we discuss the key architectural choices that enable high performance simulations.

\subsection{Asynchronous pipeline}
\label{sec_async_pipeline}
TPIC executes each time step as a set of tasks distributed over persistent host worker threads, implemented in LPFx with native C++ threads. Each worker owns dedicated device streams, its communication state, and the data it produces. The workers are the particle mover $M$, the field advance $F$, one local worker $L_s$ for each species $s$, the particle migration $C$, the moment deposition $D$, the Maxwell solver $S$, the diagnostics $Q$, and the output writer $W$. A control thread submits each time step to the workers. The host threads mainly enqueue device work and drive communication, so that the device streams execute concurrently.

Dependencies between workers are explicit. TPIC uses device events, recorded on the producer stream and waited on the consumer stream, to order device work without host synchronization. Host fences notify a host thread that a task has been issued, e.g., before it starts a dependent communication.
The pipeline fixes the order of communication operations, ensuring all ranks issue communication calls in the same sequence, a strict requirement for NCCL and RCCL backends. Specifically, halo exchange precedes migration, species are migrated in index order, and $\widehat{\mathbf{J}}$ is assembled before $\mathbf{M}$.

\begin{figure}[pos=t!]
  \centering
  \makebox[\linewidth][c]{
\begingroup
\definecolor{pipeS0}{HTML}{2F6DB5}
\definecolor{pipeS1}{HTML}{8E44AD}
\definecolor{pipeField}{HTML}{2E8B57}
\definecolor{pipeMoment}{HTML}{C68A00}
\definecolor{pipeMigrate}{HTML}{C0392B}
\definecolor{pipeOut}{HTML}{6B6B6B}

\begin{tikzpicture}[
  x=0.8cm, y=1cm,
  font=\sffamily\fontsize{8}{9.5}\selectfont,
  lane/.style={anchor=east, align=right},
  box/.style={line width=0.45pt, rounded corners=1.5pt},
  task-s0/.style={box, draw=pipeS0!80!black, fill=pipeS0!18},
  task-s1/.style={box, draw=pipeS1!80!black, fill=pipeS1!16},
  task-field/.style={box, draw=pipeField!80!black, fill=pipeField!18},
  task-moment/.style={box, draw=pipeMoment!85!black, fill=pipeMoment!20},
  task-migrate/.style={box, draw=pipeMigrate!80!black, fill=pipeMigrate!16},
  task-out/.style={box, draw=pipeOut!90!black, fill=pipeOut!16},
  boundary/.style={densely dashed, line width=0.5pt, draw=black!50},
  dep/.style={densely dotted, line width=0.6pt, draw=black!75,
              -{Stealth[length=1.5mm,width=1.1mm]}}
]
  \def\task#1#2#3#4#5{%
    \filldraw[#4] (#2+0.03,#1-0.21) rectangle (#3-0.03,#1+0.21);
    \node at ({(#2+#3)/2},#1) {#5};}

  \foreach \y/\name in {0/{$L_0$ species 0}, -0.62/{$L_1$ species 1},
      -1.24/{$M$ mover}, -1.86/{$C$ migration}, -2.48/{$D$ deposition},
      -3.10/{$F$ field advance}, -3.72/{$S$ Maxwell}, -4.34/{$Q$ diagnostics},
      -4.96/{$W$ output writer}} {
    \fill[black!3] (-2.75,\y-0.29) rectangle (17.75,\y+0.29);
    \node[lane] at (-0.15,\y) {\name};
  }

  \draw[boundary] (0,0.42) -- (0,-5.38);
  \draw[boundary] (16.8,0.42) -- (16.8,-5.38);
  \node[anchor=south] at (0,0.42) {step $n$};
  \node[anchor=south] at (16.8,0.42) {step $n{+}1$};

  \draw[dep] (0.65,-1.03) -- (0.95,-0.21);    
  \draw[dep] (1.45,-1.03) -- (1.75,-0.41);    
  \draw[dep] (0.65,-1.45) -- (0.95,-1.65);    
  \draw[dep] (1.5,-1.45) -- (3.35,-1.65);     
  \draw[dep] (2.7,-1.65) -- (2.95,-0.21);     
  \draw[dep] (4.35,-1.65) -- (4.6,-0.41);     
  \draw[dep] (2.35,-0.21) -- (2.65,-2.27);    
  \draw[dep] (3.2,-0.83) -- (4.15,-2.27);     
  \draw[dep] (4.25,-0.21) -- (5.9,-2.27);     
  \draw[dep] (5.9,-0.83) -- (7.55,-2.27);     
  \draw[dep] (1.85,-2.89) -- (2.65,-2.69);    
  \draw[dep] (9.85,-2.69) -- (10.15,-3.51);   
  \draw[dep] (10.55,-2.69) -- (11.15,-3.51);  
  \draw[dep] (15.3,-3.93) -- (15.5,-4.13);    
  \draw[dep] (16.7,-4.55) -- (16.9,-4.75);    

  \task{0}{0.8}{2.5}{task-s0}{sort local}
  \task{0}{2.8}{4.3}{task-s0}{arrivals}
  \task{-0.62}{1.6}{3.3}{task-s1}{sort local}
  \task{-0.62}{4.5}{6.0}{task-s1}{arrivals}

  \task{-1.24}{0}{0.8}{task-s0}{$s_0$}
  \task{-1.24}{0.8}{1.6}{task-s1}{$s_1$}
  \task{-1.24}{16.8}{17.6}{task-s0,opacity=0.45}{$s_0$}

  \task{-1.86}{0.8}{2.8}{task-migrate}{$s_0$}
  \task{-1.86}{2.8}{4.5}{task-migrate}{$s_1$}

  \task{-2.48}{2.5}{4.0}{task-moment}{$s_0$ local}
  \task{-2.48}{4.0}{5.5}{task-moment}{$s_1$ local}
  \task{-2.48}{5.5}{7.4}{task-moment}{$s_0$ arrivals}
  \task{-2.48}{7.4}{9.3}{task-moment}{$s_1$ arrivals}
  \task{-2.48}{9.3}{10.0}{task-moment}{$\widehat{\mathbf J}$}
  \task{-2.48}{10.0}{10.7}{task-moment}{$\mathbf M$}

  \task{-3.10}{0}{2.0}{task-field}{$\mathbf E^{n}$, $\mathbf B^{n}$}

  \task{-3.72}{10.0}{11.0}{task-field}{RHS}
  \task{-3.72}{11.0}{12.55}{task-field}{GMRES}
  \task{-3.72}{12.55}{15.4}{task-field}{smooth $\mathbf E^{n+1/2}$}

  \task{-4.34}{6.0}{7.3}{task-out}{reduce}
  \task{-4.34}{7.3}{8.75}{task-out}{capture}
  \task{-4.34}{15.4}{16.8}{task-out}{finalize}

  \task{-4.96}{0}{16.8}{task-out}{write previous snapshot to disk}
  \task{-4.96}{16.8}{17.6}{task-out,opacity=0.45}{}

  \foreach \x/\kind/\text in {-2.1/task-s0/species $s_0$, 0.6/task-s1/species $s_1$,
      3.3/task-migrate/migration, 5.9/task-field/fields, 7.85/task-moment/moments,
      10.4/task-out/{diagnostics, output}} {
    \filldraw[\kind] (\x,-5.80) rectangle (\x+0.45,-5.58);
    \node[anchor=west] at (\x+0.5,-5.69) {\text};
  }
  \draw[dep] (14.6,-5.69) -- (15.15,-5.69);
  \node[anchor=west] at (15.2,-5.69) {dependency};
\end{tikzpicture}
\endgroup}
  \caption{Asynchronous TPIC pipeline for two particle species, with one row per host worker. Dotted arrows mark the main dependencies between workers.
  Worker $C$ migrates each species once it is pushed, while $L_s$ concurrently sorts the particles that stay local. $D$ deposits local particles as soon as $ L_s$ sorts them, then deposits the arrivals, which $L_s$ finalizes once the migration completes. The diagnostics worker $Q$ overlaps its work with $D$ and $S$, and the next step starts once $S$ and $Q$ complete. The disk writer $W$ writes field and particle snapshots, and checkpoints while the next step runs.}
  \label{fig_pic_pipeline}
\end{figure}
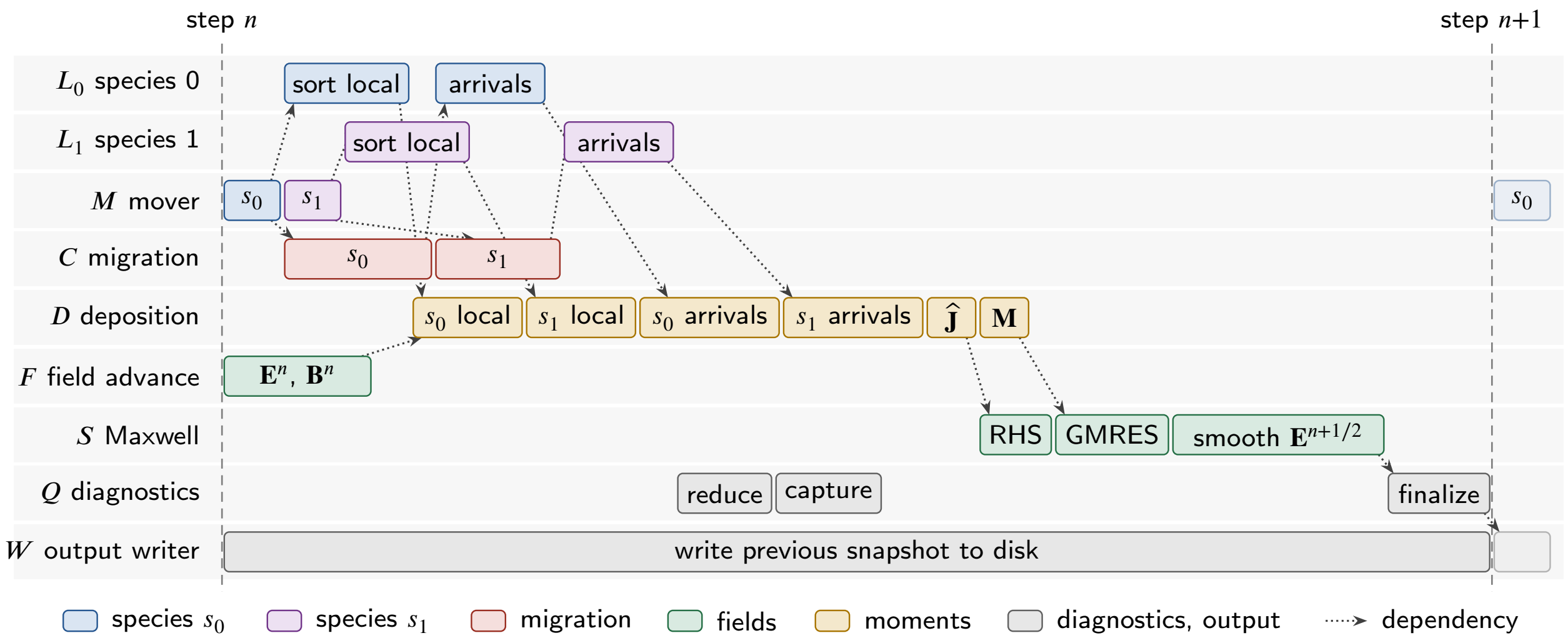

Fig.~\ref{fig_pic_pipeline} schematizes one TPIC time step with two species. The step starts from the electric field computed by the previous solve, i.e. $\mathbf{E}^{n-1/2}$. Worker $M$ pushes the species in order, while $F$ advances $\mathbf{E}$ and $\mathbf{B}$ with Eq.~\ref{eq_ecsim_field_update} and interpolates the new $\mathbf{B}$ to the nodes, later used by $D$ to evaluate $\boldsymbol{\alpha}_p$. Once species $s$ is pushed, $L_s$ separates the particles leaving the subdomain, which $C$ migrates on a high-priority stream, from those staying local, which $L_s$ sorts concurrently. With partial sorting (Section~\ref{sec_particle_sorting}), $D$ deposits $\widehat{\mathbf{J}}$ and $\mathbf{M}$ on two separate streams, first from the sorted local particles and then from the arrivals, which $L_s$ appends once the migration completes. After all species are migrated, $D$ completes $\widehat{\mathbf{J}}$ and $\mathbf{M}$ assembly with halo exchange across subdomains.

Worker $S$ starts the field solve before the deposition completes. It computes the field terms of the right-hand side as soon as the new state is available, completes it once $\widehat{\mathbf{J}}$ is assembled, and starts GMRES on the interior entries of $\mathbf{M}$ while their ghost contributions are still being exchanged. It then calculates the field $\mathcal{S}\mathbf{E}^{n+1/2}$ for the next particle push. Meanwhile, $Q$ computes local diagnostics and captures the output snapshot, and it starts its collective reductions only after $S$ completes, so the two workers never interleave communication. The next step starts when both $S$ and $Q$ finish, while $W$ writes the captured snapshot to disk. In this way, migration, moment assembly, diagnostics, and output overlap with computation, avoiding device idling and maximizing compute resources utilization. 

\subsection{Portable data structures and communication with LPFx}
\label{sec_lpfx}

\begin{figure}[pos=t!]
  \centering
\begingroup
\definecolor{stackInk}{HTML}{000000}
\definecolor{stackApp}{HTML}{3B4C9B}
\definecolor{stackMpi}{HTML}{5A5A5A}
\definecolor{stackCuda}{HTML}{76B900}
\definecolor{stackRocm}{HTML}{ED7D31}

\begin{tikzpicture}[
  font=\sffamily\fontsize{9}{11}\selectfont,
  layer/.style={line width=0.45pt, rounded corners=2.5pt},
  title/.style={text=stackInk, font=\sffamily\fontsize{10}{12}\selectfont\bfseries}
]
  \filldraw[layer, draw=stackApp, fill=stackApp!30] (0,2.2) rectangle (7.2,2.8);
  \node[title] at (3.6,2.5) {TPIC};

  \filldraw[layer, draw=stackApp, fill=stackApp!8] (0,0) rectangle (7.2,2.05);
  \node[title] at (3.6,1.78) {LPFx};
  \filldraw[layer, draw=stackCuda!60!black, fill=stackCuda!20] (0.15,0.9) rectangle (2.35,1.5);
  \node at (1.25,1.2) {NCCL};
  \filldraw[layer, draw=stackMpi, fill=stackMpi!15] (2.5,0.9) rectangle (4.7,1.5);
  \node at (3.6,1.2) {MPI};
  \filldraw[layer, draw=stackRocm!70!black, fill=stackRocm!20] (4.85,0.9) rectangle (7.05,1.5);
  \node at (5.95,1.2) {RCCL};

  \filldraw[layer, draw=stackCuda!60!black, fill=stackCuda!45] (0.15,0.15) rectangle (3.525,0.75);
  \node at (1.8375,0.45) {CUDA};
  \filldraw[layer, draw=stackRocm!70!black, fill=stackRocm!45] (3.675,0.15) rectangle (7.05,0.75);
  \node at (5.3625,0.45) {HIP/ROCm};
\end{tikzpicture}
\endgroup
  \caption{TPIC software stack. LPFx wraps the CUDA and HIP/ROCm device backends and the communication transports, NCCL for NVIDIA GPUs, RCCL for AMD GPUs, and MPI for both.}
  \label{fig_lpfx_stack}
\end{figure}
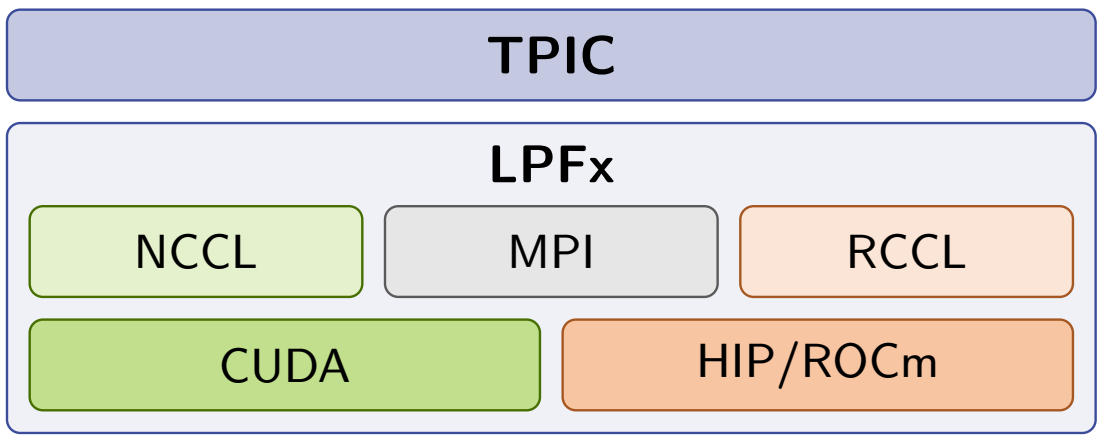

To provide performance portability across different computing and communication backends, we develop LPFx, a header-only C++20 library for distributed-memory simulations. LPFx provides host and device data structures, algorithms, and communication patterns commonly required by PIC methods, while separating numerical kernels from storage and transport details. It supports CUDA and HIP devices and provides transport implementations based on MPI, NCCL, and RCCL. TPIC leverages LPFx to ensure a single code source portable across multiple backends. Fig.~\ref{fig_lpfx_stack} shows the resulting software stack.

LPFx represents multi-component fields on uniform 3D Cartesian grids, including per-component staggering and halo regions. Field components reside at planned offsets within a contiguous allocation and may use backend-specific pitch and alignment policies. Particles are described by typed schemas and stored in a structure-of-arrays (SoA) layout, where each array holds a different particle quantity. Schemas, field locations, communication patterns, and backend choices are resolved at compile time, eliminating runtime dispatch from performance-critical paths.

LPFx provides communication patterns such as \verb|copy_to_ghost| and \verb|ghost_to_owner_add| for fields, \verb|particle_migration| for particles, collective operations, and fixed-message exchanges. Each operation is defined at compile time and prepared using runtime topology, geometry, and capacity information. The resulting prepared plan is bound to a data owner, a transport backend, and an execution stream. This modularity allows the same data structure and numerical algorithm to be used with different communication backends.

LPFx minimizes message count by aggregating data into compact communication images. Field exchange packs strided subregions into dense contiguous buffers. Particle migration uses an explicit SoA wire format containing velocities, charge, identifier, and canonical position. Both payload types are transported as uninterpreted bytes, with one message per peer.
Prepared operations retain their communication images and auxiliary storage for packing, routing, and unpacking. These resources are reused across simulation steps, avoiding repeated allocation/deallocation operations.

LPFx supports device-direct MPI, staged MPI through persistent pinned host buffers, NCCL, and RCCL. The transport backend is selected independently for each binding, and different transport strategies can coexist within one application. In the device-aware MPI configuration, TPIC uses direct device communication for fields and particles, and other device-resident operations. 
Some MPI implementations may internally stage device collectives through pageable host memory, even when device-aware MPI is enabled. Thus, for TPIC's reductions, we explicitly select LPFx's staged-MPI backend, which uses persistent pinned host buffers and avoids repeated staging-buffer allocation. In NCCL or RCCL configurations, the corresponding vendor backend is used for device payloads, while MPI remains responsible for bootstrap and control.

Communication operations expose an explicit lifecycle through \verb|begin()|, \verb|progress()| or \verb|wait()|, and \verb|conclude()|. Each binding can use a dedicated device stream for packing, transport preparation, and receive processing. This interface allows TPIC to overlap field exchange and particle migration with independent computation. 
LPFx is transparent, avoiding automatic data transfer or synchronization. It provides the required stream and event mechanisms for explicit synchronization, while TPIC retains full control over pipeline ordering and the data lifetime.

Portability between NVIDIA and AMD accelerators is implemented through a lightweight compile-time abstraction over CUDA and HIP. Backend traits map memory allocation, copies, streams, events, and device operations directly to the corresponding native runtime. Similarly, warp-level instructions such as \verb|lpfx::warp_shuffle_up| compile to \verb|__shfl_up_sync| on CUDA and \verb|__shfl_up| on HIP, without runtime overhead. TPIC can therefore use a single source implementation while retaining access to low-level operations required by performance-critical kernels.

\subsection{Particle sorting}
\label{sec_particle_sorting}
In TPIC, particles of each species are sorted independently by mesh cell within each subdomain. In the SoA layout, this groups particles belonging to the same cell contiguously in memory within each particle-attribute array. Particle sorting is key to achieving high performance, as it enables optimized cell-based kernels that leverage coalesced memory accesses, cache field values reused across multiple particles, and accumulate particle quantities within each cell with atomic-free deposition.

The sorting algorithm is implemented in the LPFx library as a direct counting sort using the integer cell index stored with each particle. The entire sorting pipeline runs on GPU.
Let $N_p$ denote the number of particles of one species to sort and $N_c$ the number of cells in the local index space. The sorting operation first builds a histogram $h_c$ by atomically counting particles in cell $c$. An exclusive prefix sum then determines the beginning of each cell's range. Let $o_c$ be the $c$ cell's offset,
\begin{equation}
  o_0=0,\qquad
  o_c=\sum_{b=0}^{c-1}h_b\quad (1\leq c\leq N_c),
  \qquad o_{N_c}=N_p.
  \label{eq_particle_cell_offsets}
\end{equation}
Cell $c$ occupies the half-open interval $[o_c,o_{c+1})$, whose length is
exactly its particle population. Empty cells have coincident offsets.

LPFx computes the offsets with a hierarchical Blelloch
scan~\citep{blelloch1993prefixsum} implemented with GPU kernels. 
The scan requires $O(N_c)$ work and $O(\log N_c)$ parallel depth. 
For each cell $c$, a separate index is initialized to $o_c$. Particles atomically advance their cell's index to obtain unique destination indices, defining a permutation that is applied to every particle attribute. Particles' attributes are then scattered into new buffers according to the permutation to obtain the sorted storage. Thus, the complete sort requires $O(N_p+N_c)$ work. 

The attributes permutation uses a rotating storage scheme to reduce the overall memory footprint. Each attribute is scattered in turn into a new available buffer, and its previous buffer becomes the destination for the next attribute. The attribute traversal order accounts for their different storage widths. Thus, despite the scattering method transfers the same amount of data as a conventional double buffer approach, in TPIC the additional particle storage is only one buffer sized for the widest attribute at the allocated particle capacity. 

TPIC supports both full and partial sorting pipelines. In full sorting, particle migration across subdomains completes before all locally owned particles, including the new incoming ones, are sorted. All subsequent stages then process particles using the optimized cell-based kernels. 
In partial sorting, particles that remain on the local rank are compacted into a
resident prefix and sorted while migration of outgoing/incoming particles concurrently progresses. As soon as local sorting completes, next stages (current and mass-matrix deposition) can consume this sorted prefix while
remote particles are still in transit. Sorted particles are processed with optimized cell-based kernels, while arriving particles are appended as an unsorted tail and processed separately with unsorted kernels.

Partial sorting therefore overlaps resident sorting and deposition with communication, at the cost of processing the received tail without cell grouping during that step. Sorting is repeated at each cycle since particle push updates particle position, invalidating the previous cell ranges. In partial sorting, particles in the unsorted tail at time step $n$ become sorted particles at time step $n+1$, provided they remain within the local subdomain. As a result, the number of unsorted particles remains approximately constant throughout the simulation and is equal to the number of migrated particles, thus being only a small fraction of the total particle population.

\subsection{Mass-matrix deposition on tensor-cores}
\label{sec_tensor_mass_deposition}
Particle-to-grid scatter operations are among the most expensive operations in PIC algorithms since they involve fine-grained reductions that hinder parallelization. In ECSIM, each mass-matrix block $M_{gg'}^{ij}$ couples field component $j$ at node $g'$ to current component $i$ at node $g$. Its assembly is therefore more expensive than standard particle-to-node deposition, since each particle contributes to every node pair in its interpolation stencil. For three-dimensional cloud-in-cell (CIC) shape functions, this yields $8\times8\times9=576$ scalar contributions per particle, or $36\times9=324$ distinct coefficients after exploiting spatial symmetry. Since multiple particles update the same global entries, a naive GPU implementation parallelized over particles requires extensive atomic operations in main memory. The resulting fine-grained atomic traffic makes mass-matrix assembly a major computational bottleneck.

Following~\citet{pennati2026massmatrix}, we express the mass-matrix assembly as a tensor contraction over the particle index, and use particle sorting to decompose it into dense cell-local products.~\citet{pennati2026massmatrix} derive this decomposition for general shape functions. With CIC, all particles in a cell couple to the same eight corners, so each cell contributes one dense $8\times8$ product. Let $\mathcal{P}_c$ contain the particles in cell $c$, $w_{ap}$ be the CIC weight of particle $p$ at the local corner $a\in\{0,\ldots,7\}$, and $\kappa_p^{ij}=\beta_{s(p)}q_p\alpha_p^{ij}$ be the particle response coefficient from Eq.~\ref{eq_ecsim_mass_matrix}, with $s(p)$ the species of $p$. Particles are processed in batches $\ell$ of four, $p_{\ell,k}$ with $k=0,\ldots,3$, which form the contraction dimension of one matrix-engine instruction. The cell-local mass matrix is then
\begin{equation}
 \begin{aligned}
  M_{ab}^{ij,c}&=\sum_{p\in\mathcal{P}_c}w_{ap}\kappa_p^{ij}w_{bp}
  =\sum_{\ell}\bigl(A^{ij,\ell}B^{\ell}\bigr)_{ab},\\
  A_{ak}^{ij,\ell}&=w_{a p_{\ell,k}}\kappa_{p_{\ell,k}}^{ij},
  \qquad B_{kb}^{\ell}=w_{b p_{\ell,k}},
 \end{aligned}
 \label{eq_cic_mma_update}
\end{equation}
where $k$ is the contraction index over particles. Each batch performs one update $D^{ij}\leftarrow D^{ij}+A^{ij,\ell}B^\ell$ of an $8\times8$ accumulator, and the deposition adds \mbox{$M_{g_c(a),g_c(b)}^{ij}\mathrel{+}=M_{ab}^{ij,c}$}, with $g_c(a)$ the global node of corner $a$.

For fixed $(i,j)$, the mass-matrix is symmetric,
\mbox{$M_{ab}^{ij,c}=M_{ba}^{ij,c}$}, thus only the 36 canonical entries of each local
$8\times8$ matrix are written.  Contributions from adjacent cells can target
the same global entry and are therefore combined atomically, but accumulation
within a cell is completed in the registers before this flush.  Thus, the number of global atomic updates is bounded by $36\times9$ per occupied cell and does not scale with its particle population.

On NVIDIA GPUs, TPIC follows~\citet{pennati2026massmatrix} and computes each $8\times8$ update with one FP64 WMMA instruction with tile $(M_t,N_t,K_t)=(8,8,4)$. Here, we improve upon~\citet{pennati2026massmatrix} by adding AMD support, so TPIC executes mass-matrix assembly on the matrix engines of both NVIDIA and AMD GPUs.

\begin{figure}[pos=t!]
  \centering
\begin{tikzpicture}[
  font=\footnotesize,
  frame/.style={draw=black!75,line width=0.8pt},
  grid/.style={draw=black!28,line width=0.3pt},
  half/.style={draw=black!70,line width=1.1pt},
  op/.style={font=\Large},
  note/.style={font=\scriptsize,align=center},
  qlabel/.style={font=\scriptsize,align=center,fill=white,fill opacity=0.83,
    text opacity=1,inner sep=1pt}
]
  \colorlet{halfL}{black!9}
  \colorlet{halfH}{black!26}
  \colorlet{blk0}{cyan!24}
  \colorlet{blk1}{violet!22}
  \colorlet{blk2}{green!24}
  \colorlet{blk3}{orange!28}

  \begin{scope}[shift={(0,0)}]
    \fill[halfL] (0,1.8) rectangle (1.8,3.6);
    \fill[halfH] (0,0) rectangle (1.8,1.8);
    \foreach \k in {1,2,3}
      \draw[grid] ({0.45*\k},0) -- ({0.45*\k},3.6);
    \foreach \a in {1,...,7}
      \draw[grid] (0,{0.45*\a}) -- (1.8,{0.45*\a});
    \draw[frame] (0,0) rectangle (1.8,3.6);
    \draw[half] (0,1.8) -- (1.8,1.8);
    \node[qlabel] at (0.9,2.7) {$A_L$};
    \node[qlabel] at (0.9,0.9) {$A_H$};
    \foreach \k in {0,...,3}
      \node[font=\tiny,text=black!60] at ({0.45*\k+0.225},-0.16) {\k};
    \node[above=3pt,font=\small\bfseries] at (0.9,3.6) {$A^{ij,\ell}$};
    \node[note] at (0.9,-0.5) {$8$ corners $\times$ $4$ particles $k$};
  \end{scope}

  \node[op] at (2.35,1.8) {$\times$};

  \begin{scope}[shift={(2.9,0.9)}]
    \fill[halfL] (0,0) rectangle (1.8,1.8);
    \fill[halfH] (1.8,0) rectangle (3.6,1.8);
    \foreach \b in {1,...,7}
      \draw[grid] ({0.45*\b},0) -- ({0.45*\b},1.8);
    \foreach \k in {1,2,3}
      \draw[grid] (0,{0.45*\k}) -- (3.6,{0.45*\k});
    \draw[frame] (0,0) rectangle (3.6,1.8);
    \draw[half] (1.8,0) -- (1.8,1.8);
    \node[qlabel] at (0.9,0.9) {$B_L$};
    \node[qlabel] at (2.7,0.9) {$B_H$};
    \foreach \k in {0,...,3}
      \node[font=\tiny,text=black!60] at (-0.16,{1.575-0.45*\k}) {\k};
    \node[above=3pt,font=\small\bfseries] at (1.8,1.8) {$B^{\ell}$};
    \node[note] at (1.8,-1.4) {$4$ particles $k$ $\times$ $8$ corners};
  \end{scope}

  \node[op] at (7.1,1.8) {$\longrightarrow$};

  \begin{scope}[shift={(7.7,0)}]
    \fill[blk0] (0,1.8) rectangle (1.8,3.6);
    \fill[blk1] (1.8,1.8) rectangle (3.6,3.6);
    \fill[blk2] (0,0) rectangle (1.8,1.8);
    \fill[blk3] (1.8,0) rectangle (3.6,1.8);
    \foreach \t in {1,...,7} {
      \draw[grid] ({0.45*\t},0) -- ({0.45*\t},3.6);
      \draw[grid] (0,{0.45*\t}) -- (3.6,{0.45*\t});
    }
    \draw[frame] (0,0) rectangle (3.6,3.6);
    \draw[half] (1.8,0) -- (1.8,3.6);
    \draw[half] (0,1.8) -- (3.6,1.8);
    \node[qlabel] at (0.9,2.7) {$\beta=0$\\$A_LB_L$};
    \node[qlabel] at (2.7,2.7) {$\beta=1$\\$A_LB_H$};
    \node[qlabel] at (0.9,0.9) {$\beta=2$\\$A_HB_L$};
    \node[qlabel] at (2.7,0.9) {$\beta=3$\\$A_HB_H$};
    \node[above=3pt,font=\small\bfseries] at (1.8,3.6) {$D^{ij}$};
    \node[note] at (1.8,-0.5) {$8\times8$ corner pairs};
  \end{scope}

  \node[font=\small] at (5.6,4.75)
    {$D^{ij}\leftarrow D^{ij}+A^{ij,\ell}B^\ell$ \quad (one MFMA, four $4\times4\times4$ blocks)};
\end{tikzpicture}
  \caption{Cell-local FP64 MFMA update on AMD GPUs for one tensor component $(i,j)$
  and four-particle batch $\ell$. Each column of $A^{ij,\ell}$ contains a
  particle's eight CIC weights multiplied by its response coefficient, and the
  matching row of $B^\ell$ contains the unscaled weights. The corner sets $L=\{0,1,2,3\}$ and $H=\{4,5,6,7\}$ split the product into four independent $4\times4\times4$ blocks $\beta$ of one MFMA instruction, each updating one quadrant of the $8\times8$ accumulator $D^{ij}$.}
  \label{fig_mass_matrix_product}
\end{figure}
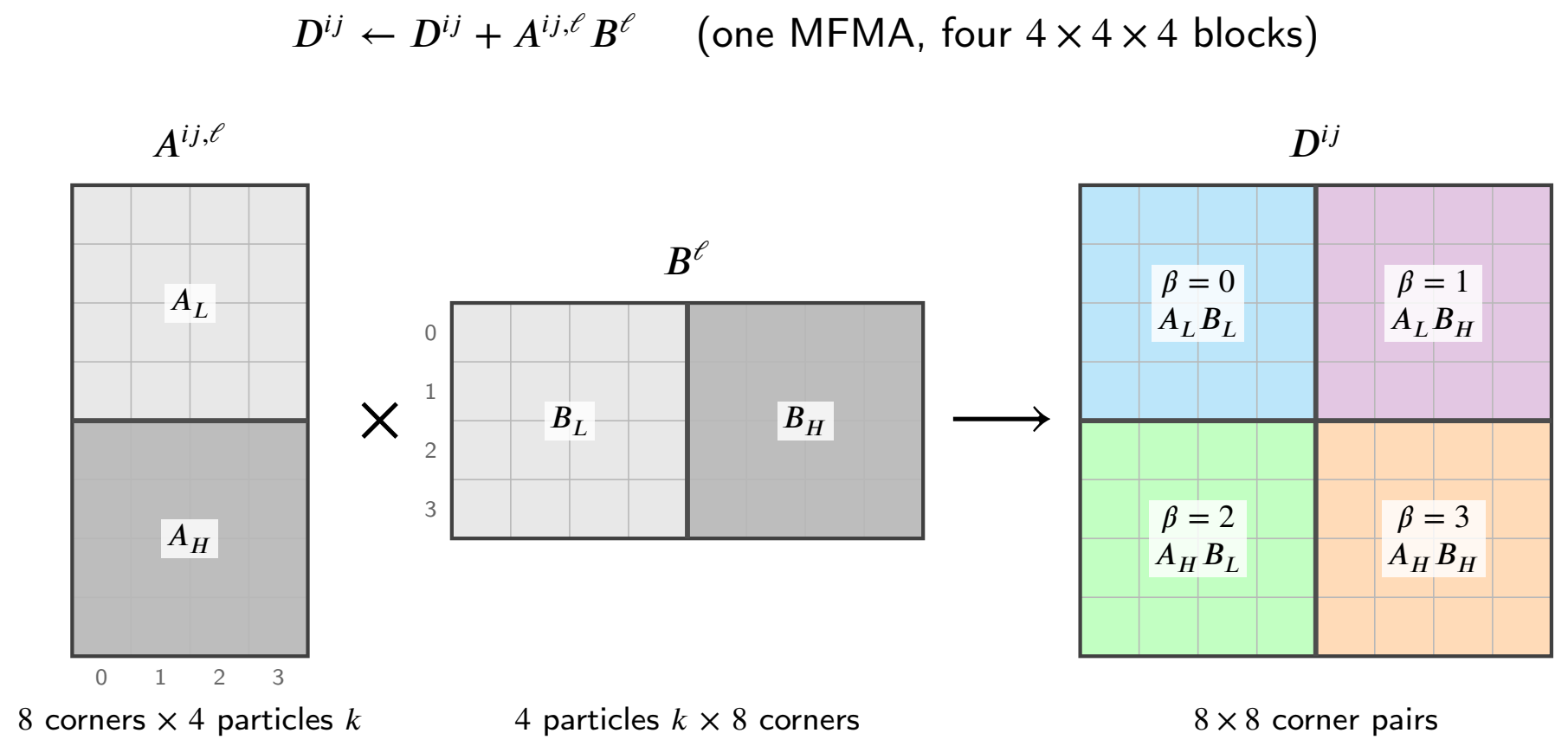

AMD GPUs do not have an instruction equivalent to the NVIDIA FP64 $8\times8\times4$ tensor core operation, but the HIP FP64 MFMA builtin \verb|__builtin_amdgcn_mfma_f64_4x4x4f64| can perform four independent products with tiles $(M_t,N_t,K_t)=(4,4,4)$ across a 64-lane wavefront in one single instruction. Thus, on HIP, TPIC partitions the eight cell corners into two sets, \mbox{$L=\{0,1,2,3\}$} and $H=\{4,5,6,7\}$, which split the rows of $A^{ij,\ell}$ into $A_L$ and $A_H$ and the columns of $B^\ell$ into $B_L$ and $B_H$. The $8\times8$ cell-local mass matrix is thus divided in four quadrants $(L,L)$, $(L,H)$, $(H,L)$, and $(H,H)$, computed by the four products $A_LB_L$, $A_LB_H$, $A_HB_L$, and $A_HB_H$, as illustrated in Fig.~\ref{fig_mass_matrix_product}. One MFMA instruction executes the four $4\times4$ matrix products concurrently, and therefore covers the same $8\times8$ logical result as one NVIDIA MMA instruction. The hardware blocks are interleaved across wavefront lanes, and a shuffle obtains the partner-corner weight for each lane's second operand.

One wavefront owns a cell, with four wavefronts in a 256-thread block.  Each
lane loads one particle per 64-particle chunk.  The wavefront processes that
chunk in four-particle batches and keeps one FP64 accumulator in register per lane and
tensor component, for nine persistent values.  The final batch is zero padded, and all 64 lanes participate in each MFMA instruction.  The operands $A^{ij,\ell}$ and $B^\ell$ of Eq.~\ref{eq_cic_mma_update} have 32 entries each, but the four blocks read 64 values of each operand, one per lane. Since $A_L$ enters the $(L,L)$ and $(L,H)$ blocks, $A_H$ the $(H,L)$ and $(H,H)$ blocks, and likewise for $B_L$ and $B_H$, each entry is held by two lanes. The register accumulators persist during the entire cell traverse, then quantities are written in main memory with an atomic operation.

\subsection{Mixed precision}
\label{sec_mixed_precision}
TPIC separates, for each quantity, the storage type used in memory from the compute type of the arithmetic that produces it. All kernels are templates on storage and compute types for particle positions and velocities, $\mathbf{E}$, $\mathbf{B}$, $\widehat{\mathbf{J}}$, and $\mathbf{M}$, and every conversion between the two is explicit. When storage and compute type coincide, no type casting instruction is issued. At runtime, the pair is selected from those compiled, by default, it matches the five levels in Table~\ref{tab_precision_levels}. Particle charges, GMRES inner products and norms, and diagnostic reductions are always computed in FP64. Single-precision storage halves the memory footprint and traffic for the corresponding arrays, which is most relevant for particles.

\begin{table}[pos=t!]
  \centering
  \caption{TPIC precision levels, with the floating-point width (64 or 32 bits) of particle positions $\mathbf{x}_p$ and velocities $\mathbf{v}_p$, fields, current, and mass matrix. Storage and compute types coincide, except for the mass matrix of the fourth level, stored in FP32 and computed in FP64 (32/64). Particle charges and all reductions are always FP64.}
  \label{tab_precision_levels}
  \footnotesize
  \begin{tabular}{@{}lccccc@{}}
    \toprule
    Level & $\mathbf{x}_p$ & $\mathbf{v}_p$ & $\mathbf{E},\mathbf{B}$ & $\widehat{\mathbf{J}}$ & $\mathbf{M}$ \\
    \midrule
    \texttt{double} & 64 & 64 & 64 & 64 & 64 \\
    \texttt{single-position} & 32 & 64 & 64 & 64 & 64 \\
    \texttt{single-position-velocity} & 32 & 32 & 64 & 64 & 64 \\
    \texttt{single-particles-moments-fields-double-mass} & 32 & 32 & 32 & 32 & 32/64 \\
    \texttt{single-particles-moments-fields} & 32 & 32 & 32 & 32 & 32 \\
    \bottomrule
  \end{tabular}
\end{table}

Particle positions are stored as cell-relative coordinates in $[0,1)$, together with an integer cell index. FP32 positions thus have a resolution of at least $2^{-24}$ of the cell size, independently of the domain size. Since positions enter the energy balance only through the interpolation weights, which deposition and push compute in FP64 from the same stored coordinates, single-precision positions preserve exact energy conservation (Section~\ref{sec_reconnection}). Lower precision for velocities, fields, or moments instead introduces single-precision round-off in the energy balance.

In GMRES, the Krylov basis and the operator applications use the compute type of $\mathbf{E}$, while the solution, the right-hand side, and the residual are kept in FP64. Thus, with FP32 fields, each restart cycle is an iterative-refinement step, and we always test convergence on the FP64 residual \mbox{$\mathbf{b}-\mathcal{A}\mathbf{E}^{n+1/2}$}. Since GMRES reads $\mathbf{M}$ at every iteration, its storage type determines a large part of the memory traffic of the solve. The fourth level of Table~\ref{tab_precision_levels} therefore stores $\mathbf{M}$ in FP32 but computes it in FP64, which enables the FP64 matrix-engine deposition of Section~\ref{sec_tensor_mass_deposition}, required by this kernel. The matrix-engine kernel accumulates the contributions of each cell in FP64 registers and rounds them only once, at the atomic flush.

\subsection{Normalization of physical quantities}
\label{sec_normalization}

In TPIC, physical quantities are expressed in dimensionless plasma units based on the Gaussian (CGS) system. Let $n_0$ be the reference particle density, $m_i$ the reference ion mass, $e$ the elementary charge, and $c_{\mathrm{phys}}$ the physical speed of light. The reference frequency (the ion plasma frequency $\omega_{pi}$), the reference length (the ion skin depth $d_i$), and the reference electromagnetic-field strength $F_\star$ are
\begin{equation}
  \omega_{pi}=\sqrt{\frac{4\pi n_0e^2}{m_i}},
  \qquad d_i=\frac{c_{\mathrm{phys}}}{\omega_{pi}},
  \qquad F_\star=\frac{m_i c_{\mathrm{phys}}\omega_{pi}}{e}.
  \label{eq_reference_units}
\end{equation}
Using the superscript ``phys'' for dimensional quantities, the normalized variables are
\begin{equation}
  \begin{aligned}
    t&=\omega_{pi}t^{\mathrm{phys}},
    &\mathbf{x}&=\frac{\mathbf{x}^{\mathrm{phys}}}{d_i},
    &\mathbf{v}&=\frac{\mathbf{v}^{\mathrm{phys}}}{c_{\mathrm{phys}}},\\
    \mathbf{E}&=\frac{\mathbf{E}^{\mathrm{phys}}}{F_\star},
    &\mathbf{B}&=\frac{\mathbf{B}^{\mathrm{phys}}}{F_\star}.
  \end{aligned}
  \label{eq_normalized_variables}
\end{equation}
Thus, the normalized speed of light is $c=1$. Species charge-to-mass ratios are expressed in units of $e/m_i$. Singly charged ions have $q/m=+1$, while electrons have $q/m=-\mu$, with $\mu=m_i/m_e$. At the reference density, $\omega_{pe}/\omega_{pi}=\sqrt{\mu}$ and $d_e/d_i=\mu^{-1/2}$. The normalized magnetic-field magnitude equals $\Omega_{ci}/\omega_{pi}$.

Charge and current densities retain the Gaussian convention of the Maxwell equations:
\begin{equation}
  \rho=\frac{\rho^{\mathrm{phys}}}{4\pi e n_0},
  \qquad
  \mathbf{J}=\frac{\mathbf{J}^{\mathrm{phys}}}
                   {4\pi e n_0c_{\mathrm{phys}}}.
  \label{eq_normalized_sources}
\end{equation}
With these scalings, the Maxwell equations take the form
\begin{equation}
  \nabla\cdot\mathbf{E}=4\pi\rho,
  \qquad
  \frac{\partial\mathbf{E}}{\partial t}=\nabla\times\mathbf{B}-4\pi\mathbf{J},
  \qquad
  \frac{\partial\mathbf{B}}{\partial t}=-\nabla\times\mathbf{E},
  \qquad
  \nabla\cdot\mathbf{B}=0,
  \label{eq_normalized_maxwell}
\end{equation}
and the particle equation of motion
\begin{equation}
  \frac{d\mathbf{v}}{dt}=\frac{q}{m}\left(\mathbf{E}+\mathbf{v}\times\mathbf{B}\right).
  \label{eq_normalized_motion}
\end{equation}
We choose the factor $4\pi$ in Eq.~\ref{eq_normalized_sources} so the sources enter the field equations as $4\pi\rho$ and $4\pi\mathbf{J}$, as in the Gaussian system.

Let $Z_s e$ be the charge of a particle of species $s$ and let
\begin{equation}
  \hat\rho_s=\frac{|Z_s|\,n_s}{n_0}
\end{equation}
be the charge density of species $s$ in units of $e\,n_0$. For singly charged species $\hat\rho_s=n_s/n_0$. The normalized charge density is then $\rho_s=\operatorname{sign}(q/m)_s\,\hat\rho_s/(4\pi)$, where $\operatorname{sign}(q/m)_s=\pm1$ is the sign of the charge of species $s$. The charge-to-mass ratio in units of $e/m_i$ is $(q/m)_s=\pm Z_s\,m_i/m_s$, and the plasma frequency of species $s$ in normalized units is $\omega_{ps}^2=4\pi\rho_s\,(q/m)_s=\hat\rho_s\,|q/m|_s$. For the reference ions ($\hat\rho=1$, $q/m=1$) this gives $\omega_{pi}=1$.

Pressure and energy density are measured in units of
\begin{equation}
  P_\star=F_\star^2=4\pi n_0m_i c_{\mathrm{phys}}^2,
  \qquad
  \mathcal{E}_\star=P_\star d_i^3,
  \label{eq_normalized_energy_units}
\end{equation}
where $\mathcal{E}_\star$ is the unit of integrated energy. With these scales, the field energy density is $(|\mathbf{E}|^2+|\mathbf{B}|^2)/(8\pi)$ and the particle kinetic energy is $\sum_p m_p|\mathbf{v}_p|^2/2$, with the computational particle mass $m_p$ of Section~\ref{sec:ecsim_algorithm} in units of $4\pi n_0m_id_i^3$. Thermal velocities denote one-dimensional standard deviations: $v_{\mathrm{th},s}=\sqrt{k_{\mathrm B}T_s/m_s}/c_{\mathrm{phys}}$.

\section{Experimental setup}
\label{sec_experimental_setup}
We run the experiments on a single node system equipped with an NVIDIA Grace Hopper 200 GPU, where the host is an ARM Neoverse-V2 72 cores CPU. The device has 96 GB of HBM, while the host has 586 GB of RAM. 

Large-scale simulations are performed on the accelerated partition of LUMI (LUMI-G), where each compute node has one 64-core AMD EPYC 7A53 CPU with 512~GB of DDR4 memory and four AMD MI250x accelerators. Each MI250x comprises two independent Graphics Compute Dies (GCDs), each with 64~GB of HBM. Thus, each node provides eight GCDs, and a total of 512~GB of GPU memory. The accelerators and CPU are interconnected through AMD Infinity Fabric, while inter-node communication is provided by four 200~Gbit/s HPE Cray Slingshot-11 network interfaces.

Table~\ref{tab_software} lists the software used to build and run TPIC on the two systems. Both builds use CMake Release mode. On LUMI-G, we trace MPI with CrayPat, GPU kernels with rocprofv3, and the full timeline with rocprof-sys. Kernel metrics come from Nsight Compute on GH200 and rocprof-compute on MI250x.

If not reported otherwise, we run all experiments with TPIC partial sorting, matrix engines mass matrix assembly, and the \verb|single-position| setup. Thus, all quantities are in FP64, except particle positions, which are in FP32. We use device-aware MPI as the communication backend.

\begin{table}[pos=t!]
  \centering
  \caption{Software used to build and run TPIC.}
  \label{tab_software}
  \small
  \setlength{\tabcolsep}{4pt}
  \begin{tabular}{@{}lll@{}}
    \toprule
    & GH200 & LUMI-G (MI250x) \\
    \midrule
    C++ compiler & \texttt{nvc++} (NVHPC 26.5) & \texttt{hipcc}, AMD Clang 19.0.0 \\
    Device compiler & \texttt{nvcc}, GCC 14.2.1 host & AMD Clang 19.0.0 \\
    GPU toolkit & CUDA 13.2 & ROCm/HIP 6.4.4 \\
    Target & \texttt{sm\_90} & \texttt{gfx90a} \\
    MPI & Open MPI 5.0.10rc2 (HPC-X 2.50) & Cray MPICH 9.0.1 \\
    RCCL & -- & 2.22.3, aws-ofi-nccl 1.21.1 \\
    Profilers & Nsight Compute & CrayPat 25.09, rocprofv3, rocprof-sys, rocprof-compute \\
    \bottomrule
  \end{tabular}
\end{table}

\section{Numerical results}
\label{sec_numerica_results}

\subsection{Finite grid instability removal}
\label{sec_fgi}
We use a cold drifting plasma to verify that TPIC is free of the finite-grid instability~\citep{lapenta2017ecsim}. In explicit PIC, a plasma whose Debye length $\lambda_{De}$ is not resolved by the grid is heated numerically by the aliasing of the particle density on the grid. We initialize cold electrons and ions of density $n_0$, with mass ratio $\mu=2000$, both drifting along $x$ with $u_0=0.2$. The plasma carries no current and is in equilibrium, i.e., a plasma at rest seen from a moving frame, with no physical instability. We use a periodic domain of length $L_x=128\,d_e=2.86\,d_i$, discretized with $128\times1\times1$ cells, $\Delta x=d_e=2.24\times10^{-2}\,d_i$, so that the grid does not resolve $\lambda_{De}=0$. We initialize 256 particles per cell per species, with electrons and ions at the same position, so that the initial charge density vanishes, and $\mathbf{E}=\mathbf{B}=0$. We run the simulation for $40000$ steps with $\theta=1/2$ and $\Delta t=5.59\times10^{-4}\,\omega_{pi}^{-1}=0.025\,\omega_{pe}^{-1}$, up to $T=1000\,\omega_{pe}^{-1}$. We compare TPIC with a 1D explicit, momentum-conserving electrostatic PIC code, with CIC shape functions, a second-order finite difference Poisson solver and a leapfrog mover, which starts from the same particles and uses the same $\Delta x$ and $\Delta t$.

\begin{figure}
    \centering
    \includegraphics[width=0.5\linewidth]{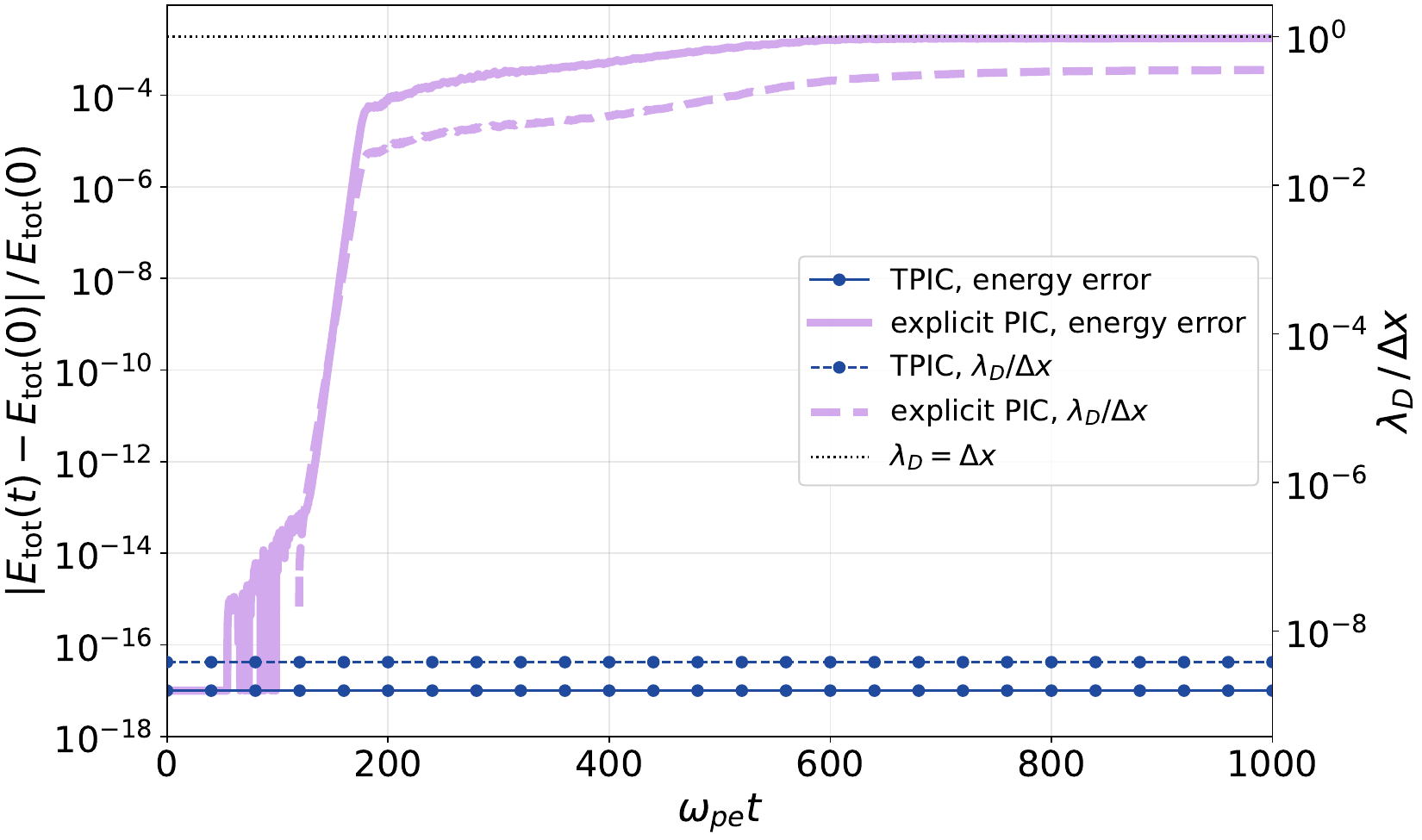}
    \caption{Finite-grid instability test. Relative error of the total energy (solid, left axis) and electron Debye length in units of the cell size (dashed, right axis) in TPIC and in the explicit PIC code. TPIC values are at round-off level; a zero energy error is drawn at $10^{-17}$. The dotted line marks $\lambda_{De}=\Delta x$.}
    \label{fig_fgi_energy}
\end{figure}

Fig.~\ref{fig_fgi_energy} shows the relative error of the total energy and the electron Debye length $\lambda_{De}=v_{\mathrm{th},e}/\omega_{pe}$, with $v_{\mathrm{th},e}$ the velocity spread about the drift. TPIC removes the finite-grid instability and keeps the initial equilibrium, with a GMRES solver tolerance of $1\times10^{-16}$, the total energy is conserved to machine precision, the electric energy stays at $10^{-31}$ of the initial electron kinetic energy, and the electrons remain cold. In the explicit code, the round-off perturbations of the charge density grow exponentially at $\gamma\simeq0.21\,\omega_{pe}$ up to $\omega_{pe}t\simeq180$, and heat the plasma. The heating continues until the Debye length is comparable to the grid spacing, $\lambda_{De}$ grows from zero and saturates at $\lambda_{De}\simeq0.36\,\Delta x$. 

\begin{figure}
    \centering
    \includegraphics[width=0.7\linewidth]{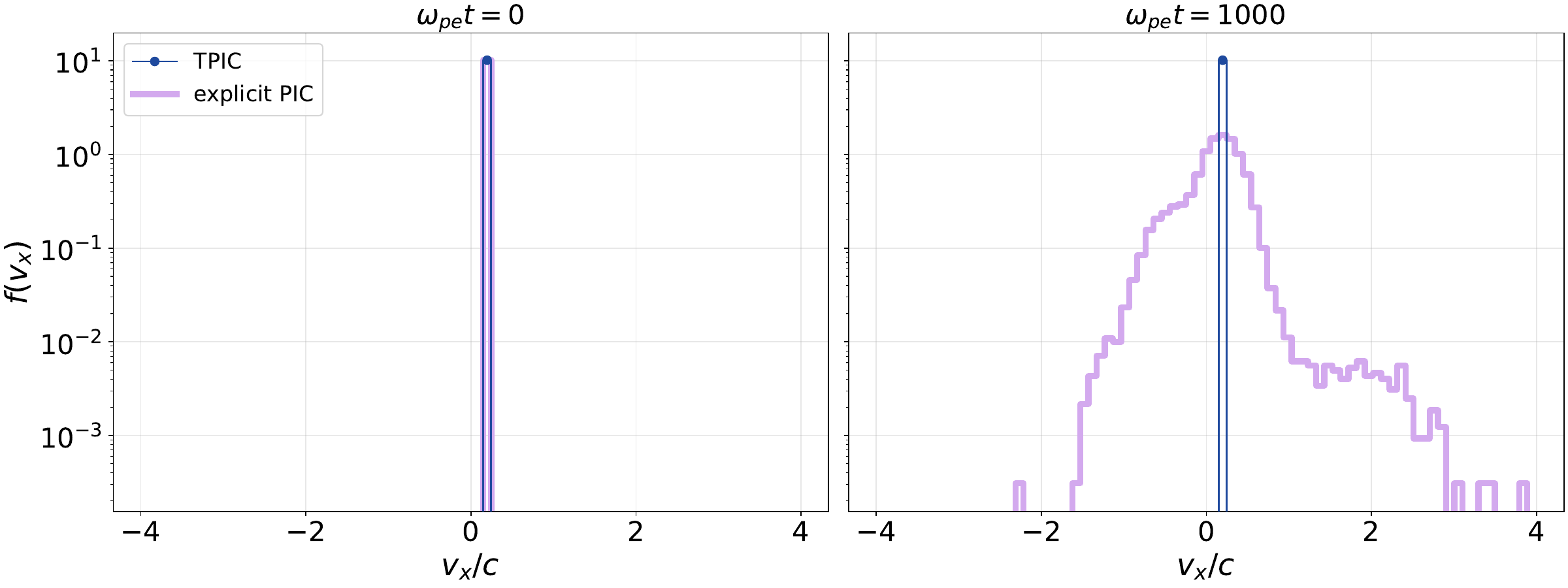}
    \caption{Finite-grid instability test. Electron velocity distribution in TPIC and in the explicit PIC code at $t=0$ (left) and $t=T$ (right).}
    \label{fig_fgi_distribution}
\end{figure}

\begin{figure}
    \centering
    \includegraphics[width=0.7\linewidth]{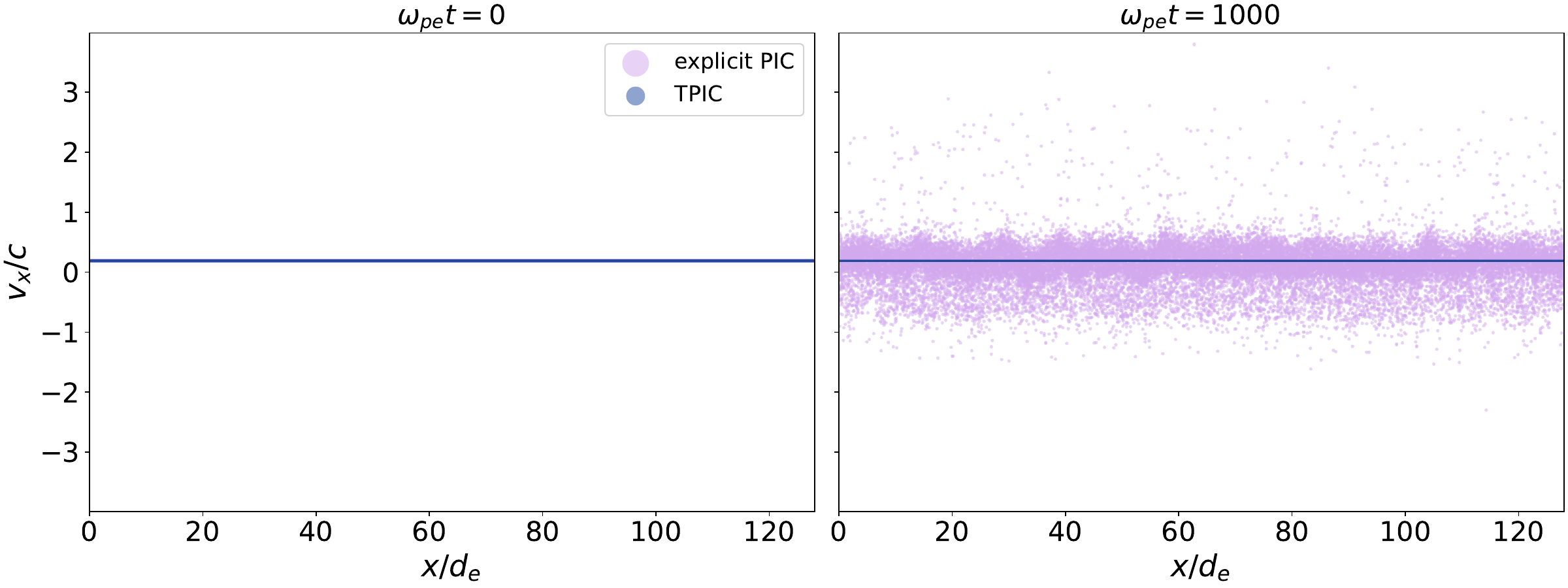}
    \caption{Finite-grid instability test. Electron phase space $(x,v_x)$ of TPIC (dark blue) and the explicit PIC code (light purple), at $t=0$ (left) and $t=T$ (right).}
    \label{fig_fgi_phase_space}
\end{figure}

Figs.~\ref{fig_fgi_distribution} and~\ref{fig_fgi_phase_space} show the electron velocity distribution and phase space at $t=0$ and $t=T$. In TPIC, the electrons remain a cold beam at $v_x=u_0$. In the explicit code, the distribution broadens, and the numerical heating accelerates electrons to unphysical superluminal velocities. Electrons exceed $|v_x|=c$ after $\omega_{pe}t\simeq500$, and at $T$, $1.2\%$ of them are superluminal, with tails up to $v_x\simeq3.8\,c$.

\subsection{Two-stream instability}
\label{sec_two_stream}
We use the two-stream instability test to verify the electrostatic response of TPIC~\citep{lapenta2017ecsim}. In this experiment, we analyze the linear growth rate of a seeded mode and the exchange of energy between particles and the $\mathbf{E}$ field. We initialize the system with two cold electron beams, each of density $n_0/2$, drifting along $x$ with velocities $\pm u_b$, $u_b=0.2$, with a cold ion background of density $n_0$, with mass ratio $\mu=2000$. We use a periodic domain of length $L_x=2\pi\,d_e=0.1405\,d_i$, discretized with $128\times1\times1$ cells, $\Delta =1.10\times10^{-3}\,d_i=4.9\times10^{-2}\,d_e$. We initialize 256 particles per cell per species. We seed the mode $m=3$, which has $k=2\pi m/L_x=3\,d_e^{-1}$ introducing a perturbation in particle densities $n=n_0(1+\delta\cos(kx))$, with $\delta=10^{-4}$. The electric field is initialized from Gauss's law, $E_x=-(\delta/k)\sin(kx)$. We run the simulation for $2000$ steps with $\theta=1/2$ and $\Delta t =5.59\times10^{-4}\,\omega_{pi}^{-1}=0.025\,\omega_{pe}^{-1}$, up to $T=50\,\omega_{pe}^{-1}$.

For cold beams and immobile ions, a perturbation $\propto\exp[i(kx-\omega t)]$ follows the dispersion relation
\begin{equation}
  1=\frac{\omega_{pe}^2/2}{(\omega-ku_b)^2}+\frac{\omega_{pe}^2/2}{(\omega+ku_b)^2},
  \label{eq_two_stream_disp}
\end{equation}
whose solutions are $\omega^2/\omega_{pe}^2=a^2+\tfrac12\pm\tfrac12\sqrt{1+8a^2}$, with $a=ku_b/\omega_{pe}$. For $a<1$, we have a growing mode with growth rate
\begin{equation}
  \gamma=\omega_{pe}\left(\tfrac12\sqrt{1+8a^2}-a^2-\tfrac12\right)^{1/2}.
  \label{eq_two_stream_gamma}
\end{equation}
In the experiment, the seeded mode has $a=0.6$, thus $\gamma=0.3534\,\omega_{pe}$. 

\begin{figure}
    \centering
    \subfloat{\includegraphics[width=0.5\linewidth]{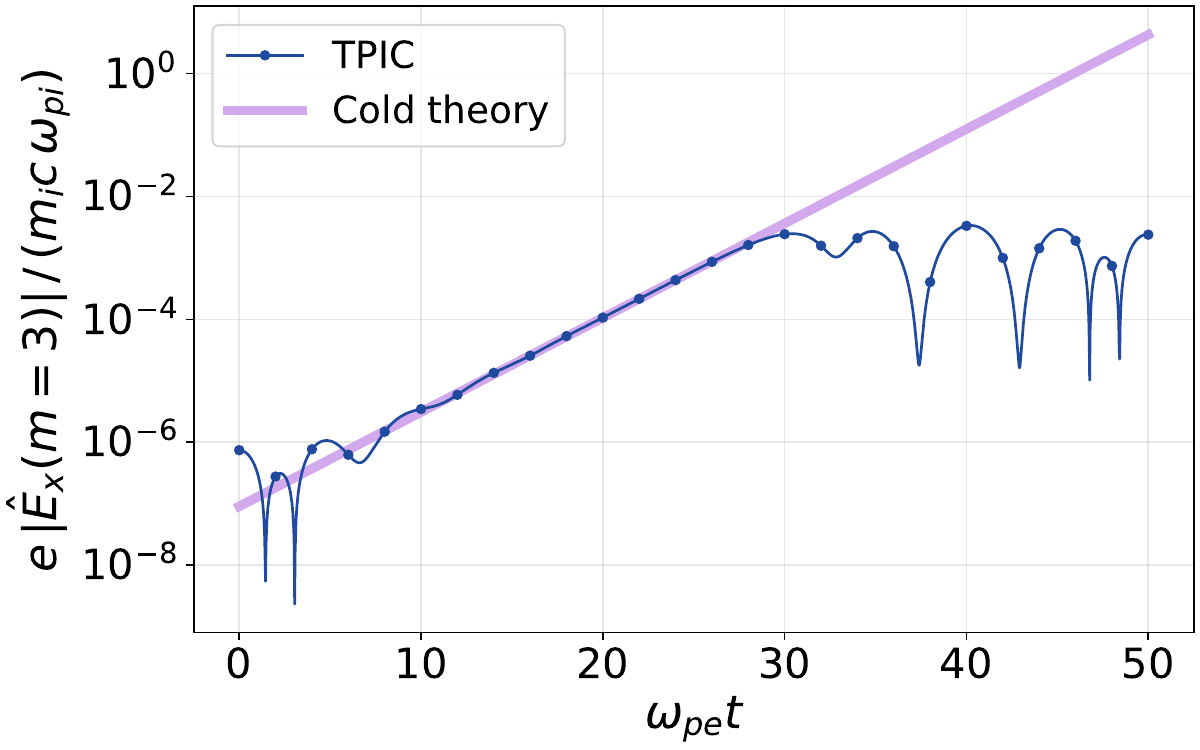}}
    \subfloat{\includegraphics[width=0.5\linewidth]{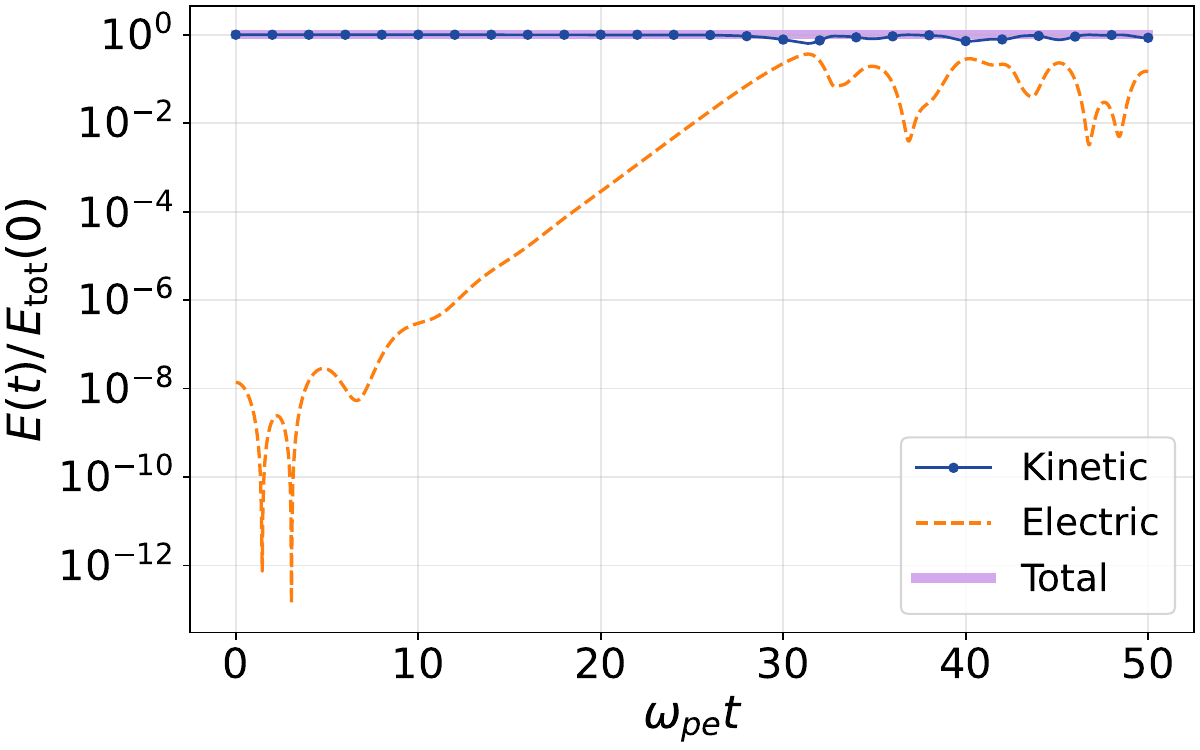}}
    \caption{Two-stream instability. Left: amplitude of the seeded Fourier mode $m=3$ of $E_x$ in TPIC and exponential growth at the cold rate $\gamma$ of Eq.~\ref{eq_two_stream_gamma}. Right: kinetic, electric and total energy, in units of the initial total energy.}
    \label{fig_two_stream}
\end{figure}

Fig.~\ref{fig_two_stream} (left) shows the amplitude $|\hat E_x(m=3)|$ of Fourier transform of the seeded mode, which grows exponentially. A fit of $\ln|\hat E_x|$ over $10\le\omega_{pe}t\le25$ gives $\gamma=0.3539\,\omega_{pe}$, $0.13\%$ above the theoretical growth rate.

\begin{figure}
    \centering
    \includegraphics[width=\linewidth]{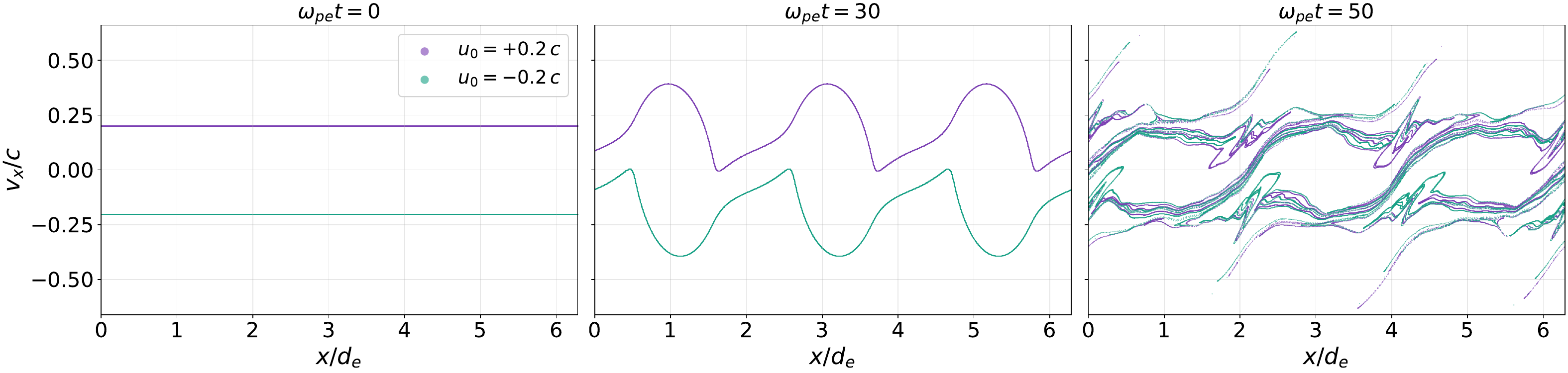}
    \caption{Two-stream instability. Electron phase space $(x,v_x)$ of the two beams at $\omega_{pe}t=0$, $30$ and $50$.}
    \label{fig_two_stream_phase_space}
\end{figure}

The growth saturates at $\omega_{pe}t\simeq30$, due to the trapping of the beam electrons. Fig.~\ref{fig_two_stream_phase_space} shows the electron phase space. At $\omega_{pe}t=30$, the two beams reach $v_x=0$ and are trapped by the wave, and at $T$ the trapped electrons form phase-space vortices. Fig.~\ref{fig_two_stream} (right) shows the kinetic, electric, and total energy, normalized to the initial total system energy. The electric energy grows from $1.4\times10^{-8}$ at the rate $2\gamma$ and peaks at $36\%$ of the total energy at $\omega_{pe}t=31.4$, where the kinetic energy has dropped by the same amount. Then, the two exchange energy periodically. The magnetic energy remains below $10^{-28}$ of the total. With a GMRES solver tolerance of $1\times10^{-16}$, the total system energy is conserved to a relative error of $5.6\times10^{-16}$ over the run, including the nonlinear phase.

\subsection{Quasineutral plasma expansion}

We use the expansion of a quasineutral plasma into vacuum to verify TPIC solver in the electrostatic limit, in three dimensions and in a closed domain. In particular, this nonlinear test investigates the energy exchange between electrons and ions through the ambipolar electric field. We initialize electrons and ions with the same Gaussian density $n=n_0\exp(-r^2/2\sigma_0^2)$, where $r$ is the distance from the domain center and $\sigma_0=0.5\,d_i$. The electrons are Maxwellian with $v_{\mathrm{th},e}=0.05$, the ions are cold, and the mass ratio is $\mu=100$. Thus, $\Lambda=\sigma_0/\lambda_{De0}=100$, with $\lambda_{De0}=v_{\mathrm{th},e}/\omega_{pe}=5\times10^{-3}\,d_i$ the Debye length at the center. The domain of $14\times14\times14\,d_i$ is discretized with $256\times256\times256$ cells, $\Delta x=\Delta y=\Delta z=5.47\times10^{-2}\,d_i=10.9\,\lambda_{De0}$, with PEC walls and reflecting particle boundaries on all faces. We initialize both species at the same positions, $8\times8\times4$ per cell within $r\le4.3\,\sigma_0$, for $1.30\times10^{8}$ particles in total. Thus, the initial charge density vanishes, and we start from $\mathbf{E}=\mathbf{B}=0$. We apply six passes of the filter of Eq.~\ref{eq_smoothing_s6} with $\alpha=0.2$, and run the simulation for $1200$ steps with $\theta=1/2$ and $\Delta t=0.25\,\omega_{pi}^{-1}$ ($\omega_{pe}\Delta t=2.5$, $c\Delta t/\Delta x=4.57$), up to $T=300\,\omega_{pi}^{-1}$. We use 8 AMD MI250x GCDs of LUMI-G with a $2\times2\times2$ domain decomposition.

For $\Lambda\gg1$, the expansion is quasineutral, and the Vlasov equation admits an exact self-similar solution~\citep{Dorozhkina1998quasineutral_expansion,Dorozhkina1998quasineutral_exp_finite_mass}. The density remains Gaussian, the flow is linear in $r$, and the electrons cool adiabatically:
\begin{equation}
  \begin{gathered}
  n(r,t)=n_0\frac{\sigma_0^3}{\sigma^3}\,e^{-r^2/2\sigma^2},
  \qquad \sigma^2=\sigma_0^2\Big(1+\frac{t^2}{\tau^2}\Big),\\
  \mathbf{u}=\frac{t}{t^2+\tau^2}\,\mathbf{r},
  \qquad \frac{T_e}{T_{e0}}=\frac{\sigma_0^2}{\sigma^2},
  \end{gathered}
  \label{eq_expansion}
\end{equation}
with $\tau=\sigma_0/c_s$, and $c_s=v_{\mathrm{th},e}(1+\mu)^{-1/2}$ the initial ion sound speed, which includes the electron inertia. Here, $c_s=4.98\times10^{-3}$ and $\tau=100.4\,\omega_{pi}^{-1}$, so the run ends at $t=2.99\,\tau$. The ambipolar field is radial, so $\mathbf{B}$ remains zero, and the electron thermal energy is converted into ion kinetic energy, $K_i/E_0=\mu/(1+\mu)\,t^2/(t^2+\tau^2)$ and $K_e=E_0-K_i$, with $E_0$ the initial energy, which is only kinetic since $\mathbf{E}=\mathbf{B}=0$.

\begin{figure}
    \centering
    \includegraphics[width=\linewidth]{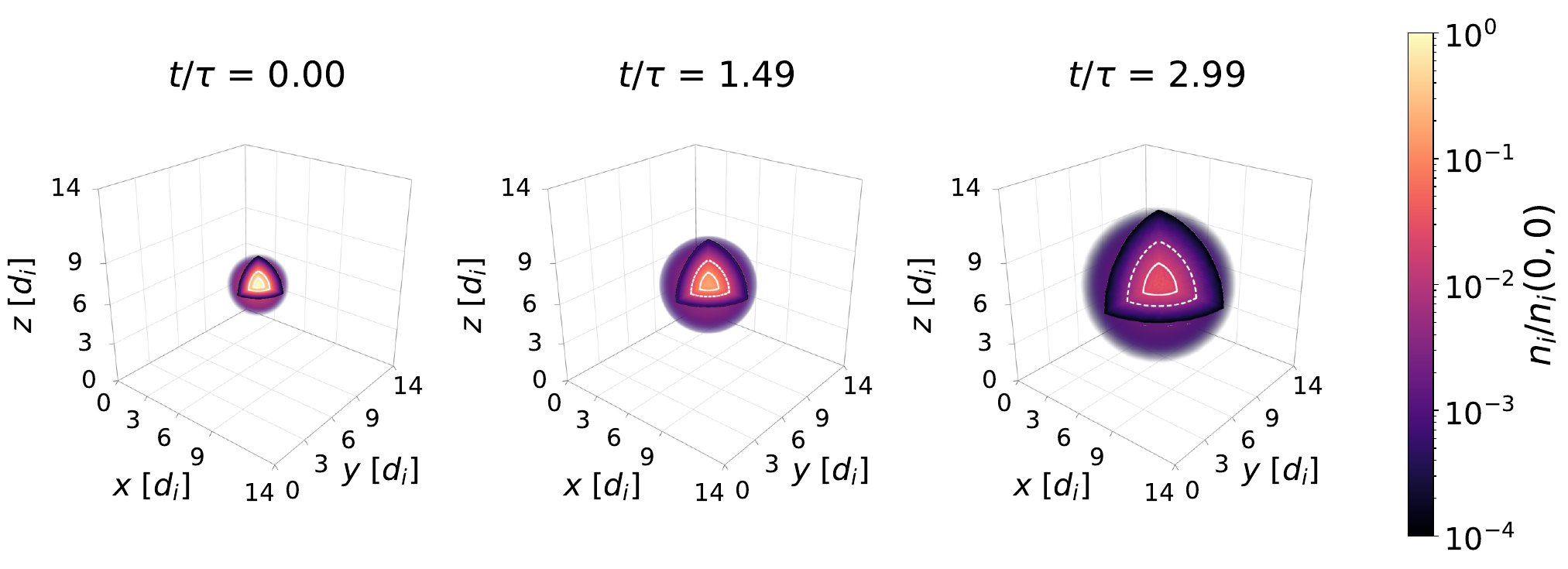}
    \caption{Quasineutral plasma expansion. Ion density at $t/\tau=0$, $1.49$, and $2.99$, in units of the initial central density. The octant facing the viewer is cut away. The white circles on the cut faces mark the analytic $r=\sigma$ (solid) and $r=2\sigma$ (dashed) of Eq.~\ref{eq_expansion}.}
    \label{fig_expansion_density}
\end{figure}

\begin{figure}
    \centering
    \subfloat{\includegraphics[width=0.42\linewidth]{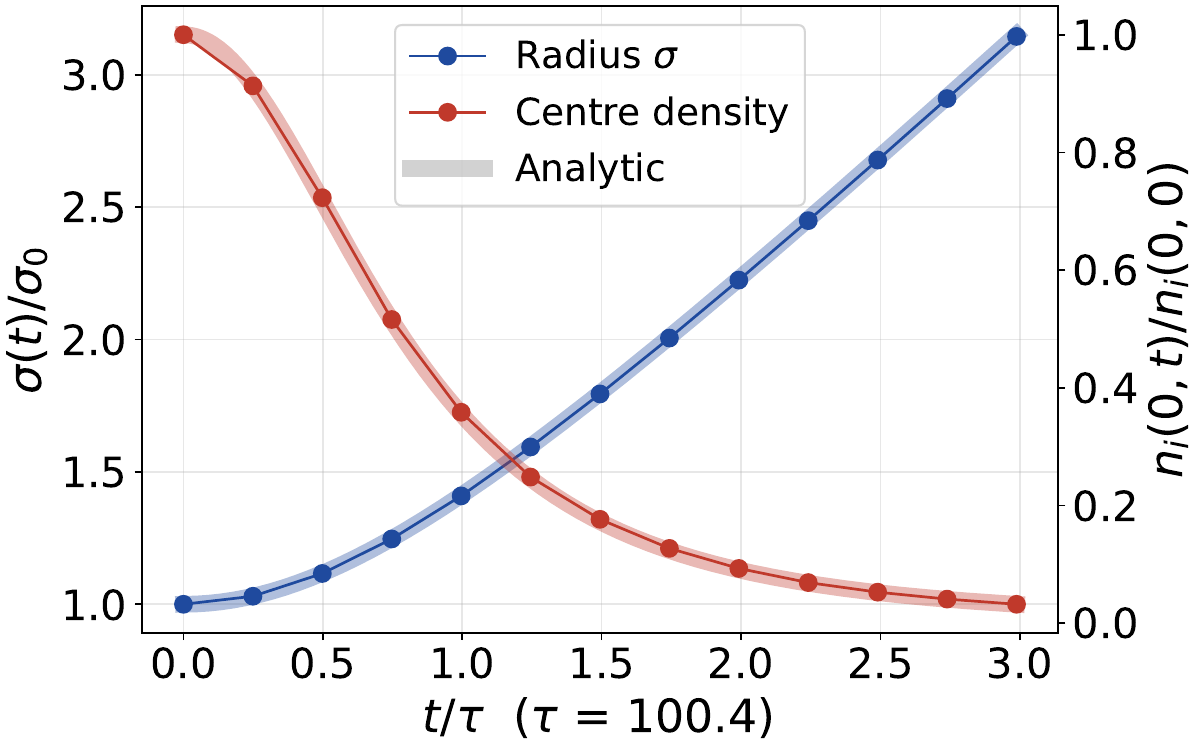}}
    \hspace{4pt}
    \subfloat{\includegraphics[width=0.42\linewidth]{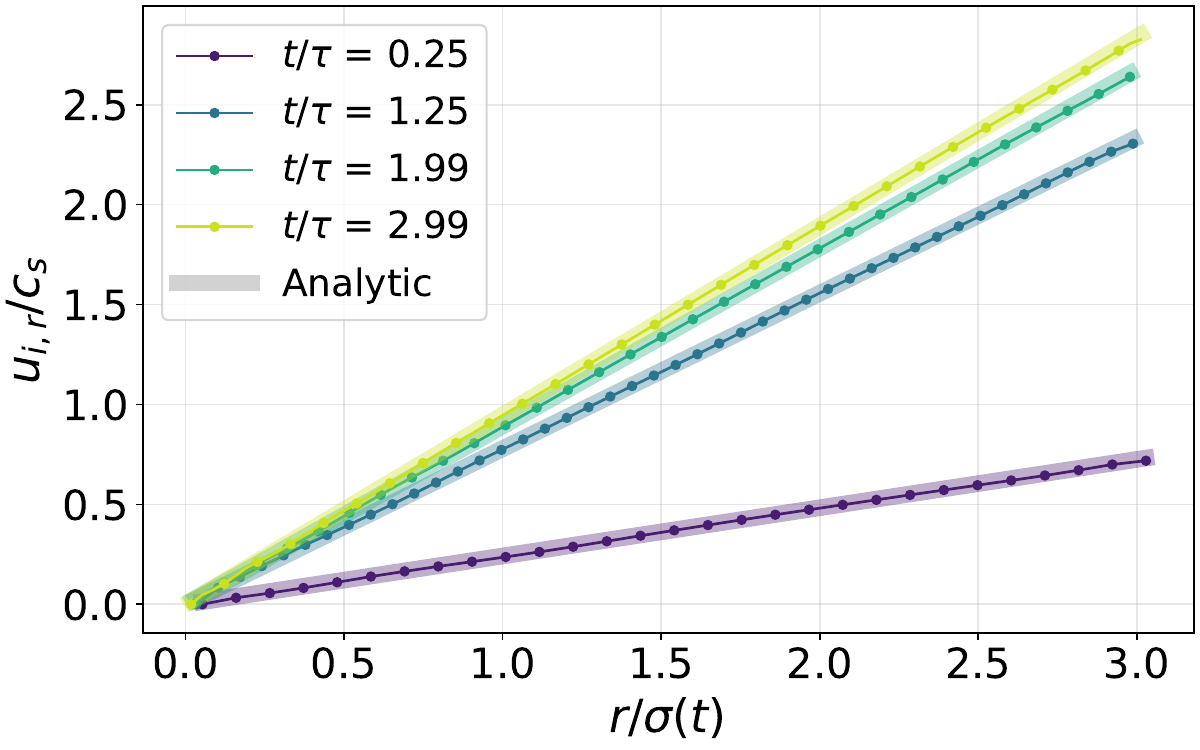}}
    \caption{Quasineutral plasma expansion in TPIC (dotted lines) and Eq.~\ref{eq_expansion} (wide light lines). Left: width of the ion density $\sigma$, averaged over the three axes, and ion density at the center. Right: radial ion flow, averaged over spherical shells, in units of the initial ion sound speed $c_s$, at four times.}
    \label{fig_expansion_profiles}
\end{figure}

\begin{figure}
    \centering
    \subfloat{\includegraphics[width=0.42\linewidth]{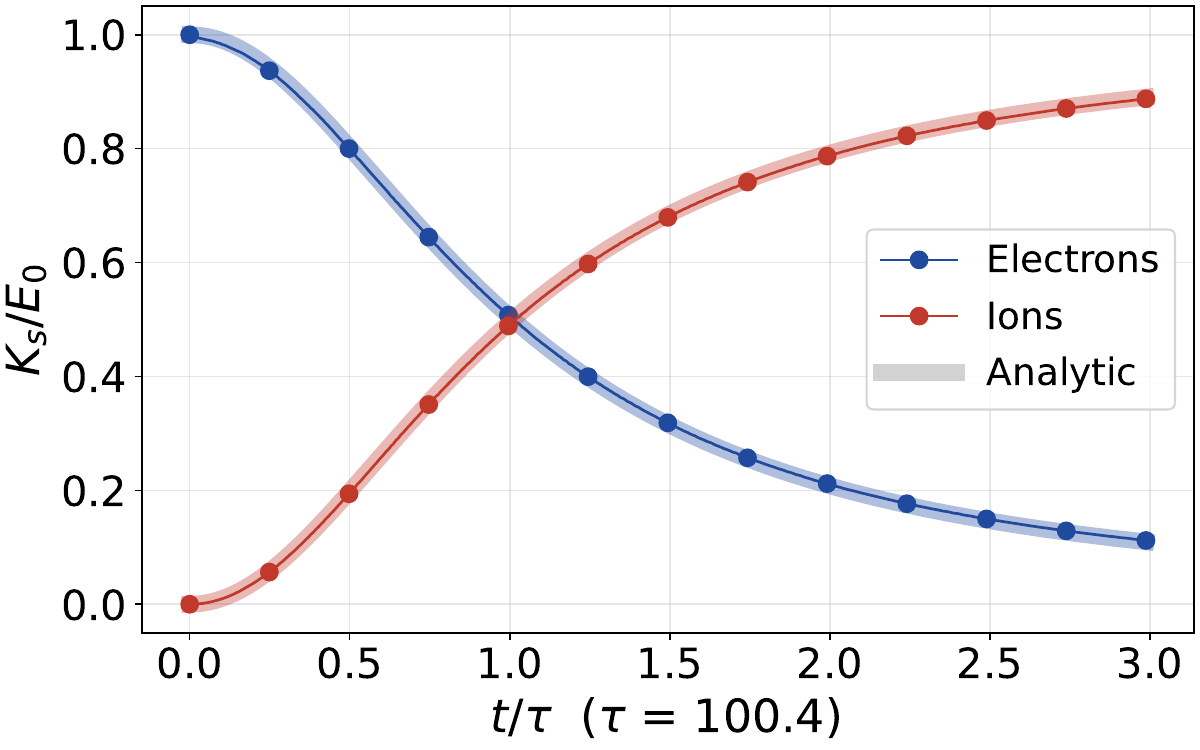}}
    \hspace{4pt}
    \subfloat{\includegraphics[width=0.42\linewidth]{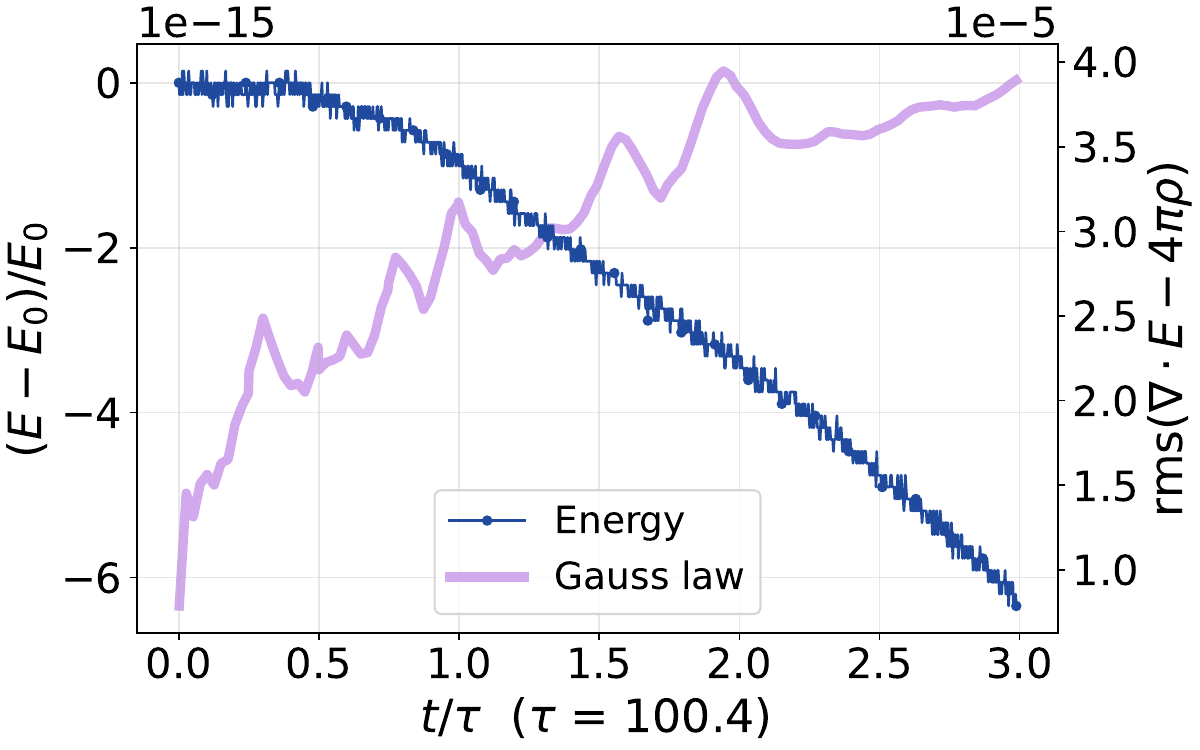}}
    \caption{Quasineutral plasma expansion. Left: kinetic energy of electrons and ions in units of the initial energy $E_0$, in TPIC (dotted lines) and analytic (wide light lines). Right: relative change of the total energy (left axis) and rms residual of Gauss's law (right axis).}
    \label{fig_expansion_energy}
\end{figure}

Fig.~\ref{fig_expansion_density} shows the ion density at three times. The plasma remains spherical, the rms widths of the ion density along $x$, $y$, and $z$ differ by less than $0.06\%$. Fig.~\ref{fig_expansion_profiles} (left) compares the width, averaged over the three axes, and the ion density at the center with Eq.~\ref{eq_expansion}. The width agrees within $0.22\%$, and the center density within $3.0\%$. Fig.~\ref{fig_expansion_profiles} (right) shows the radial ion flow, averaged over spherical shells, which follows the linear profile of Eq.~\ref{eq_expansion} up to $r=3\,\sigma$. For $t>\tau/4$, the fitted flow slope is within $0.6\%$ of the analytic one. Fig.~\ref{fig_expansion_energy} (left) shows the kinetic energy of the two species. At $t=2.99\,\tau$, the ions carry $88.7\%$ of $E_0$, $0.3\%$ below the analytic value. The electric field energy, neglected in the analytic solution, reaches at most $0.78\%$ of $E_0$, at $t=0.13\,\tau$, and the magnetic energy at most $1.5\times10^{-5}$.

Fig.~\ref{fig_expansion_energy} (right) shows the relative change of the total energy and the rms residual of Gauss's law. With a GMRES solver tolerance of $1\times10^{-15}$, the total system energy is conserved to a relative error of $6.3\times10^{-15}$ over the run, with PEC walls on all faces. Since ECSIM does not conserve charge exactly, the Gauss's law residual is not zero, but it remains below $4\times10^{-5}$.

\pagebreak
\subsection{Whistler waves}
\label{sec_whistler}
Whistler waves are low frequency waves propagating in magnetized plasmas~\citep{yu2018whistlerPIC,ballerini2024whistler}. We use them to verify the dispersion of electromagnetic waves propagating along a uniform magnetic field, with a setup adapted from the whistler test of~\citet{gonzalezherrero2018ecsim} to a mass ratio $\mu=1000$. We use a periodic 2D domain of $20\,d_i\times1.25\,d_i$ ($632\,d_e\times39.5\,d_e$), discretized with $1024\times64$ cells, $\Delta x=\Delta y=1.95\times10^{-2}\,d_i=0.62\,d_e$. The background field is $\mathbf{B}_0=(0.015,0,0)$, which gives $|\Omega_{ce}|=15\,\omega_{pi}$, $\Omega_{ci}=0.015\,\omega_{pi}$ and $\omega_{pe}/|\Omega_{ce}|=2.11$. The electrons are bi-Maxwellian, with thermal velocities $v_{\mathrm{th},e\parallel}=0.01$ along $\mathbf{B}_0$ and $v_{\mathrm{th},e\perp}=0.03$ across it ($T_{e\perp}/T_{e\parallel}=9$). The ions are Maxwellian with $v_{\mathrm{th},i}=9.49\times10^{-4}$, i.e., $T_i=T_{e\perp}$. We initialize 256 particles per cell per species. We advance the simulation for $6000$ steps with $\theta=1/2$ and $\Delta t=0.015\,\omega_{pi}^{-1}$ ($|\Omega_{ce}|\Delta t=0.225$), up to $T=90\,\omega_{pi}^{-1}=1350\,|\Omega_{ce}|^{-1}$. We do not seed any wave, the modes are excited by the particle noise.

For $\mathbf{k}\parallel\mathbf{B}_0$, the cold-plasma dispersion relation separates into a right-handed and a left-handed relation, with the handedness being the one of the field rotation about $\mathbf{B}_0$:
\begin{align}
  \frac{k^2 c^2}{\omega^2} = R &= 1
    - \frac{\omega_{pe}^2}{\omega\,(\omega - \lvert\Omega_{ce}\rvert)}
    - \frac{\omega_{pi}^2}{\omega\,(\omega + \Omega_{ce})},
    \label{eq_disp_R}\\
  \frac{k^2 c^2}{\omega^2} = L &= 1
    - \frac{\omega_{pe}^2}{\omega\,(\omega + \lvert\Omega_{ce}\rvert)}
    - \frac{\omega_{pi}^2}{\omega\,(\omega - \Omega_{ci})},
    \label{eq_disp_L}
\end{align}
where $\Omega_{ce}$ ($\Omega_{ci}$) is the electron (ion) gyrofrequency and $\omega_{pe}$ ($\omega_{pi}$) the electron (ion) plasma frequency. Each relation has two propagating solutions. Eq.~\ref{eq_disp_R} gives the whistler wave, with $0<\omega<|\Omega_{ce}|$, and a high-frequency electromagnetic mode named the R light wave. Eq.~\ref{eq_disp_L} gives the ion-cyclotron wave, with $0<\omega<\Omega_{ci}$, and the high-frequency L light wave. 

\begin{figure}
    \centering
    \includegraphics[width=0.7\linewidth]{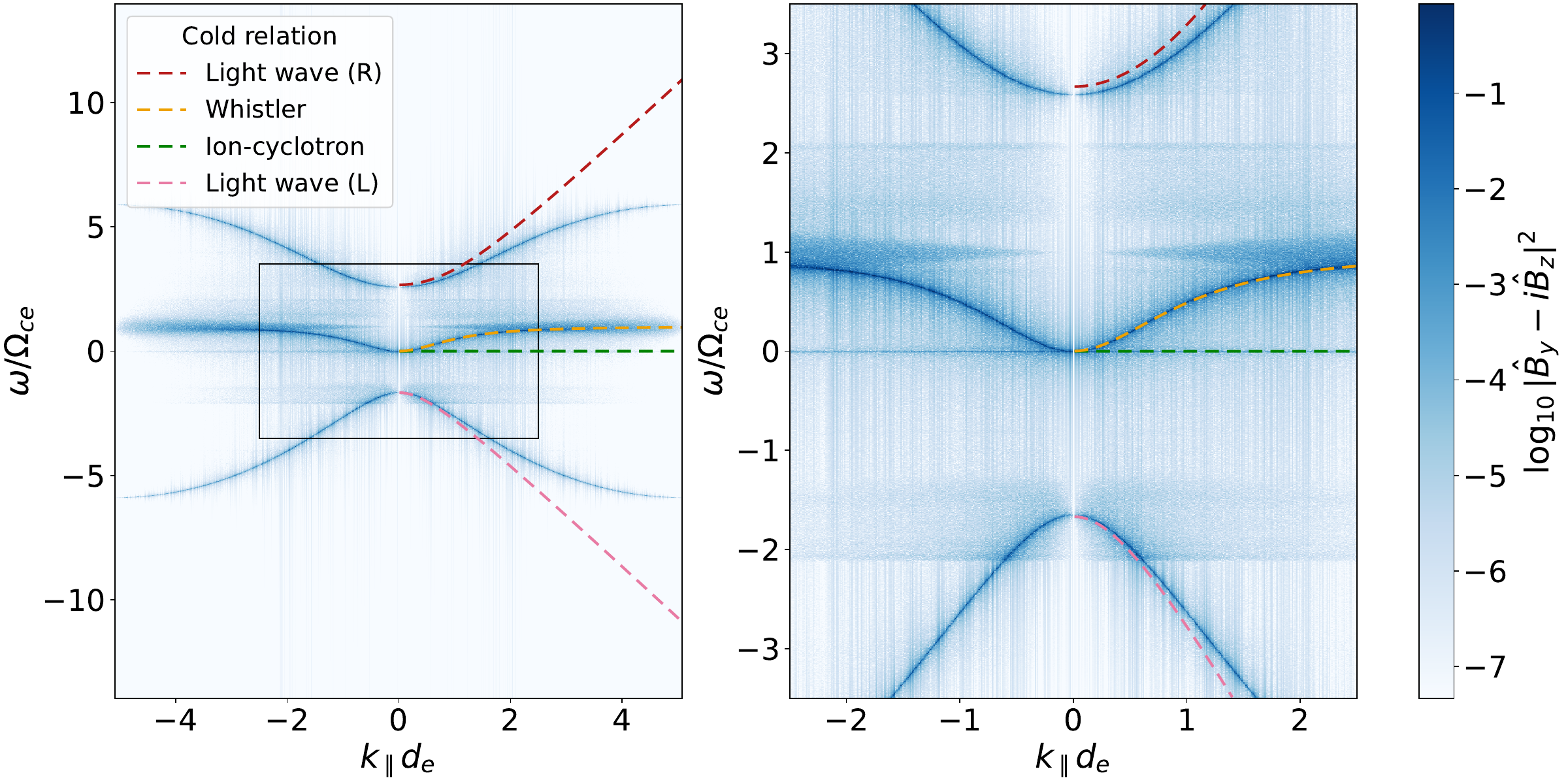}
    \caption{Whistler waves test. Power spectrum $\log_{10}|\hat B_y-i\hat B_z|^2$ of the transverse magnetic field. Positive (negative) frequencies are right-handed (left-handed) waves. Left panel: all resolved wavenumbers, with frequencies up to the Nyquist frequency $\pi/\Delta t$. The box indicates the region enlarged on the right. The dashed lines represent the solutions of the cold dispersion relations Eq.~\ref{eq_disp_R} and Eq.~\ref{eq_disp_L}, drawn for $k>0$ only.}
    \label{fig_whistler}
\end{figure}

Fig.~\ref{fig_whistler} shows the power spectrum $|\hat B_y-i\hat B_z|^2$, where the hat denotes the Fourier transform, in space $x$ of the field averaged over $y$, and in time, as shown in~\citet{MatsumotoOmura1985particle_sim}. With this combination, positive (negative) frequency corresponds to a right-handed (left-handed) wave, and the sign of $\omega/k$ gives its direction of propagation. The spectral resolution is $\Delta k=9.9\times10^{-3}\,d_e^{-1}$ and $\Delta\omega=4.7\times10^{-3}\,|\Omega_{ce}|$. The Nyquist limits are $|k|d_e=5.09$ and $\pi/\Delta t=13.96\,|\Omega_{ce}|$.

The TPIC whistler mode agrees with Eq.~\ref{eq_disp_R}. Over the 200 modes with $0.5\le|k|d_e\le1.5$, the spectral peak lies, on average, $1.9\%$ above the cold dispersion relation and, at most, $4.4\%$. This discrepancy is explained by the discretization and the finite electron temperature.

\begin{figure}
    \centering
    \includegraphics[width=0.7\linewidth]{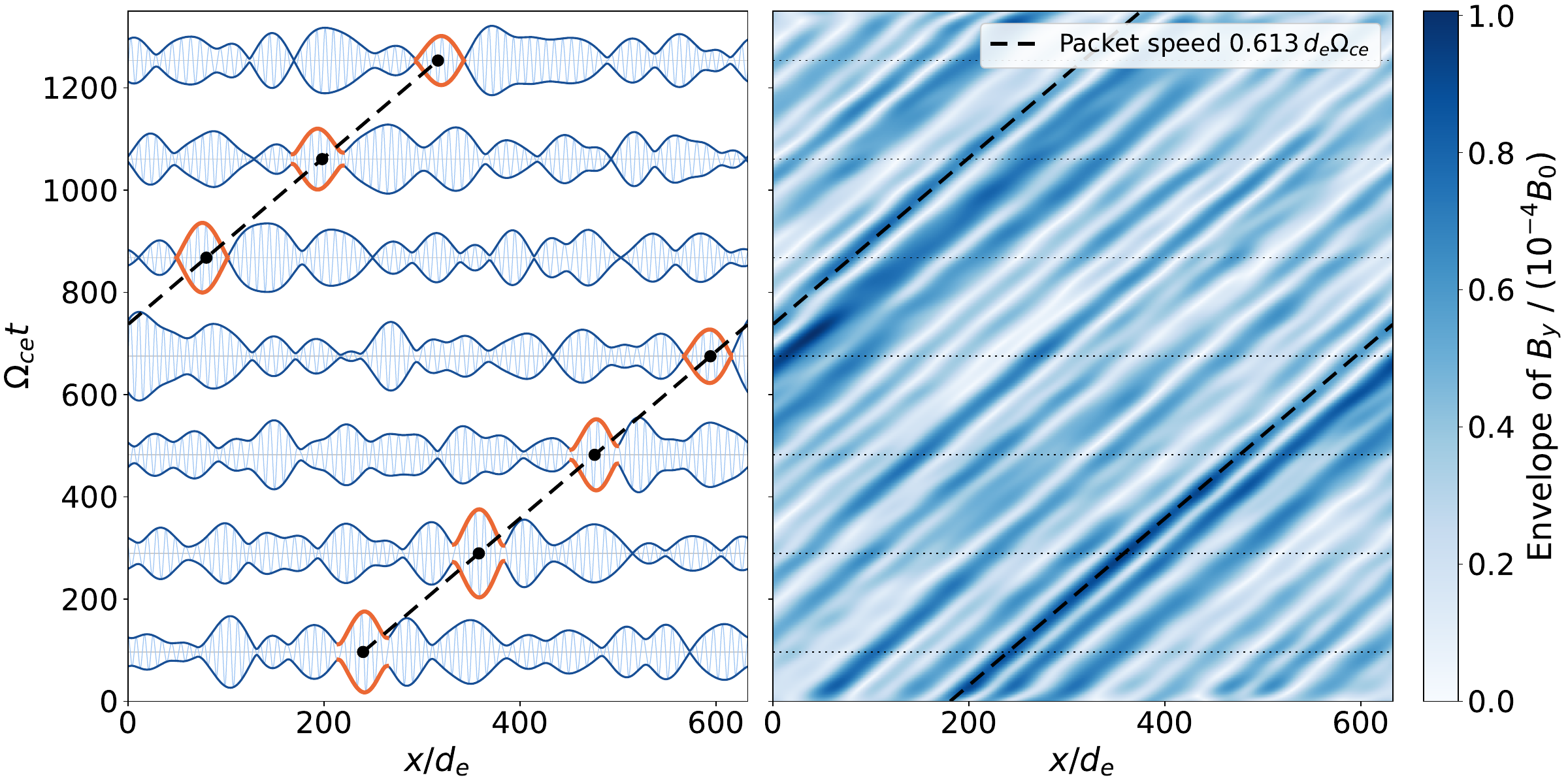}
    \caption{Whistler waves test. Whistler wave packets with $0.6\le kd_e\le0.8$ travelling towards $+x$. Left: $B_y$ (light blue) and its envelope (dark blue) at seven different times. The orange line marks one packet. The dots lie on the line fitted through its peak. Right: envelope evolution over $x$ and $t$, the dashed line is the path of the tracked packet.}
    \label{fig_whistler_crossing}
\end{figure}

To follow the propagation of the whistlers, we keep in the spectrum of $\hat B_y-i\hat B_z$ the 20 modes with $0.6\le kd_e\le0.8$ and the frequencies within $30\%$ of Eq.~\ref{eq_disp_R}, and transform back to $x$ and $t$. This isolates the whistlers traveling towards $+x$, which form into wave packets, as shown in Fig.~\ref{fig_whistler_crossing}, that travel at the group speed. The cold relation, Eq.~\ref{eq_disp_R} without its ion term, gives a group speed of $0.608\,d_e|\Omega_{ce}|$, while the tracked packet, in orange in Fig.~\ref{fig_whistler_crossing}, moves at $0.613\,d_e|\Omega_{ce}|$.

The light waves have frequencies below the cold relation prediction, with the discrepancy increasing as $|k|$ grows. The difference is $1$-$6\%$ for $|k|d_e<1$ and up to $46\%$ at the largest wavenumber. This is due to the numerical dispersion of the ECSIM scheme, which, for $\theta=1/2$, does not remove unresolved high-frequency modes, but it shifts their frequencies with the map \mbox{$\omega_{num}=(2/\Delta t)\arctan(\omega\Delta t/2)<\pi/\Delta t$}.
For $\theta>1/2$ the scheme is dissipative and damps these modes~\citep{gonzalezherrero2018ecsim}.

The ion-cyclotron mode is not resolved in frequency. The run lasts $0.21$ ion gyroperiods, so that $\Delta\omega=4.7\,\Omega_{ci}$, and the wave collapses onto the line $\omega=0$. Resolving it requires a run about two orders of magnitude longer. With a GMRES solver tolerance of $1\times 10^{-16}$, the total system energy is conserved to a relative error of $9.7\times10^{-15}$ over the run.

\subsection{Weibel instability}
\label{sec_weibel}
The Weibel instability generates magnetic fields from the temperature anisotropy of an unmagnetized plasma~\citep{weibel1959instability}. We use it to verify the kinetic electromagnetic response of TPIC. We initialize bi-Maxwellian electrons with thermal velocities $v_{\mathrm{th},e,x}=v_{\mathrm{th},e,z}=0.05$ and $v_{\mathrm{th},e,y}=0.1$, so that $T_{e,y}/T_{e,x}=4$, and cold ions of the same density, with mass ratio $\mu=100$. We use a periodic 2D domain of $4\,d_i\times4\,d_i$ ($40\,d_e\times40\,d_e$), discretized with $448\times448$ cells, $\Delta x=\Delta y=8.93\times10^{-3}\,d_i=8.93\times10^{-2}\,d_e$. We initialize 256 particles per cell per species. We seed the mode $m=4$, with $k=2\pi m/L_x=0.628\,d_e^{-1}$, through the magnetic field $B_z=\delta B\cos(kx)$, with $\delta B=10^{-5}$ and $\mathbf{E}=0$. We run the simulation for $12000$ steps with $\theta=1/2$ and $\Delta t=2.5\times10^{-3}\,\omega_{pi}^{-1}=0.025\,\omega_{pe}^{-1}$, up to $T=300\,\omega_{pe}^{-1}$.

For $\mathbf{k}$ along the cold direction $x$, the transverse mode with components $E_y$ and $B_z$ follows the kinetic dispersion relation~\citep{Davidson1972instability_anisotropic}
\begin{equation}
  \omega^2-k^2c^2-\omega_{pe}^2-\omega_{pi}^2+\omega_{pe}^2\,\frac{T_{e,y}}{T_{e,x}}\left[1+\xi Z(\xi)\right]=0,
  \qquad \xi=\frac{\omega}{\sqrt{2}\,k\,v_{\mathrm{th},e,x}},
  \label{eq_weibel_disp}
\end{equation}
where $Z$ is the plasma dispersion function,
\begin{equation}
  Z(\xi)=\frac{1}{\sqrt{\pi}}\int_{-\infty}^{\infty}\frac{e^{-s^2}}{s-\xi}\,ds,\qquad \operatorname{Im}\xi>0.
  \label{eq_plasma_dispersion_function}
\end{equation}
$Z$ describes the electron response. The instability requires $|\xi|\lesssim1$, where the thermal motion along $\mathbf{k}$ couples the anisotropy to the mode. For a purely growing mode, $\omega=i\gamma$, $\xi=i\eta$ with $\eta=\gamma/(\sqrt{2}\,k\,v_{\mathrm{th},e,x})$, and $Z(i\eta)=i\sqrt{\pi}\,e^{\eta^2}\operatorname{erfc}(\eta)$, so that Eq.~\ref{eq_weibel_disp} becomes
\begin{equation}
  \gamma^2+k^2c^2+\omega_{pi}^2=\omega_{pe}^2\left[A-(A+1)\sqrt{\pi}\,\eta\,e^{\eta^2}\operatorname{erfc}(\eta)\right],
  \label{eq_weibel_gamma}
\end{equation}
with $A=T_{e,y}/T_{e,x}-1$. Modes with \mbox{$kd_e<(A-1/\mu)^{1/2}=1.73$} are unstable. The seeded mode grows at $\gamma=0.0307\,\omega_{pe}$, close to the fastest-growing mode.

\begin{figure}
    \centering
    \includegraphics[width=0.8\linewidth]{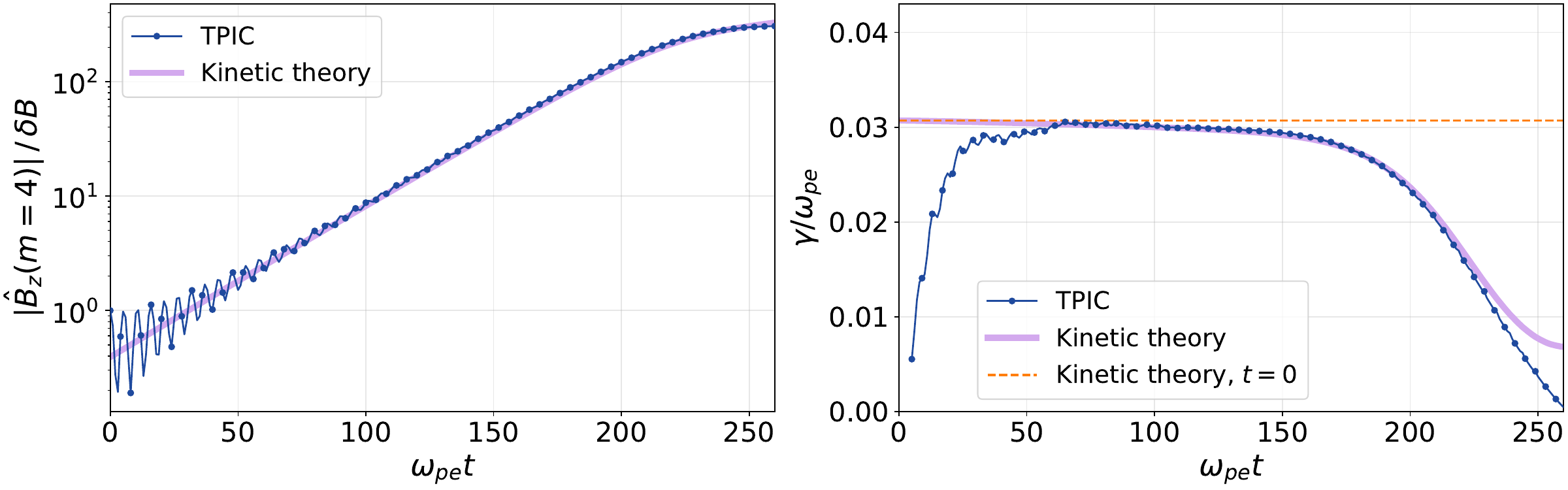}
    \caption{Weibel instability. Left: amplitude of the seeded mode $m=4$ of $B_z$, in units of the seed $\delta B$, and growth at the kinetic rate of Eq.~\ref{eq_weibel_gamma} with the measured electron temperatures. Right: instantaneous growth rate of the seeded mode and Eq.~\ref{eq_weibel_gamma} with the measured and with the initial temperatures.}
    \label{fig_weibel}
\end{figure}

Fig.~\ref{fig_weibel} (left) shows the amplitude of the seeded mode $|\hat B_z(m=4)|$ averaged over $y$. The magnetic seed also excites a light wave at $\omega=1.19\,\omega_{pe}$, undamped for $\theta=1/2$, which superposes with the mode and remains clearly visible up to $\omega_{pe}t\simeq60$. The electron anisotropy relaxes during the run, so we evaluate Eq.~\ref{eq_weibel_gamma} using temperatures derived from the electron pressure at each time. A fit of $\ln|\hat B_z|$ over $60\le\omega_{pe}t\le160$ gives $\gamma=0.0301\,\omega_{pe}$, $0.9\%$ above the theoretical rate averaged over the same interval. Fig.~\ref{fig_weibel} (right) shows the instantaneous growth rate, measured over windows of $10\,\omega_{pe}^{-1}$. It follows the theory within $2.3\times10^{-3}\,\omega_{pe}$ for $60\le\omega_{pe}t\le240$, while $T_{e,y}/T_{e,x}$ decreases from $3.9$ to $1.9$. The mode saturates at $\omega_{pe}t\simeq260$, with amplitude $306\,\delta B$ and a magnetic energy of $3.3\%$ of the total energy. With a GMRES solver tolerance of $1\times10^{-15}$, the total system energy is conserved to a relative error of $5.4\times10^{-14}$ over the run.

\begin{figure}
    \centering
    \includegraphics[width=0.7\linewidth]{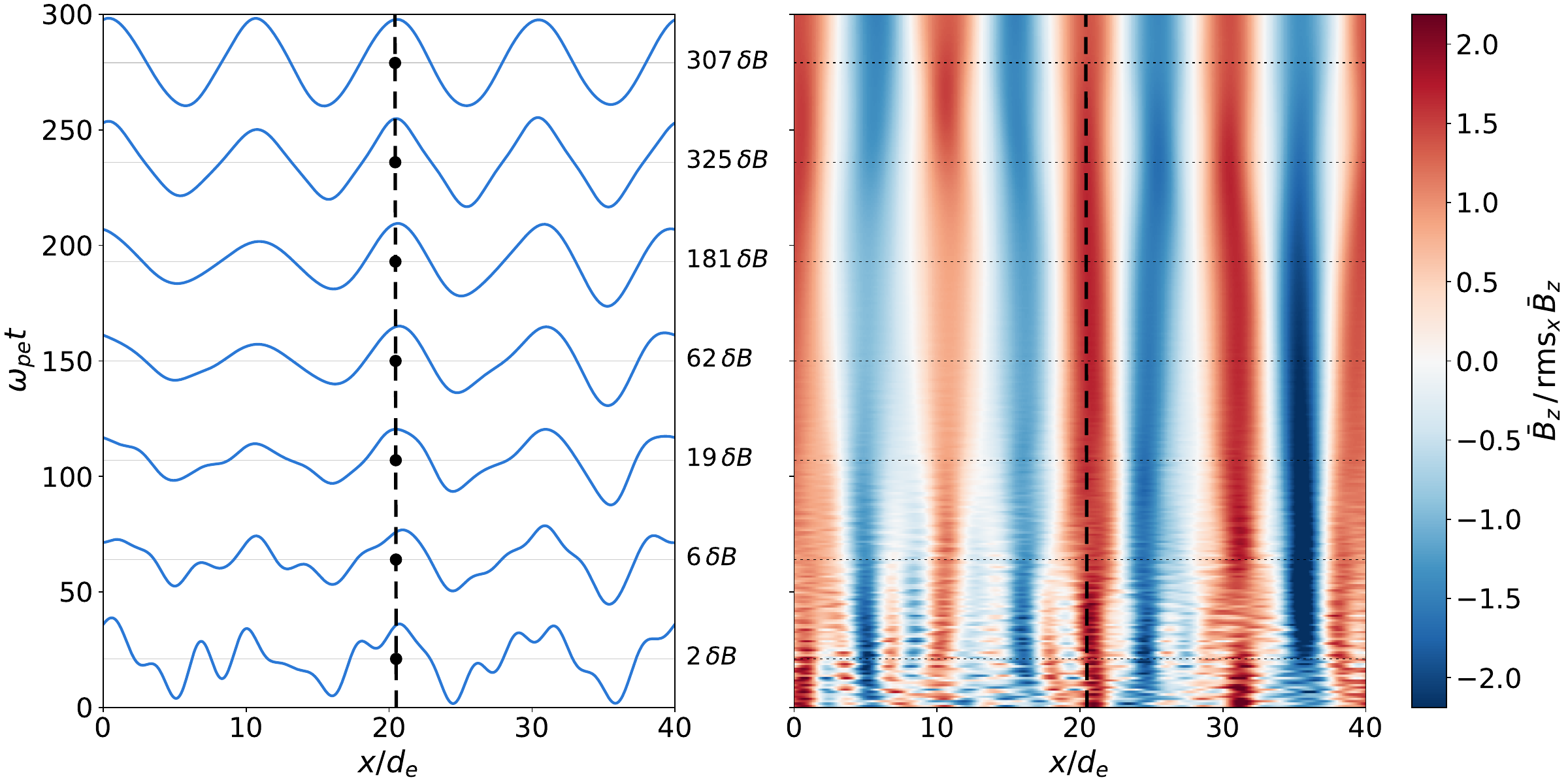}
    \caption{Weibel instability: $B_z$ averaged over $y$, with the Fourier modes $m\le16$. Left: $B_z$ at seven different times, each profile rescaled to its own maximum $|B_z|$; the right axis gives this maximum in units of the seed $\delta B$. Right: evolution over $x$ and $t$, with $B_z$ at each time divided by its root mean square over $x$. The dashed line follows a crest of the seeded mode, the dots mark it at the times of the left panel.}
    \label{fig_weibel_space_time}
\end{figure}

Fig.~\ref{fig_weibel_space_time} shows the evolution of $B_z$ averaged over $y$ in space and time. The seeded mode holds more than $80\%$ of the power of the averaged field from $\omega_{pe}t\simeq80$ and $99\%$ at saturation. The mode grows without propagating, as expected for a purely unstable mode with $\omega=i\gamma$. 

\pagebreak
\subsection{Whistler turbulence}
\label{sec_whistler_turbulence}
We run a three-dimensional whistler turbulence simulation to investigate the energy cascade from the seeded modes to small electron scales. We use a setup adapted from~\citet{Gary2012whistlerTurbulence, saito2008whistlerTurbulence}. The plasma is homogeneous, with mass ratio $\mu=1836$ and a uniform magnetic field $\mathbf{B}_0=B_0\hat{\mathbf{z}}$, with $\omega_{pe}/|\Omega_{ce}|=\sqrt5$. Electrons and ions are Maxwellian, with $T_i=T_e$ and $\beta_e=\beta_i=0.1$, so that $v_{\mathrm{th},e}=0.1$ and $\lambda_{De}=0.1\,d_e$. The periodic cubic domain of side $L=1.40\,d_i=60\,d_e$, is discretized with $300\times300\times300$ cells, $\Delta x=\Delta y=\Delta z=0.2\,d_e=2\,\lambda_{De}$. We seed 150 helical whistler modes of equal magnetic energy, with random phases and the cold-plasma frequency, at the wavevectors $\mathbf{k}=2\pi(n_x,n_y,n_z)/L$ with $1\le|n_z|\le3$ and $\mathbf{k}_\perp$ along $x$, $y$, or the diagonals, with $|n_x|,|n_y|\le3$, so that $k_\parallel d_e\le0.31$ and $k_\perp d_e\le0.44$. The electric field follows from Faraday's law and the electron flow from Amp\`ere's law, while the ions are at rest. The initial fluctuation energy is $\epsilon_0=\langle|\delta\mathbf{B}|^2\rangle/B_0^2=0.1$. We initialize 256 particles per cell per species, $1.38\times10^{10}$ particles in total. We apply one pass of the filter of Eq.~\ref{eq_smoothing_s6} with $\alpha=0.5$, and run the simulation for 4000 steps with $\theta=1/2$ and $\Delta t=5.83\times10^{-3}\,\omega_{pi}^{-1}=0.25\,\omega_{pe}^{-1}$ ($|\Omega_{ce}|\Delta t=0.11$, $c\Delta t\sqrt3/\Delta x=2.17$), up to $T=23.3\,\omega_{pi}^{-1}=1000\,\omega_{pe}^{-1}=447\,|\Omega_{ce}|^{-1}$. We use single-precision particle positions and 32 AMD MI250x GCDs on 4 LUMI-G nodes, with a $4\times4\times2$ domain decomposition. The 4000 steps take $1.6$~h of wall-clock time, including the periodic output of fields, particles, and checkpoints, and the median time step takes 779~ms.

\begin{figure}
    \centering
    \includegraphics[width=\linewidth]{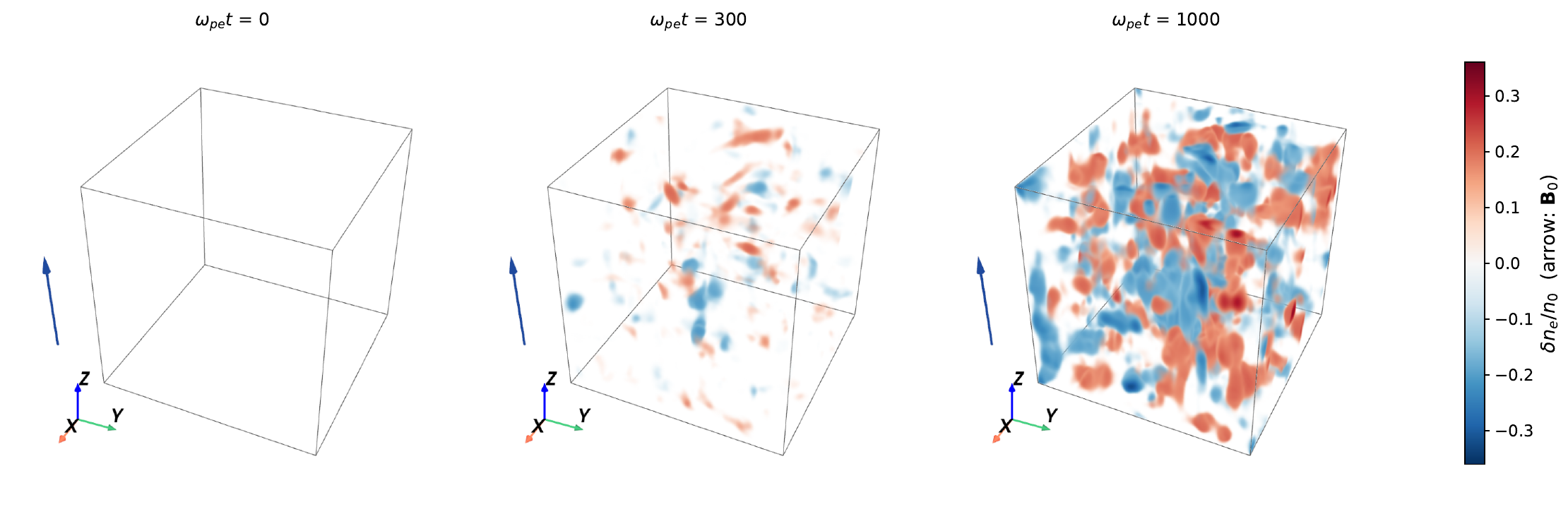}
    \caption{Whistler turbulence. Volume rendering of the electron density fluctuation $\delta n_e/n_0$ at $\omega_{pe}t=0$, $300$, and $1000$. The arrow marks the direction of $\mathbf{B}_0$.}
    \label{fig_whistler_turbulence_density}
\end{figure}

\begin{figure}
    \centering
    \includegraphics[width=0.8\linewidth]{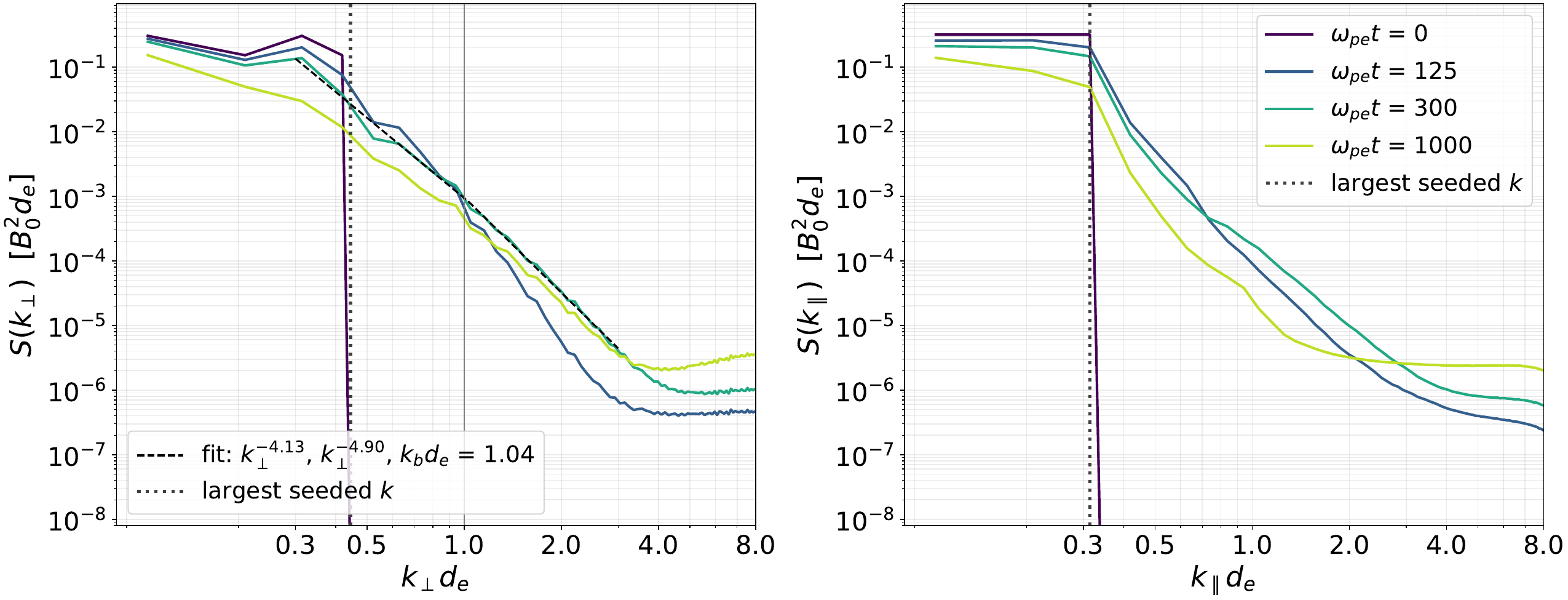}
    \caption{Whistler turbulence. Reduced magnetic energy spectra as a function of $k_\perp$ (left) and $k_\parallel$ (right) at four times. The dotted lines mark the largest seeded wavenumber, the dashed line is the broken power-law fit at $\omega_{pe}t=300$.}
    \label{fig_whistler_turbulence_spectra}
\end{figure}

Fig.~\ref{fig_whistler_turbulence_density} shows the electron density fluctuations, absent in the initial state, which develop during the cascade.
Fig.~\ref{fig_whistler_turbulence_spectra} shows the reduced magnetic energy spectra. At $t=0$, the energy is only in the seeded modes. The nonlinear interaction of the whistlers transfers energy to the unseeded modes at shorter wavelengths, which hold up to $10\%$ of the initial fluctuation energy at $\omega_{pe}t\simeq125$. The spectra extend to $kd_e\simeq3$, where they reach the particle noise level. The cascade is anisotropic, with more energy at $k_\perp$ than at $k_\parallel$. 
At $\omega_{pe}t=300$ ($|\Omega_{ce}|t=134$), $S(k_\perp)$ follows a broken power law, with slopes $-4.1$ and $-4.9$ below and above $k_\perp d_e=1.04$, close to the electron-scale break at $k_\perp d_e\simeq1$ found by~\citet{Gary2012whistlerTurbulence}. 

\begin{figure}
    \centering
    \includegraphics[width=0.8\linewidth]{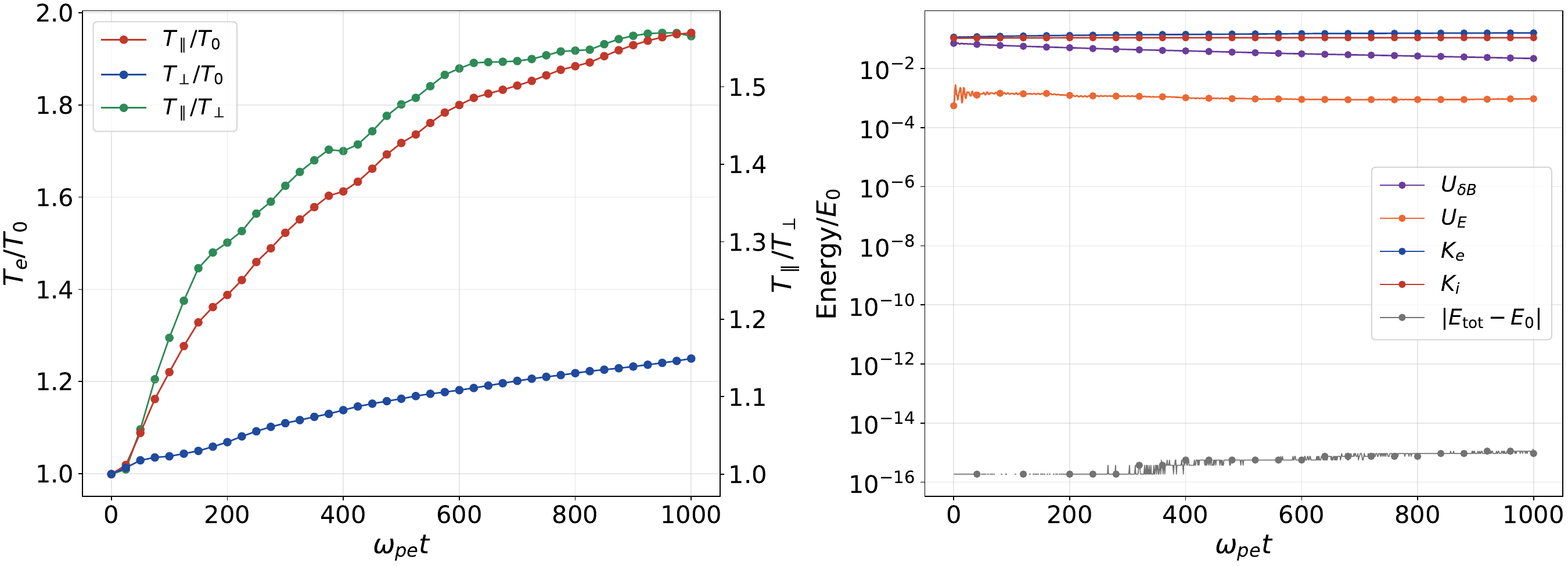}
    \caption{Whistler turbulence. Left: parallel and perpendicular electron temperatures in units of the initial temperature $T_0$, and their ratio (right axis). Right: magnetic fluctuation energy $U_{\delta B}$, electric energy $U_E$, kinetic energy of electrons $K_e$ and ions $K_i$, and change of the total energy, in units of the initial total energy $E_0$.}
    \label{fig_whistler_turbulence_energy}
\end{figure}

Fig.~\ref{fig_whistler_turbulence_energy} (left) shows that the electrons heat anisotropically. The parallel electron temperature, computed without the local bulk flow, grows to $1.96\,T_0$ and the perpendicular one to $1.25\,T_0$, with $T_0=m_ev_{\mathrm{th},e}^2$, so that $T_\parallel/T_\perp=1.57$ at $T$, close to the $1.6$ found by~\citet{Gary2012whistlerTurbulence}. 
Fig.~\ref{fig_whistler_turbulence_energy} (right) details the energy exchange. At $T$, the magnetic fluctuation energy has decayed to $31\%$ of its initial value. The electrons absorb $64\%$ of the initial fluctuation energy and the ions $5\%$, while the electric energy is $1.3\%$. With a GMRES solver tolerance of $1\times10^{-15}$, the total system energy is conserved to a relative error of $1.3\times10^{-15}$ over the run.

\subsection{Magnetic reconnection}
\label{sec_reconnection}
We perform a three-dimensional magnetic reconnection simulation with TPIC, following the GEM challenge setup~\citep{birn2001gem}, and we compare the results against the well-established iPIC3D~\citep{markidis2010_ipic3d} code. The system is a single Harris current sheet of thickness $\delta=0.5\,d_i$, where the magnetic field is $B_x=B_0\tanh[(y-L_y/2)/\delta]$, with $B_0=0.0195$, so that $\Omega_{ci}=0.0195\,\omega_{pi}$ and $v_A=0.0195\,c$. The plasma in the current sheet has density $n_0\,\mathrm{sech}^2[(y-L_y/2)/\delta]$, while the uniform background plasma has density $0.1\,n_0$. We use Maxwellian electrons and ions with mass ratio $\mu=64$ and thermal velocities $v_{\mathrm{th},e}=0.045$ and $v_{\mathrm{th},i}=0.0126$, so that $T_i/T_e=5$. A perturbation of the magnetic flux function, localized at the center of the sheet with width $\delta$ and uniform along $z$, seeds the reconnection at the X-line $x=L_x/2$, $y=L_y/2$. The drift of the sheet species along $z$, computed from $\nabla\times\mathbf{B}$, carries the current of the initial field.
The domain of $30\times20\times20\,d_i$ is discretized with $300\times200\times200$ cells, $\Delta x=\Delta y=\Delta z=0.1\,d_i=0.8\,d_e$. It is periodic in $x$ and $z$, with PEC walls and reflecting particle boundaries along  $y$. We initialize 256 particles per cell for each of the four species (sheet and background particles), $1.23\times10^{10}$ particles in total. We use $\Delta t=0.1\,\omega_{pi}^{-1}$ ($\Omega_{ci}\Delta t=1.95\times10^{-3}$, $\omega_{pe}\Delta t=0.8$), so that $c\Delta t\sqrt{3}/\Delta x=1.73$ exceeds the 3D CFL limit of explicit PIC. 

Both TPIC and iPIC3D apply six passes of the filter of Eq.~\ref{eq_smoothing_s6} with $\alpha=0.2$, and use a GMRES tolerance of $10^{-15}$. TPIC uses $\theta=1/2$ and single-precision particle positions, while iPIC3D, which implements the implicit moment method~\citep{brackbill1982_imm}, uses $\theta=1$. We run both codes on 32 AMD MI250x GCDs on 4 LUMI-G nodes, with a $4\times4\times2$ domain decomposition for 8000 steps, up to $\Omega_{ci}t=15.6$.

\begin{figure}
    \centering
    \subfloat{\includegraphics[width=0.42\linewidth]{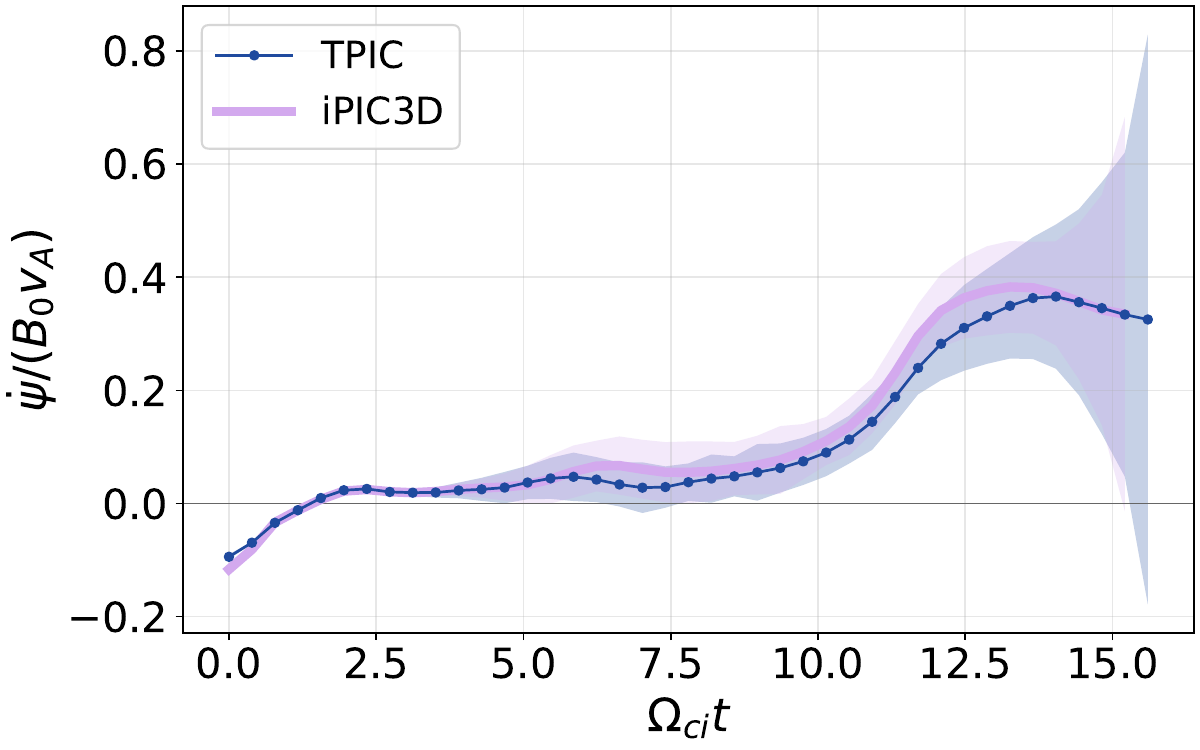}}
    \hspace{4pt}
    \subfloat{\includegraphics[width=0.42\linewidth]{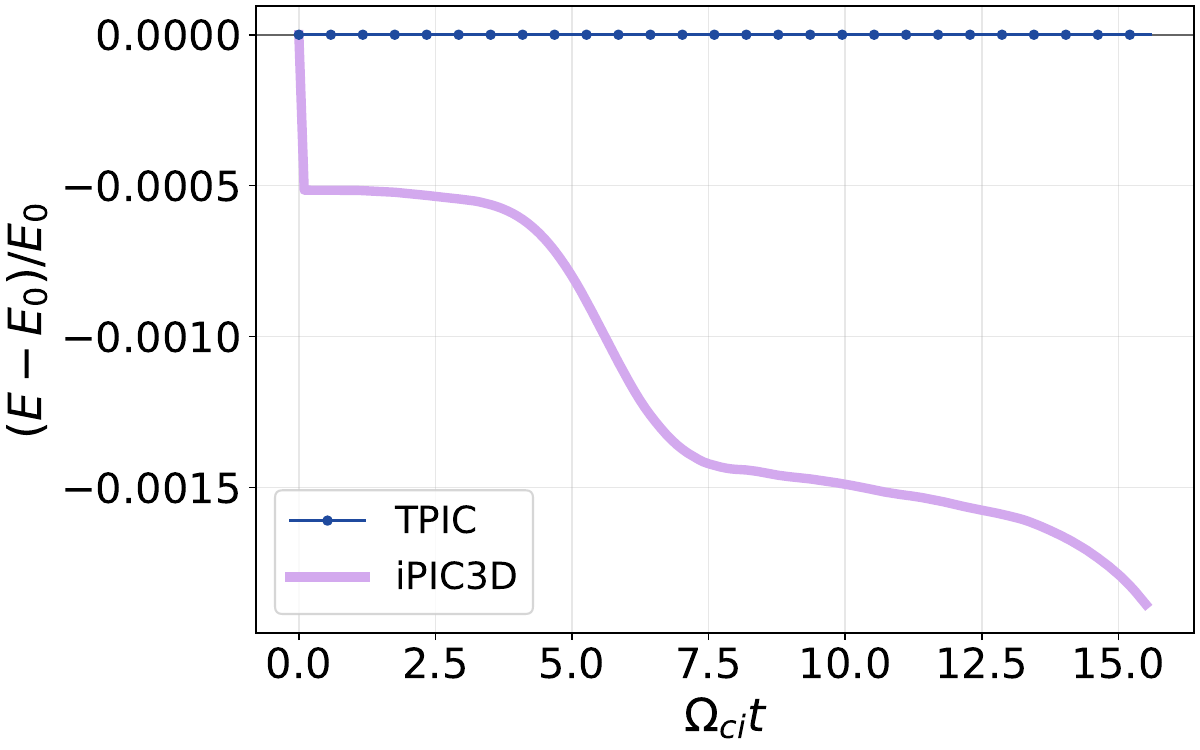}}
    \caption{Magnetic reconnection in TPIC and iPIC3D. Left: reconnection rate $\dot\psi/(B_0v_A)$, with $\psi$ given by Eq.~\ref{eq_reconnected_flux}. The lines show the mean along the X-line (averaged along $z$), and the shaded bands one standard deviation along it. Right: relative change of the total energy.}
    \label{fig_reconnection_ipic3d}
\end{figure}

We measure the reconnected flux along the X-line,
\begin{equation}
  \psi(z,t)=\frac12\left[\int_{L_x/2}^{L_x}B_y\,dx-\int_{0}^{L_x/2}B_y\,dx\right]_{y=L_y/2},
  \label{eq_reconnected_flux}
\end{equation}
and compute the reconnection rate $\dot\psi/(B_0v_A)$ from field snapshots taken every 200 steps. In Fig.~\ref{fig_reconnection_ipic3d}, we report its mean and standard deviation along the X-line in $z$, comparing TPIC and iPIC3D results. The left panel reports the reconnection rate. In both codes the rate is initially negative, as the initial perturbation decays and part of the seeded flux is lost. The rate becomes positive at $\Omega_{ci}t\simeq1.4$, remains at $0.02$-$0.03$ for $2\lesssim\Omega_{ci}t\lesssim4.7$, and then increases to a first maximum, $0.047$ at $\Omega_{ci}t=5.85$ in TPIC and $0.067$ at $\Omega_{ci}t=6.63$ in iPIC3D. From $\Omega_{ci}t\simeq9$, the reconnection accelerates in both codes. Over the whole run, the mean rates differ by at most $0.059$, at $\Omega_{ci}t=12.09$, comparable to the standard deviation along the X-line. The rate peaks at $0.366$ at $\Omega_{ci}t=14.04$ in TPIC and at $0.383$ at $\Omega_{ci}t=13.26$ in iPIC3D, and it decreases to $0.33$ in both codes at $\Omega_{ci}t=15.21$. The standard deviation along the X-line is below $0.01$ up to $\Omega_{ci}t\simeq3.5$. Then it grows, as the lower-hybrid drift instability develops on the flanks of the current sheet and the reconnection becomes non-uniform along the X-line, and at the end, when the drift kink instability develops, it is comparable to the mean rate.

The right panel in Fig.~\ref{fig_reconnection_ipic3d} shows the relative change of the total energy. With a GMRES solver tolerance of $1\times10^{-15}$, TPIC conserves the total energy to a relative error of $1.7\times10^{-14}$ over the 8000 steps. The iPIC3D scheme is dissipative, it loses $5.1\times10^{-4}$ of the initial energy within the first 50 steps, mainly through the damping of the seeded perturbation. The iPIC3D energy loss remains nearly constant up to $\Omega_{ci}t\simeq3$, and then it increases as instabilities develop to $1.9\times10^{-3}$ at $\Omega_{ci}t=15.5$.

\begin{figure}
    \centering
    \includegraphics[width=0.7\linewidth]{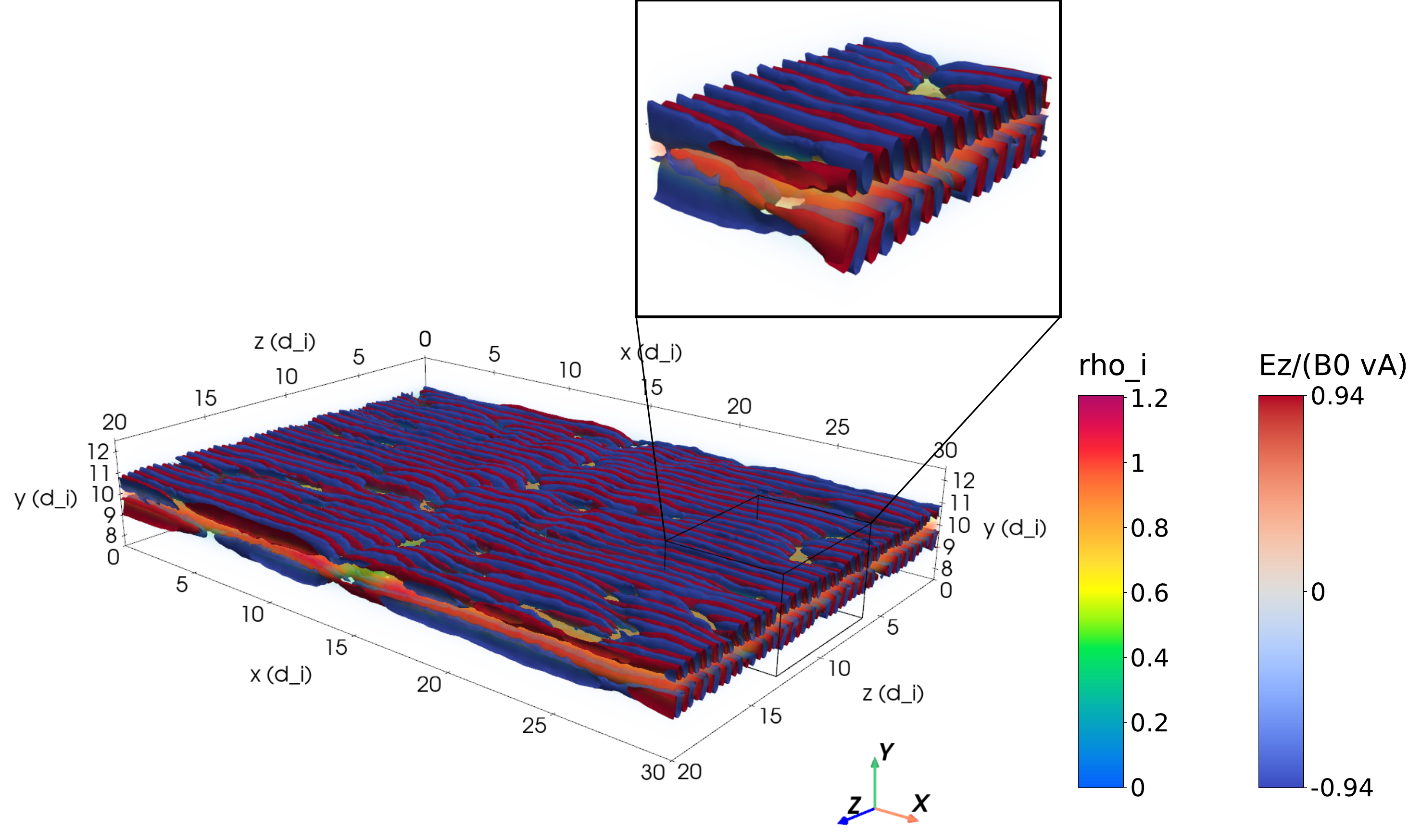}
    \caption{Lower-hybrid drift instability in TPIC at $\Omega_{ci}t=4.68$. Isosurfaces of positive (red) and negative (blue) $E_z/(B_0v_A)$, and volume rendering of the density of the current-sheet ions, in units of $n_0$, for $8\le y/d_i\le12$.}
    \label{fig_reconnection_lhdi}
\end{figure}

The lower-hybrid drift instability (LHDI) develops first, driven by the diamagnetic drift on the flanks of the current sheet, where the density gradient is large. We observe it through fluctuations in $E_z$ along $z$, as shown in Fig.~\ref{fig_reconnection_lhdi}. On both flanks of the sheet, the wave fronts extend along $x$ over the whole box, and $E_z$ alternates in sign along $z$ with wavelength $0.71\,d_i$.

\begin{figure}
    \centering
    \includegraphics[width=0.7\linewidth]{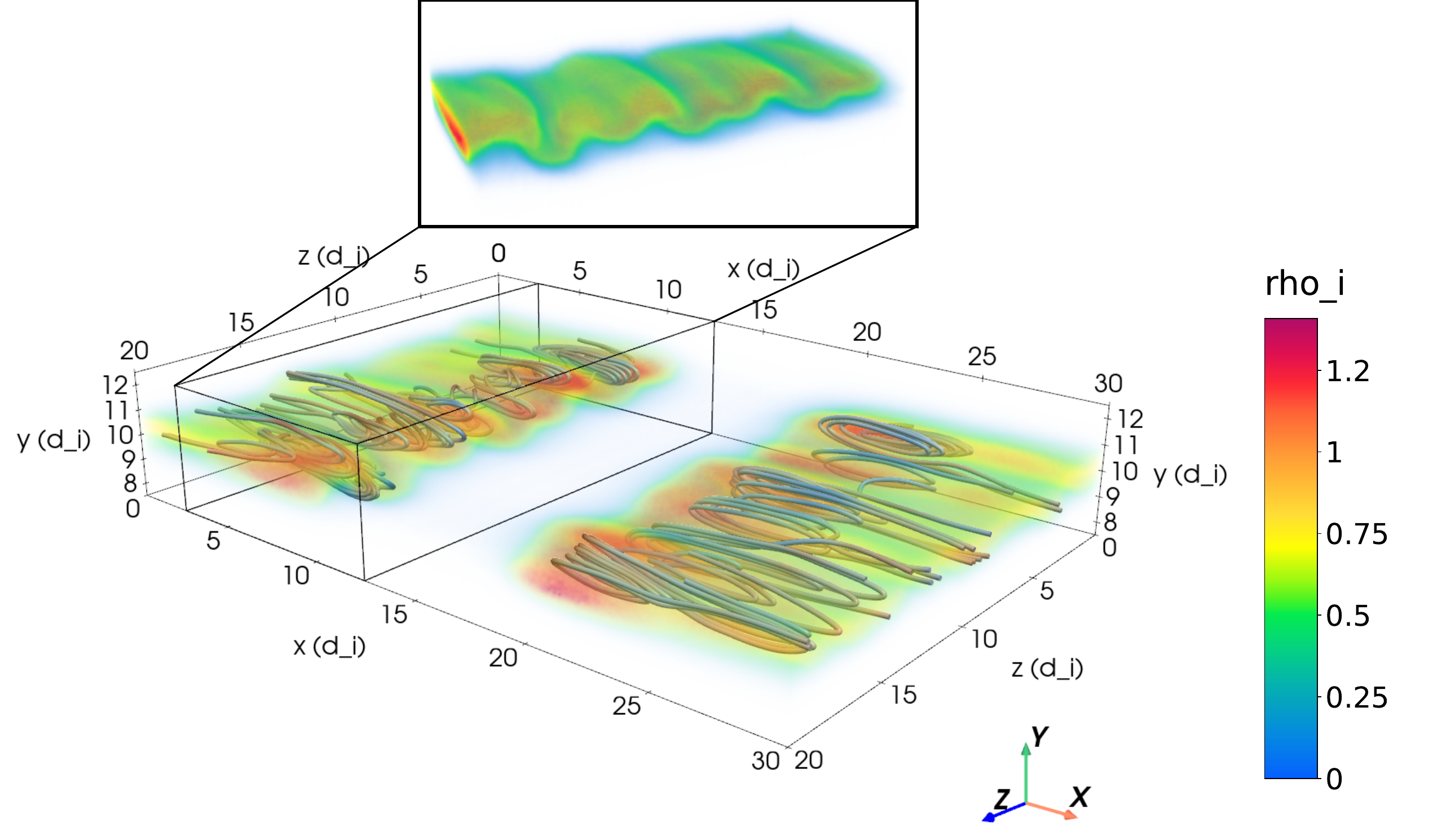}
    \caption{Flux ropes and kink of the current sheet in TPIC at $\Omega_{ci}t=15.21$. Volume rendering of the density of the current-sheet ions, in units of $n_0$, and magnetic field lines on both sides of the X-line, for $8\le y/d_i\le12$.}
    \label{fig_reconnection_ropes}
\end{figure}

Flux ropes form in the outflow regions from $\Omega_{ci}t\simeq5.5$, while the reconnection rate increases. They are secondary magnetic islands in the $x$-$y$ planes that extend along $z$~\citep{markidis2013plasmoidchains}. On longer timescales, the current sheet kinks along $z$ due to the onset of the drift kink instability~\citep{Daughton1998kinkKinetic, lapenta2002lhdi_kink}. Fig.~\ref{fig_reconnection_ropes} shows the flux ropes on both sides of the X-line at $\Omega_{ci}t=15.21$, and the kink of the rope at $x\simeq L_x/4$.

\begin{figure}
    \centering
    \includegraphics[width=0.7\linewidth]{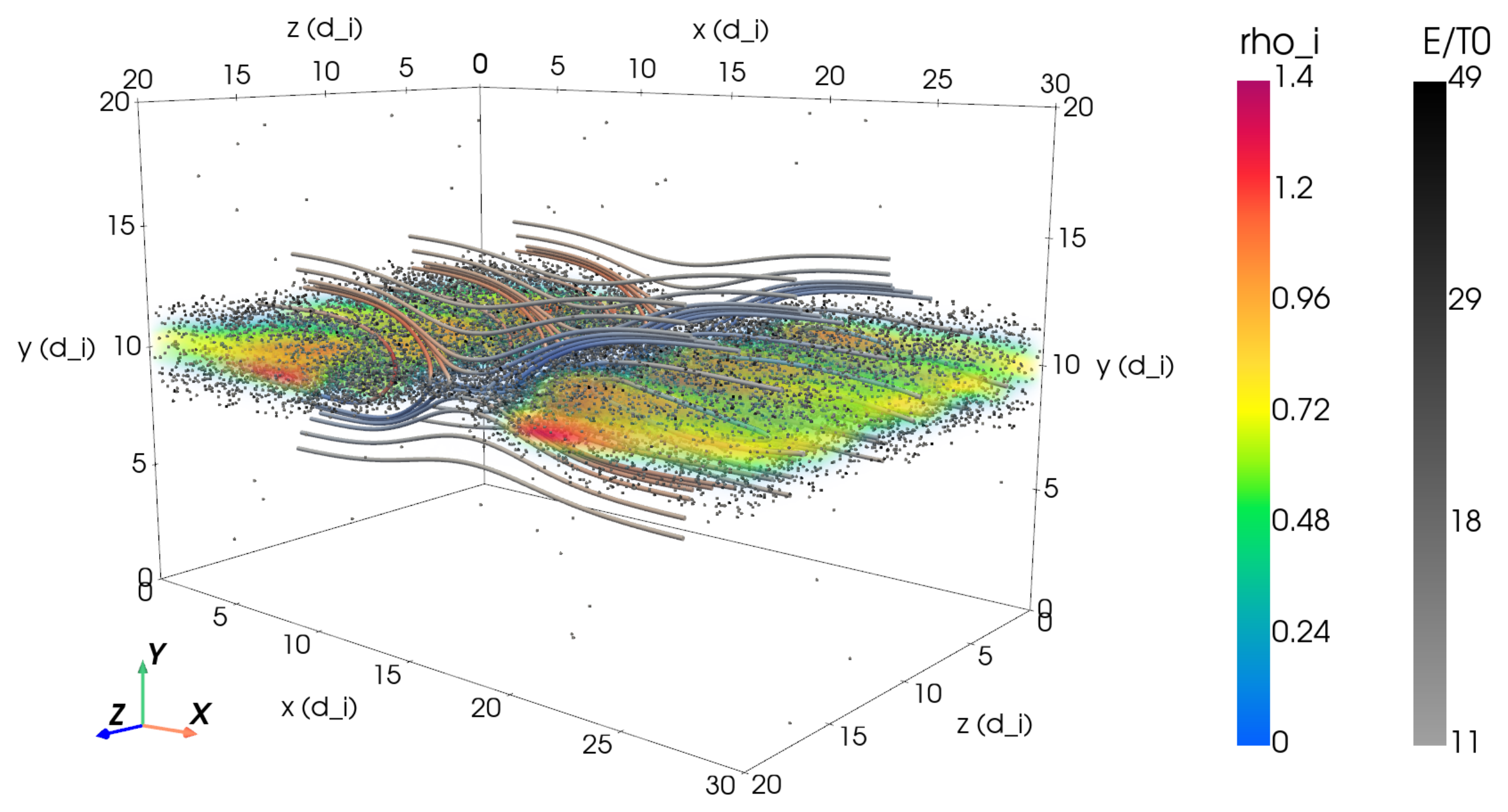}
    \caption{Magnetic reconnection in TPIC at $\Omega_{ci}t=15.6$. Volume rendering of the density of the current-sheet ions, in units of $n_0$, reconnected magnetic field lines, and a random sample of 20000 electrons with kinetic energy above the 98th percentile, $E\ge10.6\,T_0$, colored by $E/T_0$. The ion density is shaped by the drift kink instability.}
    \label{fig_reconnection_tpic}
\end{figure}

Fig.~\ref{fig_reconnection_tpic} shows the density of the sheet ions and the magnetic field lines at the end of the TPIC run, at $\Omega_{ci}t=15.6$. The field lines reconnect across the sheet at the X-line, and the sheet ions accumulate in the outflow regions. The current sheet is no longer uniform along $z$, since the drift kink instability has developed. The figure also shows the most energetic electrons, with kinetic energy $E=m_e|\mathbf{v}|^2/2\ge10.6\,T_0$, where $T_0=m_ev_{\mathrm{th},e}^2$ is the initial temperature of the sheet electrons, corresponding to a speed of $|\mathbf{v}|\ge0.207\,c=4.61\,v_{\mathrm{th},e}=10.6\,v_A$.

TPIC runs this simulation in $2.2$ hours, with a median time step, including fields and particle output, of 995~ms. On the same 32 GCDs, the median iPIC3D time step is 1330~ms $1.33\times$ slower. Part of the difference comes from the field solve. The implicit moment method of iPIC3D yields a much worse-conditioned Maxwell system, and its GMRES solver requires on average 147 iterations per step to converge to the same $1\times10^{-15}$ tolerance, against 18 in TPIC.

\FloatBarrier
\section{Performance analysis}
\label{sec_performance_analysis}

\subsection{Strong \& weak scaling}
We run strong and weak scaling tests using the same GEM setup described in Section~\ref{sec_reconnection} without smoothing, with $\theta=1/2$, and $\Delta t=0.01\,\omega_{pi}^{-1}$. We use full double precision, partial sorting, and the matrix-engine mass-matrix deposition. We use device-aware MPI as the communication backend, and we map one rank per GPU/GCD. Table~\ref{tab_scaling_baseline} lists the baseline of the scaling tests. Weak scaling keeps $64^3$ cells per GCD, up to $1024^3$ cells and $1.9\times10^{12}$ particles on 4,096 GCDs, while strong scaling keeps the initial global grid fixed and increases only the computational resources.
We measure the execution time for 100 time steps, excluding start-up and finalization time, and repeat each test three times. The parallel efficiency is $N_0T_{N_0}/(N\,T_N)$ in strong and $T_{N_0}/T_N$ in weak scaling tests, where $T_N$ is the time on $N$ GCDs and $N_0$ the baseline.

\begin{figure}[pos=t!]
    \centering
    \subfloat{\includegraphics[width=0.45\linewidth]{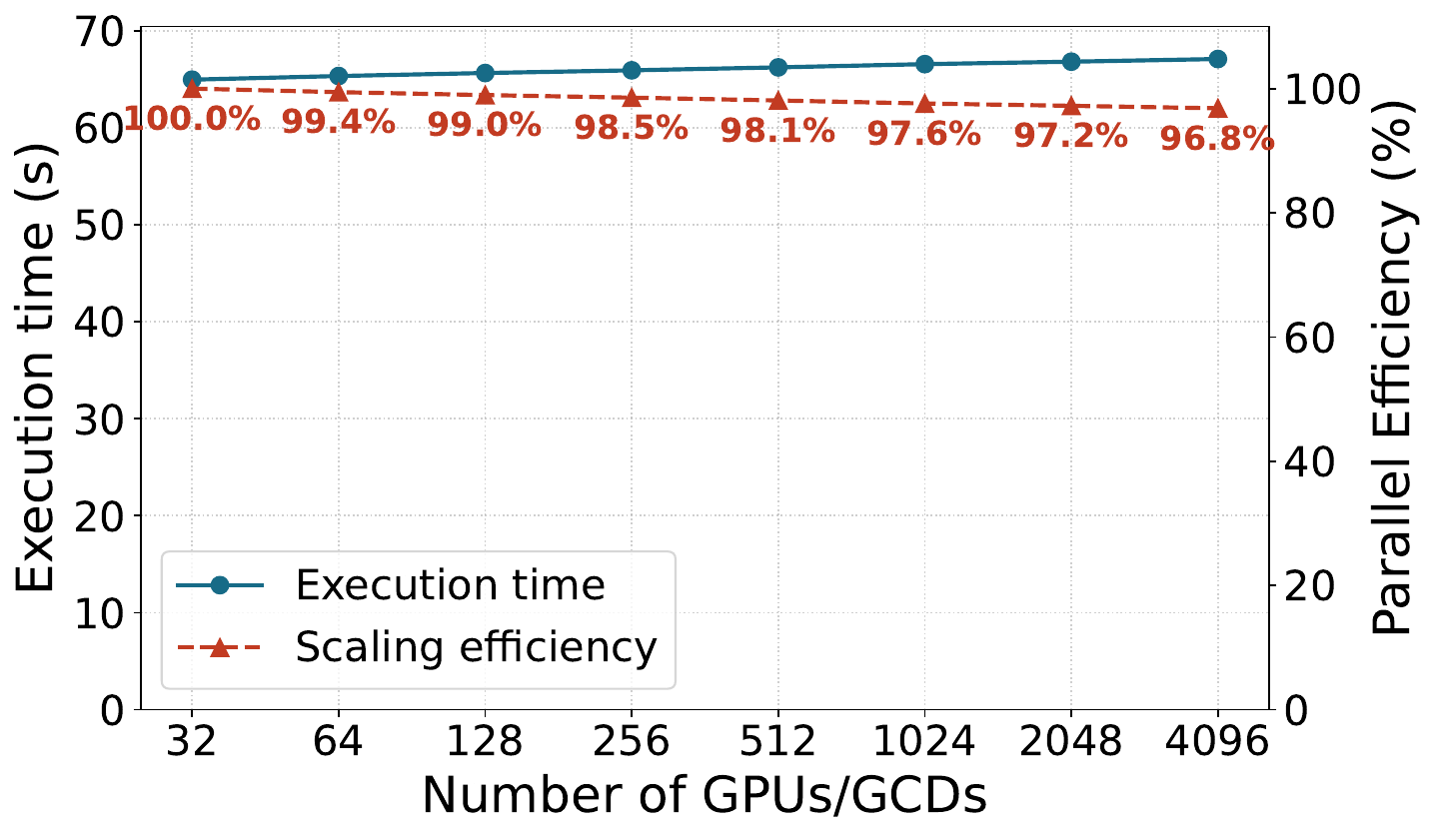}}
    \hspace{4pt}
    \subfloat{\includegraphics[width=0.45\linewidth]{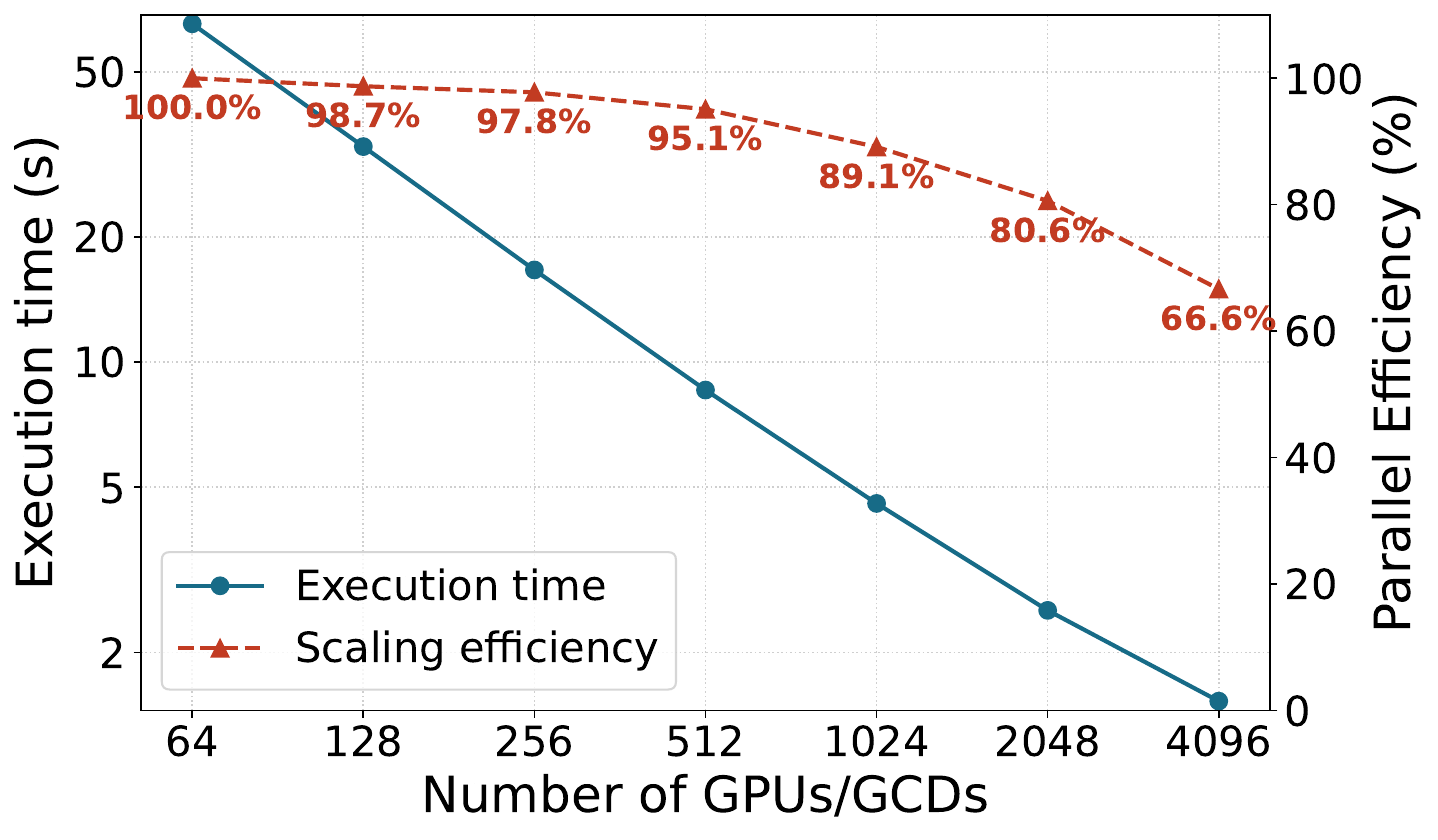}}
    \caption{Weak and strong scaling on LUMI-G, where each node is equipped with $4\times$ AMD MI250x GPUs, for a total of 8 GCDs per node. Wall-clock time of 100 steps (left axis) and parallel efficiency (right axis). Left: weak scaling from 32 to 4,096 GCDs. Right: strong scaling from 64 to 4,096 GCDs, with the time axis in logarithmic scale. Plots report the average time and efficiency measured over three runs per point. Standard deviation is always less than $0.5\%$.}
    \label{fig_scaling}
\end{figure}

Fig.~\ref{fig_scaling} (left) shows the weak scaling results. At constant work per GCD, the execution time per time step increases from 650~ms at 32 GCDs to 671~ms on 4,096 GCDs, i.e., an efficiency of 96.8\%. This 3\% increase is due to communication.

In strong scaling, reported in Fig.~\ref{fig_scaling} (right), the time per step decreases from 652~ms on 64 GCDs to 15.3~ms on 4,096 GCDs, a $42.7\times$ speedup on $64\times$ more GCDs. The efficiency remains above 95\% up to 512 GCDs, with $32^3$ cells per GCD, and drops to 89.1\%, 80.6\%, and 66.6\% at 1024, 2048, and 4,096 GCDs. At 4,096 GCDs, each subdomain holds only $16^3$ cells, with four times the surface-to-volume ratio of the baseline, making the communication cost from halo exchanges, particle migration, and GMRES global reductions more significant.

\begin{table}[pos=t!]
  \centering
  \caption{Baseline of the scaling tests on LUMI-G. Particles per cell (ppc) are given as species $\times$ particles per species.}
  \label{tab_scaling_baseline}
  \small
  \begin{tabular}{@{}l c c c c@{}}
    \toprule
    Test & Cells & ppc & GMRES tol & Subdomains \\
    \midrule
    Weak   & $256\times256\times128$ & $4\times448$ & $10^{-5}$ & $4\times4\times2=32$ \\
    Strong & $256\times256\times256$ & $4\times448$ & $10^{-5}$ & $4\times4\times4=64$ \\
    \bottomrule
  \end{tabular}
\end{table}

\subsection{Mass-matrix deposition on tensor-cores}
\label{sec_perf_tensor_mass}

\begin{figure}
    \centering
    \subfloat{\includegraphics[width=0.42\linewidth]{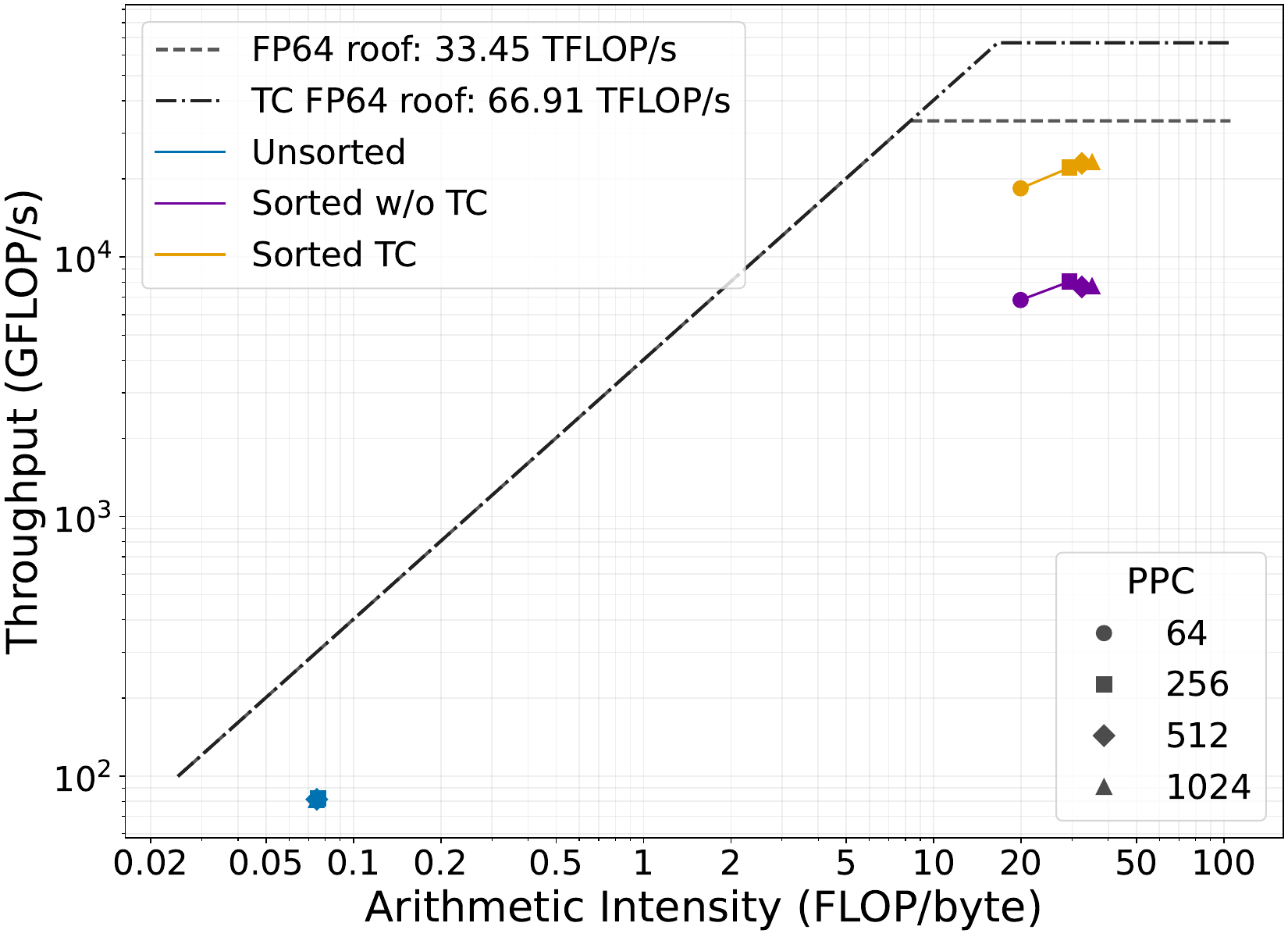}}
    \hspace{4pt}    \subfloat{\includegraphics[width=0.42\linewidth]{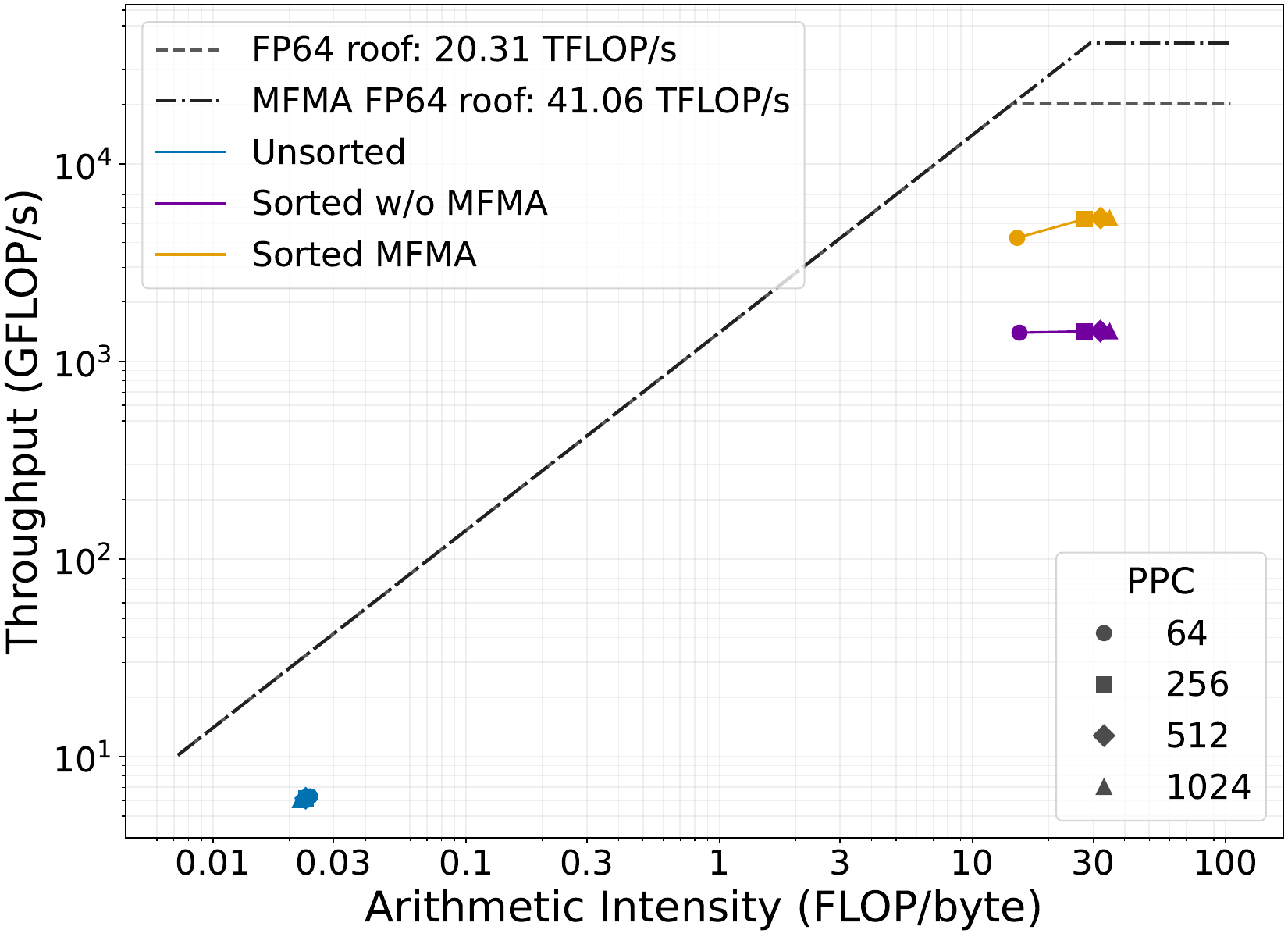}}
    \caption{Roofline analysis of the mass-matrix deposition kernels, including unsorted deposition, sorted w/o matrix engines, and sorted with matrix engines. Left panel: roofline on NVIDIA GH200 GPU; right panel: roofline on AMD MI250x single GCD. Kernels are executed in isolation on a $ 64^3$-cell domain, varying the number of particles per cell from 64 to 1024. Arithmetic intensity and throughput of the tensor core kernel combine FLOPs from both conventional and TC operations, counted with Nsight Compute and rocprof-compute.}
    \label{fig_roof_gh200_mi250x}
\end{figure}

We compare the matrix-engine mass matrix deposition kernel of Section~\ref{sec_tensor_mass_deposition} with two FP64 baselines, the unsorted per-particle kernel with global atomics, and a sorted kernel with vector FMAs. We test varying the number of particles per cell from 64 to 1024, with a fixed $64^3$ cell domain. The two sorted kernels thus execute the same FLOPs and move the same data. We do not include the sorting cost here. Execution time and DRAM traffic are measured on each GPU (Nsight Compute on GH200, rocprof-compute on one MI250x GCD). We always count FLOPs with Nsight Compute on GH200, using only the FP64 vector operations of active threads plus tensor-core operations. We use the same counts for MI250x, whose counters report wavefront instructions and not per-thread operations. FP64 atomic additions in the main memory are treated as data transfer and are not counted as FLOPs. Fig.~\ref{fig_roof_gh200_mi250x} show the rooflines for GH200 and MI250x, Table~\ref{tab_mass_deposition_time} details the execution times.

The unsorted kernel has an arithmetic intensity (AI) of 0.075~FLOP/B on GH200 and 0.02~FLOP/B on MI250x, more than two orders of magnitude below the ridge points (8.3 and 16.6~FLOP/B for the FP64 and tensor-core (TC) roofs on GH200, 14.5 and 29.3~FLOP/B for the FP64 and MFMA roofs on MI250x).
At the same FLOPs, the lower AI on MI250x is due to higher DRAM traffic, consistent with the smaller L2 cache of a GCD (8~MB vs. 60~MB).
The kernel reaches a throughput of 81 and 6~GFLOP/s on the two architectures, respectively. It is memory-bound, but it sustains only 27\% and 19\% of peak bandwidth because atomic throughput limits it.

Sorting reduces the DRAM traffic per particle by $92$-$163\times$ on GH200 and $216$-$548\times$ on MI250x, to about 40~B on both GPUs at 1024~ppc, as the per-cell traffic is amortized with more particles.
The full $8\times8$ tiles increase the FLOPs per particle by $2.9\times$, and the AI grows by more than two orders of magnitude, to 20-35~FLOP/B on GH200 and 15-35~FLOP/B on MI250x. Thus, the sorted kernels move from the memory-bound to the compute-bound region. Without matrix engines, the sorted kernel uses less than 9\% of the DRAM bandwidth and is $29$-$34\times$ (GH200) and $77$-$82\times$ (MI250x) faster than the unsorted kernel.

With the same AI, matrix engines only raise the throughput, executing 83\% of the FLOPs. On GH200, throughput increases from 6.8-8.0 to 18.4-23.2~TFLOP/s (up to 27-35\% of the TC peak), a $2.7$-$3.0\times$ speedup. On MI250x, it increases from 1.4 to 4.2-5.3~TFLOP/s (10-13\% of the MFMA peak), a $3.0$-$3.8\times$ speedup. The gain grows with ppc as the per-cell work outside the matrix engines is amortized. 
Overall, sorting and matrix engines accelerate deposition by $78$-$99\times$ on GH200 and $233$-$306\times$ on MI250x, with the latter suffering particularly from atomic contention.

\begin{table}[t]
  \centering
  \caption{Execution time (ms) of the mass-matrix deposition kernels for different numbers of particles per cell (ppc): unsorted ($t_\mathrm{U}$), sorted without ($t_\mathrm{S}$) and with ($t_\mathrm{M}$) matrix engines, and speedups. Profiled kernel durations, averaged over two runs.}
  \label{tab_mass_deposition_time}
  \small
  \begin{tabular}{@{}l r rrr rrr@{}}
    \toprule
    & & \multicolumn{3}{c}{Time (ms)} & \multicolumn{3}{c}{Speedup} \\
    \cmidrule(lr){3-5} \cmidrule(l){6-8}
    Device & ppc & $t_\mathrm{U}$ & $t_\mathrm{S}$ & $t_\mathrm{M}$
        & $t_\mathrm{U}/t_\mathrm{S}$ & $t_\mathrm{S}/t_\mathrm{M}$ & $t_\mathrm{U}/t_\mathrm{M}$ \\
    \midrule
    \multirow{4}{*}{GH200}
      & 64   &    98.5 &   3.40 &  1.26 & 29.0 & 2.69 &  78.0 \\
      & 256  &   390.8 &  11.56 &  4.21 & 33.8 & 2.75 &  92.9 \\
      & 512  &   789.7 &  24.20 &  8.09 & 32.6 & 2.99 &  97.6 \\
      & 1024 &  1587.7 &  48.06 & 15.98 & 33.0 & 3.01 &  99.3 \\
    \midrule
    \multirow{4}{*}{\shortstack[l]{MI250x\\(1 GCD)}}
      & 64   &  1278.4 &  16.59 &  5.49 & 77.0 & 3.02 & 232.7 \\
      & 256  &  5252.7 &  65.32 & 17.61 & 80.4 & 3.71 & 298.4 \\
      & 512  & 10446.7 & 130.29 & 34.86 & 80.2 & 3.74 & 299.7 \\
      & 1024 & 21306.6 & 261.18 & 69.72 & 81.6 & 3.75 & 305.6 \\
    \bottomrule
  \end{tabular}
\end{table}

\subsection{Trace \& profiling}
\begin{figure}
    \centering
    \includegraphics[width=0.95\linewidth]{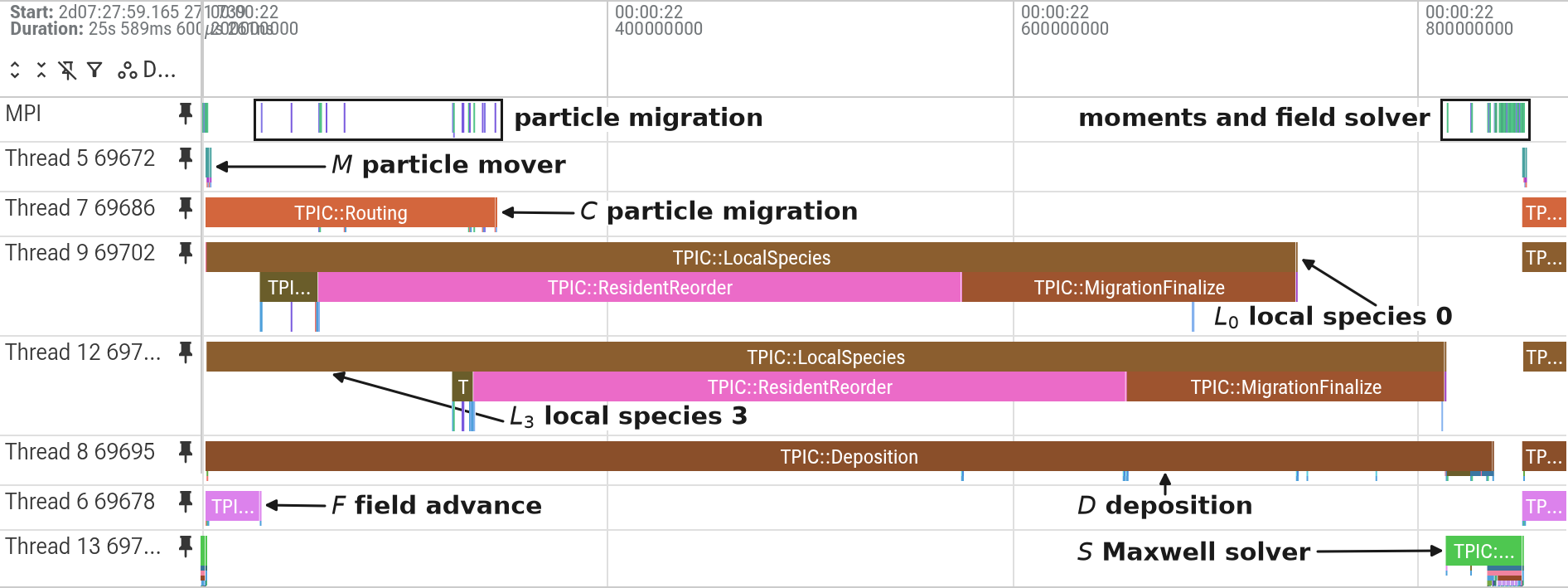}
    \caption{rocprof-sys trace of one time step on rank 0 of the 32-GCD weak scaling baseline, on four LUMI-G nodes. Each row is a host worker with its ROCTX ranges. The local workers $L_1$ and $L_2$ are omitted. The top row collects the MPI calls of all threads, except the polling calls (\texttt{MPI\_Testall}), which rocprof-sys does not record. Boxes mark the particle migration MPI messages and the moment and field solver MPI exchanges. The picture does not report GPU kernels. Given the asynchronous pipeline, host worker times do not reflect the actual GPU compute time.}
    \label{fig_trace}
\end{figure}

We profile the 32-GCD weak scaling baseline (Table~\ref{tab_scaling_baseline}) in full double precision, with GMRES tolerance $10^{-15}$, in three separate runs on LUMI-G to obtain a time breakdown and detailed information on computation and communication patterns. We use rocprof-sys to record a 10 steps trace of every rank. We collect ROCTX ranges marking each TPIC worker, GPU kernels, and MPI calls. rocprofv3 records every kernel dispatch and memory copy. CrayPat traces every MPI call over 50 steps and reports calls, messages, bytes, and time per rank and thread. We collect only data of the simulation main loop, thus excluding simulation startup and finalize. The three runs take about 647~ms per step, similar to the 650~ms measured in the weak scaling test.

The trace in Fig.~\ref{fig_trace} shows one time step of rank 0. Each row shows a host worker of Section~\ref{sec_async_pipeline} submitting, polling, or waiting, while its GPU work runs asynchronously on the device streams. Worker $M$ enqueues the four particle pushes. Worker $C$ routes the migrating particles while $L_0$ and $L_3$ sort their resident particles (\texttt{TPIC::\allowbreak ResidentReorder}). We omit workers $L_1$ and $L_2$ from the pictures for clarity. Each $L_s$ then waits for $D$ to deposit its sorted particles before appending the arrivals (\texttt{TPIC::\allowbreak MigrationFinalize}). Worker $D$ stays active for the whole step. It deposits each species once it is sorted and then assembles $\widehat{\mathbf{J}}$ and $\mathbf{M}$, after which $S$ runs the field solve. In the MPI row, the migration messages overlap the particle sorting, and they end before the first species is fully sorted. The moment halos and the GMRES exchanges close the step.

Table~\ref{tab_time_breakdown} breaks down one time step on rank 0 from the rocprofv3 run. The GPU is busy 95\% of the step on every rank, so the kernels drive the step time. Particle sorting takes half of the step, followed by the mass-matrix deposition (20\%) and the mover (15\%).

Each rank sends 35~MB per step to its neighbors in about 320 point-to-point messages, and 69\% of these bytes belong to the $\widehat{\mathbf{J}}$ and $\mathbf{M}$ halos, as detailed in Table~\ref{tab_comm_breakdown}. Only two exchanges lie on the critical path, the moment assembly and the field solve, which run between the end of the deposition and the next particle push. The GPU idles for 31~ms per step (median over ranks and steps), almost all of it in this window. This idling mainly comes from moment halo communication and GMRES reductions, whose completion depends on the slowest rank. Particle migration, instead, is fully hidden. Worker $C$ routes all species for 138~ms per step while the movers and sorts keep the GPU busy. The GPU idles only 0.05~ms in this interval, and routing completes at least 190~ms before the first species finishes sorting and needs its arrivals.

\begin{table}[pos=t!]
  \centering
\caption{GPU utilization time by kernel per time step on rank 0 of the 32-GCD weak scaling baseline (rocprofv3). Kernels of different streams run concurrently, so their raw durations overlap. Each stage is the duration of its kernels when run alone on the GPU, multiplied by the launches per step. The idle time is the step time minus the GPU busy time. Sorting and deposition differ by less than 2\% between ranks, while the idle time ranges from 25 to 41~ms.}

  \label{tab_time_breakdown}
  \small
  \begin{tabular}{@{}lrr@{}}
    \toprule
    Stage & Time [ms] & Share \\
    \midrule
    Particle sorting (4 species) & 324 & 50.1\% \\
    Mass-matrix deposition (matrix engines) & 131 & 20.2\% \\
    Particle mover & 99 & 15.3\% \\
    Current deposition & 36 & 5.6\% \\
    Field solver, migration kernels, copies & 20 & 3.1\% \\
    GPU idle & 37 & 5.7\% \\
    \midrule
    Time step & 647 & 100\% \\
    \bottomrule
  \end{tabular}
\end{table}

\begin{table}[pos=t!]
  \centering
  \caption{MPI traffic per time step of a rank with 17 neighbor ranks (CrayPat), by exchange and worker.}
  \label{tab_comm_breakdown}
  \small
  \begin{tabular}{@{}llrrc@{}}
    \toprule
    Exchange & Worker & MB & Messages & Critical path \\
    \midrule
    $\widehat{\mathbf{J}}$ and $\mathbf{M}$ halos & $D$ & 26.6 & 49 & yes \\
    GMRES halos and reductions & $S$ & 8.7 & 251 & yes \\
    Particle migration & $C$, $L_s$ & 2.7 & $21+4\times17$ & no \\
    $\mathbf{B}$ halo & $F$ & 0.6 & 17 & no \\
    \bottomrule
  \end{tabular}
\end{table}

\subsection{Communication framework comparison}
\begin{figure}[pos=t]
    \centering
    \includegraphics[width=0.45\linewidth]{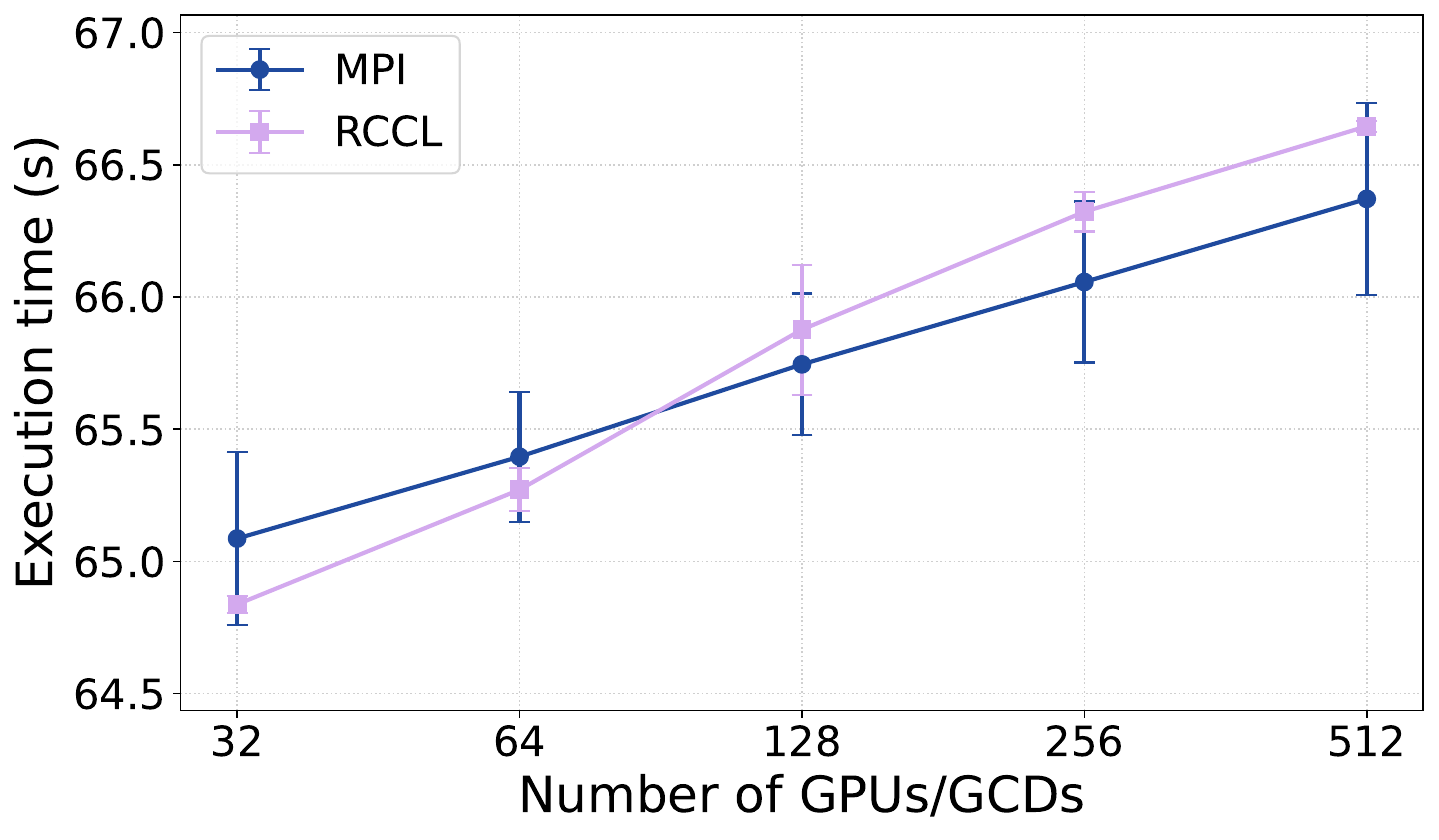}
    \caption{Weak scaling execution time of 100 steps on LUMI-G with device-aware MPI and RCCL communication. Error bars show the standard deviation over runs.}
    \label{fig_mpi_rccl}
\end{figure}

We compare the RCCL and device-aware MPI communication configurations of TPIC on the weak scaling test of Table~\ref{tab_scaling_baseline}, from 32 to 512 GCDs. With RCCL, field halos, mass-matrix exchanges, particle payloads, and all reductions use RCCL. Fig.~\ref{fig_mpi_rccl} shows the execution time, averaged over three runs per point. The two backends differ by less than 1\% at all scales, with RCCL slightly faster up to 64 GCDs, and MPI better at larger scales. Differences for all points are within the statistical standard deviation. From 32 to 512 GCDs, the time increases by about 2\% with MPI and 2.8\% with RCCL, i.e., a weak scaling efficiency of 98\% and 97.3\%. 

\section{Discussion}
\label{sec_discussion}
The ECSIM discretization introduces a spurious term in the Poynting theorem (Eq.~\ref{eq_discrete_poynting}) that hinders exact energy conservation in the case of a closed domain. This is canceled only by combining the zeroed electric components with the ghost-cell parities prescribed by the image field theory in Table~\ref{tab_wall_parities}. With this closure, the discrete Poynting flux vanishes for any combination of periodic, PEC, and PMC boundaries, edges, and corners included. All experiments reported in Section~\ref{sec_numerica_results} confirm that TPIC conserves the total energy to machine precision, including three-dimensional runs in a closed domain with PEC walls. 
Exact energy conservation also removes the finite grid instability, in a cold drifting plasma with an unresolved Debye length, TPIC keeps the initial equilibrium, while the explicit PIC algorithm heats the plasma numerically. The two-stream instability, quasineutral plasma expansion, whistler waves, and Weibel instability tests agree with the corresponding analytical linear and non-linear theories, the whistler turbulence simulation reproduces the electron-scale spectral break and the anisotropic electron heating of~\citet{Gary2012whistlerTurbulence}, and the magnetic reconnection simulation is consistent with the iPIC3D results. TPIC obtains these results with grid spacings larger than the Debye length in all tests and, in the three-dimensional ones, with time steps beyond the explicit CFL limit, while conserving the total energy to machine precision. This reduces the computational cost and keeps long runs stable, without numerical heating or damping. Although ECSIM does not exactly preserve the charge continuity equation, yielding an error in Gauss' law, the experiments show that the error stays bounded and does not significantly affect result accuracy.

Matrix engines combined with particle sorting greatly accelerate the mass-matrix deposition. The unsorted kernel is limited by atomic throughput at an arithmetic intensity below $0.1$~FLOP/B. Sorting and in-register accumulation within each cell turn the scatter into dense $8\times8$ cell-local products, reduce the DRAM traffic per particle by two orders of magnitude, and move the kernel into the compute-bound region. FP64 matrix engines then execute these products $2.7$-$3.0\times$ faster on GH200 and $3.0$-$3.8\times$ faster on MI250x than vector FMAs, for an overall speedup of up to $99\times$ and $306\times$, respectively. The same matrix engine formulation maps onto the different instructions of the two vendors, one $8\times8\times4$ MMA on NVIDIA and the four $4\times4\times4$ blocks of one MFMA on AMD. The measured throughput is below the matrix-engine peaks due to the work executed outside the matrix engines, i.e., the evaluation of $\boldsymbol{\alpha}_p$ and the CIC weights, the operand shuffles, and the final atomic flush.

The TPIC asynchronous pipeline executes particle migration, moment assembly, diagnostics, and output as independent tasks, hiding communication and yielding quasi-optimal scaling, as shown by LUMI-G results. In weak scaling, the time per step increases by only 3\% from 32 to 4,096 GCDs, where TPIC pushes $1.9\times10^{12}$ particles on a $1024^3$ grid. In strong scaling, the efficiency stays above 95\% up to 512 GCDs, and decreases to 66.6\% at 4,096 GCDs, where each subdomain holds $16^3$ cells and a time step takes 15.3~ms. In small subdomains, the calculation-to-communication ratio is less favorable, and latency due to halo exchanges and GMRES global reductions hinders the performance.

Profiling on 32 GCDs shows that TPIC is bound by GPU kernels. The GPU is busy 95\% of the time step, and particle migration is fully hidden. Particle sorting takes half of the step, resulting in the most time consuming operation of the loop. Sorting nevertheless pays off by a large margin, since it enables cell-based mover, current, and mass-matrix kernels that reuse cached cell fields, access memory coalesced, and accumulate without global atomics~\citep{pennati2026massmatrix}. On one MI250x GCD, the unsorted mass-matrix kernel alone takes 10.4~s per species at 512~ppc (Table~\ref{tab_mass_deposition_time}). TPIC instead sorts one species with 448~ppc in 81~ms and deposits its mass matrix with matrix engines in 33~ms.

Neither MPI nor vendor-specific communication backends are uniformly faster. RCCL and device-aware MPI differ by less than 1\% up to 512 GCDs, with RCCL ahead at small scale and MPI at larger scale (Fig.~\ref{fig_mpi_rccl}). TPIC's communication patterns have different requirements, from bandwidth-bound halo, mass-matrix, and particle exchanges to small, latency-bound GMRES reductions on the critical path of the field solve. Since LPFx binds the transport backend to each communication operation independently (Section~\ref{sec_lpfx}), the vendor libraries (NCCL and RCCL) and MPI can be combined within one run. This granularity enables a future investigation of hybrid xCCL+MPI configurations, where the optimal backend is selected for each communication pattern and scale, particularly in the latency-dominated strong-scaling regime.

The mixed-precision infrastructure separates storage and compute types for each quantity, and the precision level is selected at run time among the compiled pairs. Single-precision positions retain exact energy conservation, as shown by the reconnection run. Lower precision for velocities, fields, and moments introduces FP32 round-off in the energy balance, in exchange for halving the memory footprint and traffic of the corresponding arrays. The infrastructure therefore enables a systematic investigation of the accuracy and performance of each precision level. This also concerns the deposition, since most recent matrix engines hardware targets lower-precision formats. Computing the cell-local products in lower precision, or emulating FP64 products on lower-precision matrix engines, could extend the speedup to future architectures, provided that the accuracy required by the energy balance is preserved.

\section{Conclusions}
\label{sec_conclusion}

In this paper, we presented TPIC, a GPU-native implicit PIC code for massive space plasma physics simulations. TPIC is based on ECSIM, and it is built on the LPFx library for portable distributed-memory particle and field data structures and communication on NVIDIA and AMD GPUs. We extended the exact energy conservation of ECSIM from periodic to closed domains, by cancelling the spurious boundary term of the discrete Poynting theorem at PEC and PMC walls. TPIC conserves the total energy to machine precision in all tests, including a three-dimensional magnetic reconnection simulation with PEC walls and $1.23\times10^{10}$ particles, which is also consistent with iPIC3D results. We also verified TPIC on a wide range of space plasma conventional test cases, including Whistler waves, Weibel instability, quasi neutral plasma expansion, and wave turbulence, showing that the simulations match the analytical theory and results previously published in literature.
We executed the ECSIM mass-matrix deposition on the FP64 matrix engines of NVIDIA and AMD GPUs. Combined with particle sorting, the deposition is up to $99\times$ faster on GH200 and $306\times$ faster on MI250x than the unsorted per-particle kernel with global atomics, with the matrix engines contributing a factor $2.7$-$3.8\times$. TPIC scales quasi-optimally on LUMI-G, with a weak scaling efficiency of 96.8\% on 4,096 GCDs, and a strong scaling efficiency above 95\% up to 512 GCDs.

A natural algorithmic extension is to develop open boundary conditions for ECSIM with a consistent discrete energy balance. Additionally, the design of TPIC and LPFx opens three major directions for future investigation. First, particle sorting takes half of the time step, while most particles remain in their cell between steps. Incremental sorting of only the particles that change cell, or more efficient sorting algorithms, could therefore greatly reduce the simulation time. Second, LPFx binds a transport backend to each communication pattern, enabling hybrid configurations of xCCL (NCCL, RCCL) and MPI, where the optimal backend is selected for each pattern. Third, the mixed-precision infrastructure enables a systematic study of reduced precision for particles, fields, and the mass matrix, including lower-precision matrix engines for the deposition.

\section*{Acknowledgments}
This work is funded by the European Union. This work has received funding from the European High Performance Computing Joint Undertaking (JU) and Sweden, Finland, Germany, Greece, France, Slovenia, Spain, and the Czech Republic under grant agreement No. 101093261, Plasma-PEPSC.

\clearpage

\appendix
\makeatletter\renewcommand{\p@section}{Appendix~}\makeatother

\section{Discrete Poynting identity at a wall}
\label{app_discrete_poynting}

Here we derive Eq.~\ref{eq_discrete_poynting} from the stencil of
Eq.~\ref{eq_curl_stencil}. We first prove that $\mathcal{C}^{\dagger}$ is the adjoint of $\mathcal{C}$ on a periodic mesh. Then, reducing the bounded mesh to a periodic one we obtain an exact expression for the discrete Poynting theorem at boundary, valid for arbitrary ghost values and multiple walls in any direction. Evaluating it for walls normal to one direction gives Eq.~\ref{eq_discrete_poynting}.

\paragraph{Mesh and inner products}
Let $e$ denote a direction along which the system is bounded by walls. Nodes are indexed by $i=0,\dots,N_e$, and cell centers by $i+\tfrac12$, with $i=0,\dots,N_e-1$ for the interior cells. Walls are located at the node planes $i=0$ and $i=N_e$, while the ghost cells are at $-\tfrac12$ and $N_e+\tfrac12$. We assume the other directions to be periodic. We denote by $\mathcal{K}$ the interior cells, and by
$\mathcal{G}$ the ghost cells, i.e., the cells outside the domain with at least one vertex in it. We indicate wall nodes by $\partial\Omega$. The nodal control volume is $V_g=V_c\prod_e w^e_g$, with
\begin{equation}
  w^e_g=\begin{cases}1/2, & g\ \text{on a wall normal to}\ e,\\
  1, & \text{otherwise},\end{cases}
  \label{eq_app_weights}
\end{equation}
so that $V_g=V_c/2$, $V_c/4$, and $V_c/8$ at face, edge, and corner nodes.

The cell and nodal inner products are
\begin{equation}  \langle\mathbf{u},\mathbf{v}\rangle_C=V_c\sum_{k\in\mathcal{K}}\mathbf{u}_k\cdot\mathbf{v}_k,
  \qquad
  \langle\mathbf{u},\mathbf{v}\rangle_N=\sum_g V_g\,\mathbf{u}_g\cdot\mathbf{v}_g,
  \label{eq_app_inner_products}
\end{equation}
where $g$ runs over all nodes, wall nodes included.

\paragraph{Operators}
Along each direction $d\in\{x,y,z\}$, with coordinate $\xi$ and mesh spacing $\Delta_d$, we define the central-difference and mean operators
\begin{equation}
  (\delta_d f)(\xi)=f\big(\xi+\tfrac{\Delta_d}{2}\big)-f\big(\xi-\tfrac{\Delta_d}{2}\big),
  \qquad
  (\mu_d f)(\xi)=\tfrac12\Big[f\big(\xi+\tfrac{\Delta_d}{2}\big)+f\big(\xi-\tfrac{\Delta_d}{2}\big)\Big],
  \label{eq_app_operators}
\end{equation}
which map node values to cell values along $d$ and vice versa. Operators along different directions commute, and  Eq.~\ref{eq_curl_stencil} can be written as
\begin{equation}
  \partial^h_x=\frac{\delta_x\mu_y\mu_z}{\Delta x},\qquad
  \partial^h_y=\frac{\mu_x\delta_y\mu_z}{\Delta y},\qquad
  \partial^h_z=\frac{\mu_x\mu_y\delta_z}{\Delta z},
  \label{eq_app_factorization}
\end{equation}
both for the node-to-cell derivatives in $\mathcal{C}$ and for the
cell-to-node derivatives in $\mathcal{C}^{\dagger}$. Thus, it holds
\begin{equation}
  (\mathcal{C}\mathbf{E})_\lambda=\sum_{d,\nu}\epsilon_{\lambda d\nu}\partial^h_dE_\nu,
  \qquad
  (\mathcal{C}^{\dagger}\mathbf{B})_\nu=\sum_{d,\lambda}\epsilon_{\nu d\lambda}\partial^h_dB_\lambda,
  \label{eq_app_curls}
\end{equation}
with $\epsilon_{\lambda d\nu}$ the Levi-Civita symbol.
At the wall nodes, $\mathcal{C}^{\dagger}$ uses the ghost cells.

\paragraph{Adjointness on a periodic mesh}
On a periodic mesh, every direction has as many nodes as cells, and $V_g=V_c$. Here we show that operators $\mathcal{C}$ and $\mathcal{C}^\dagger$ then commute, i.e.
\begin{equation}
  \langle\mathcal{C}\mathbf{E},\mathbf{B}\rangle_C
  =\langle\mathbf{E},\mathcal{C}^{\dagger}\mathbf{B}\rangle_N
  \quad\text{for all }\mathbf{E},\mathbf{B}.
  \label{eq_app_periodic_adjoint}
\end{equation}
Along one periodic direction, let $a$ be node values and $b$ cell values, and let $F_i=a_i\,b_{i-\frac12}$ at the nodes. Expanding
Eq.~\ref{eq_app_operators} gives, at every node $i$,
\begin{equation}
 \begin{aligned}
  b_{i+\frac12}(\delta a)_{i+\frac12}+a_i(\delta b)_i
  &=F_{i+1}-F_i,\\
  b_{i+\frac12}(\mu a)_{i+\frac12}-a_i(\mu b)_i
  &=\tfrac12\big(F_{i+1}-F_i\big).
 \end{aligned}
  \label{eq_app_telescoping}
\end{equation}
Summing over one period direction, the right-hand sides telescope to zero, and
\begin{equation}
  \sum_i b_{i+\frac12}(\delta a)_{i+\frac12}=-\sum_i a_i(\delta b)_i,
  \qquad
  \sum_i b_{i+\frac12}(\mu a)_{i+\frac12}=\sum_i a_i(\mu b)_i,
  \label{eq_app_sbp}
\end{equation}
so $\delta_d$ is skew-adjoint and $\mu_d$ is self-adjoint.

These identities hold with the indices along the other directions fixed,
whatever their staggering. Applying them successively along $x$, $y$, and $z$
and using that the operators commute gives
$\sum_k b_k(\partial^h_da)_k=-\sum_g a_g(\partial^h_db)_g$ for each $d$. With
Eq.~\ref{eq_app_curls} and $\epsilon_{\lambda d\nu}=-\epsilon_{\nu d\lambda}$,
\begin{equation*}
 \begin{aligned}
  \langle\mathcal{C}\mathbf{E},\mathbf{B}\rangle_C
  &=V_c\sum_k\sum_{\lambda,d,\nu}\epsilon_{\lambda d\nu}B_{\lambda,k}(\partial^h_dE_\nu)_k\\
  &=V_c\sum_g\sum_{\lambda,d,\nu}\epsilon_{\nu d\lambda}E_{\nu,g}(\partial^h_dB_\lambda)_g
  =\langle\mathbf{E},\mathcal{C}^{\dagger}\mathbf{B}\rangle_N .
 \end{aligned}
\end{equation*}

\paragraph{Remainder for arbitrary walls} We embed the domain in a periodic mesh with $N_e+2$ cells along each wall direction $e$, so that the ghost layers at opposite walls are distinct. On this
mesh, let $\tilde{\mathbf{E}}$ equal $\mathbf{E}$ at the nodes of the domain and vanish elsewhere, and let $\tilde{\mathbf{B}}$ equal $\mathbf{B}$, ghost values included, on $\mathcal{K}\cup\mathcal{G}$ and vanish elsewhere. These extensions truncate the periodic sums but, by locality of the stencils, leave the curls unchanged where the inner products use them. Specifically, $(\mathcal{C}^{\dagger}\mathbf{B})_g$ involves the cells with vertex $g$ in the domain, i.e. cells in $\mathcal{K}\cup\mathcal{G}$, where by
definition of $\tilde{\mathbf{B}}$, $\mathcal{C}^{\dagger}\tilde{\mathbf{B}}=\mathcal{C}^{\dagger}\mathbf{B}$. Similarly, $(\mathcal{C}\mathbf{E})_k$ involves the vertices of
$k$, which for $k\in\mathcal{K}$ are nodes of the domain, so
$\mathcal{C}\tilde{\mathbf{E}}=\mathcal{C}\mathbf{E}$ on $\mathcal{K}$. On a ghost cell, $\mathcal{C}\tilde{\mathbf{E}}$ involves $\mathbf{E}$ only at the vertices of the cell that lie on the wall.

Applying Eq.~\ref{eq_app_periodic_adjoint} to $\tilde{\mathbf{E}}$ and
$\tilde{\mathbf{B}}$, the cell sum reduces to $\mathcal{K}\cup\mathcal{G}$ and
the node sum, with uniform weights $V_c$, to the nodes of the domain:
\begin{equation}
  \langle\mathcal{C}\tilde{\mathbf{E}},\mathbf{B}\rangle_C = 
  \langle\mathcal{C}\mathbf{E},\mathbf{B}\rangle_C
  +V_c\sum_{k\in\mathcal{G}}\mathbf{B}_k\cdot(\mathcal{C}\tilde{\mathbf{E}})_k
  =\langle\mathbf{E},\mathcal{C}^\dagger\tilde{\mathbf{B}}\rangle_N
  =V_c\sum_g\mathbf{E}_g\cdot(\mathcal{C}^{\dagger}\mathbf{B})_g .
\end{equation}
The inner product $\langle\mathbf{E},\mathcal{C}^\dagger\tilde{\mathbf{B}}\rangle_N$ defined on the periodic domain differs from $\langle\mathbf{E},\mathcal{C}^{\dagger}\mathbf{B}\rangle_N$ only through the
weights of the wall nodes, so
\begin{equation}
    \langle\mathbf{E},\mathcal{C}^\dagger\tilde{\mathbf{B}}\rangle_N - 
    \langle\mathbf{E},\mathcal{C}^{\dagger}\mathbf{B}\rangle_N = \sum_{g\in\partial\Omega}(V_c-V_g)\,\mathbf{E}_g\cdot(\mathcal{C}^{\dagger}\mathbf{B})_g.
    \label{eq_inner_prod_curlB}
\end{equation}
We then express $\langle\mathcal{C}\mathbf{E},\mathbf{B}\rangle_C$ as
\begin{equation}
    \langle\mathcal{C}\mathbf{E},\mathbf{B}\rangle_C = \langle\mathcal{C}\tilde{\mathbf{E}},\mathbf{B}\rangle_C - 
    V_c\sum_{k\in\mathcal{G}}\mathbf{B}_k\cdot(\mathcal{C}\tilde{\mathbf{E}})_k 
    = \langle\mathbf{E},\mathcal{C}^\dagger\tilde{\mathbf{B}}\rangle_N -
    V_c\sum_{k\in\mathcal{G}}\mathbf{B}_k\cdot(\mathcal{C}\tilde{\mathbf{E}})_k.
    \label{eq_inner_prod_curlE}
\end{equation}
Combining Eq.~\ref{eq_inner_prod_curlB} and Eq.~\ref{eq_inner_prod_curlE} we obtain
\begin{equation}
  \langle\mathcal{C}\mathbf{E},\mathbf{B}\rangle_C
  -\langle\mathbf{E},\mathcal{C}^{\dagger}\mathbf{B}\rangle_V
  =\sum_{g\in\partial\Omega}(V_c-V_g)\,\mathbf{E}_g\cdot(\mathcal{C}^{\dagger}\mathbf{B})_g
  -V_c\sum_{k\in\mathcal{G}}\mathbf{B}_k\cdot(\mathcal{C}\tilde{\mathbf{E}})_k.
  \label{eq_app_remainder}
\end{equation}
The first term is the part of the nodal control volume outside the domain, i.e. 
$V_c-V_g=V_c/2$, $3V_c/4$, and $7V_c/8$ at face, edge, and corner nodes, while the
second is the ghost-cell contribution, which the uniform cell sum counts and
$\langle\cdot,\cdot\rangle_C$ does not. No assumption on the ghost values or
on the arrangement of walls has been made. On a periodic mesh, $\partial\Omega$
and $\mathcal{G}$ are empty and Eq.~\ref{eq_app_remainder} reduces to
Eq.~\ref{eq_app_periodic_adjoint}.
Eq.~\ref{eq_app_remainder} is the discrete counterpart of
$\int(\nabla\times\mathbf{E})\cdot\mathbf{B}-\int\mathbf{E}\cdot(\nabla\times\mathbf{B})
=\oint(\mathbf{E}\times\mathbf{B})\cdot\hat{\mathbf{n}}\,dS$.

\paragraph{Evaluation at a wall}
Consider walls normal to $x$, with $\hat{n}_x=-1$ at $i=0$, $\hat{n}_x=+1$ at
$i=N_x$, $\hat{\mathbf{n}}=\hat{n}_x\mathbf{e}_x$, $\Delta_n=\Delta x$, and
$A_g=\Delta y\,\Delta z$. Then $\mathcal{G}$ consists of the two ghost layers
and $V_c-V_g=V_c/2=A_g\Delta_n/2$ at the wall nodes. Let $I$ and $G$ denote the
first interior and the ghost layer, $(I,G)=(\tfrac12,-\tfrac12)$ at $i=0$ and
$(I,G)=(N_x-\tfrac12,N_x+\tfrac12)$ at $i=N_x$. For a cell quantity $h$, the
mean and the jump across the wall,
\begin{equation}
  \overline{h}=\tfrac12\left(h_I+h_G\right),\qquad
  [\![h]\!]=h_I-h_G,
  \label{eq_app_mean_jump}
\end{equation}
are defined at the centers of the wall faces, with the jump taken as interior
minus ghost at both walls. Symbols $\overline{h}_g$, $[\![h]\!]_g$, and $h_{G,g}$
denote $\overline{h}$, $[\![h]\!]$, and $h_G$ interpolated to the wall node $g$
with $\mu_y\mu_z$. The tangential derivatives
$\partial^h_{t,y}=\delta_y\mu_z/\Delta y$ and
$\partial^h_{t,z}=\mu_y\delta_z/\Delta z$ map face values to wall nodes and
back, and
$(\nabla^h_t\times\mathbf{a})_\nu=\sum_{d\in\{y,z\}}\sum_\lambda\epsilon_{\nu d\lambda}\partial^h_{t,d}a_\lambda$.
Since the wall is periodic, the same argument leads to
Eq.~\ref{eq_app_periodic_adjoint}, in two dimensions, for face-centred $\mathbf{a}$ and wall-node $\mathbf{u}$, with both sums over
one wall, gives
\begin{equation}
  \sum_{\text{faces}}\mathbf{a}\cdot(\mu_y\mu_z\mathbf{u})
  =\sum_g(\mu_y\mu_z\mathbf{a})_g\cdot\mathbf{u}_g,
  \qquad
  \sum_{\text{faces}}\mathbf{a}\cdot(\nabla^h_t\times\mathbf{u})
  =\sum_g\mathbf{u}_g\cdot(\nabla^h_t\times\mathbf{a})_g.
  \label{eq_app_tangential_adjoint}
\end{equation}

At a wall node, $\delta_x\mathbf{B}=-\hat{n}_x[\![\mathbf{B}]\!]$ and
$\mu_x\mathbf{B}=\overline{\mathbf{B}}$. On the adjacent ghost layer, where
$\tilde{\mathbf{E}}$ vanishes at the outer nodes,
$\delta_x\tilde{\mathbf{E}}=-\hat{n}_x\mathbf{E}_g$ and
$\mu_x\tilde{\mathbf{E}}=\tfrac12\mathbf{E}_g$. The stencils in
Eq.~\ref{eq_app_remainder} therefore reduce to
\begin{equation}
 \begin{aligned}
  (\mathcal{C}^{\dagger}\mathbf{B})_g
  &=-\frac{1}{\Delta_n}\,\hat{\mathbf{n}}\times[\![\mathbf{B}]\!]_g
  +(\nabla^h_t\times\overline{\mathbf{B}})_g,\\
  (\mathcal{C}\tilde{\mathbf{E}})_G
  &=-\frac{1}{\Delta_n}\,\hat{\mathbf{n}}\times(\mu_y\mu_z\mathbf{E})
  +\frac12\,\nabla^h_t\times\mathbf{E}.
 \end{aligned}
  \label{eq_app_wall_stencils}
\end{equation}
We insert them into Eq.~\ref{eq_app_remainder}, use
$\mathbf{E}\cdot(\hat{\mathbf{n}}\times\mathbf{a})=-(\mathbf{E}\times\mathbf{a})\cdot\hat{\mathbf{n}}$
and
$\mathbf{a}\cdot(\hat{\mathbf{n}}\times\mathbf{E})=(\mathbf{E}\times\mathbf{a})\cdot\hat{\mathbf{n}}$,
and move the tangential operators of the ghost sum from $\mathbf{E}$ to
$\mathbf{B}_G$ with Eq.~\ref{eq_app_tangential_adjoint}:
\begin{equation*}
 \begin{aligned}
  \langle\mathcal{C}\mathbf{E},\mathbf{B}\rangle_C
  -\langle\mathbf{E},\mathcal{C}^{\dagger}\mathbf{B}\rangle_V
  =\sum_{g\in\partial\Omega}A_g\Big[
   &\big(\mathbf{E}_g\times(\tfrac12[\![\mathbf{B}]\!]_g+\mathbf{B}_{G,g})\big)\cdot\hat{\mathbf{n}}\\
   &+\frac{\Delta_n}{2}\,\mathbf{E}_g\cdot
   \big(\nabla^h_t\times(\overline{\mathbf{B}}-\mathbf{B}_G)\big)_g\Big].
 \end{aligned}
\end{equation*}
Since $\tfrac12[\![\mathbf{B}]\!]+\mathbf{B}_G=\overline{\mathbf{B}}$ and
$\overline{\mathbf{B}}-\mathbf{B}_G=\tfrac12[\![\mathbf{B}]\!]$,
\begin{equation*}
  \langle\mathcal{C}\mathbf{E},\mathbf{B}\rangle_C
  -\langle\mathbf{E},\mathcal{C}^{\dagger}\mathbf{B}\rangle_V
  =\sum_{g\in\partial\Omega}A_g\Big[
   (\mathbf{E}_g\times\overline{\mathbf{B}}_g)\cdot\hat{\mathbf{n}}
   +\frac{\Delta_n}{4}\,\mathbf{E}_g\cdot
   \big(\nabla^h_t\times[\![\mathbf{B}]\!]\big)_g\Big],
\end{equation*}
which is Eq.~\ref{eq_discrete_poynting}. With walls in several directions,
Eq.~\ref{eq_app_remainder} remains exact, but Eq.~\ref{eq_app_tangential_adjoint} leaves terms on the boundary of each face, which add to those of the edge and corner nodes and ghosts and do not vanish for arbitrary ghost values. For the TPIC closure, the whole right-hand side of Eq.~\ref{eq_app_remainder} vanishes, so its explicit form is not needed.

\paragraph{Consistency}
Assuming a smooth $\mathbf{B}$ field across the wall, with a Taylor expansion we find $\overline{\mathbf{B}}=\mathbf{B}|_{\partial\Omega}+O(\Delta_n^2)$ and $[\![\mathbf{B}]\!]=-\Delta_n\partial_n\mathbf{B}+O(\Delta_n^3)$.
The first term of Eq.~\ref{eq_discrete_poynting} then converges with second order accuracy to the continuous flux density $(\mathbf{E}\times\mathbf{B})\cdot\hat{\mathbf{n}}$,
while the second is an $O(\Delta_n^2)$ discretization artifact, equal to $-\tfrac{\Delta_n^2}{4}\,\mathbf{E}\cdot(\nabla_t\times\partial_n\mathbf{B})$ to leading order.

\paragraph{TPIC closure}
At a wall normal to $e$, let $p_e=+1$ for PEC and $p_e=-1$ for PMC, and
\begin{equation}
  \sigma_e(B_\lambda)=\begin{cases}-p_e, & \lambda=e,\\
  \phantom{-}p_e, & \lambda\neq e,\end{cases}
  \qquad
  \tau_e(E_\nu)=-\sigma_e(B_\nu).
  \label{eq_app_parities}
\end{equation}
The closure of Table~\ref{tab_wall_parities} sets each ghost value of
$B_\lambda$ to the mirror image of the corresponding interior value times $\sigma_e(B_\lambda)$, with edge and corner ghosts carrying the product of the parities of the adjacent walls, and sets to zero, at the wall nodes, the components $E_\nu$ with $\tau_e(E_\nu)=-1$.


\bibliographystyle{cas-model2-names}

\bibliography{bibliography}

\end{document}